\documentclass[journal,twoside,romanappendices,10pt]{./Templates/IEEEtran/IEEEtran}

\usepackage{cite}\ifCLASSINFOpdf
  \usepackage[pdftex]{graphicx}
\else
\fi

\usepackage[cmex10]{amsmath}
\usepackage{algorithm}
\usepackage{array}

\ifCLASSOPTIONcompsoc
  \usepackage[caption=false,font=normalsize,labelfont=sf,textfont=sf]{subfig}
\else
  \usepackage[caption=false,font=footnotesize]{subfig}
\fi

\usepackage{fixltx2e}
\usepackage{stfloats}
\usepackage{url}

\usepackage{amsfonts, amssymb, amsmath}	
\usepackage{tabularx}					
\usepackage[table,usenames]{xcolor}
\usepackage{xfrac}
\usepackage{paralist}		
\usepackage{breqn}			
\usepackage{multicol}
\usepackage{enumitem}
\usepackage{multirow}

\usepackage{algorithm}
\usepackage{algpseudocode}

\usepackage[normalem]{ulem} 

\algblock{ParFor}{EndParFor}
\algnewcommand\algorithmicparfor{\textbf{parfor}}
\algnewcommand\algorithmicpardo{\textbf{do}}
\algnewcommand\algorithmicendparfor{\textbf{end\ parfor}}
\algrenewtext{ParFor}[1]{\algorithmicparfor\ #1\ \algorithmicpardo}
\algrenewtext{EndParFor}{\algorithmicendparfor}

\usepackage[printonlyused,withpage]{acronym}

\newcommand{\eq}[1]{\eqref{#1}}

\newcommand{\fig}[1]{Fig.~\ref{#1}}

\newcommand{\tabref}[1]{Table~\ref{#1}}
\newcommand{\sect}[1]{Section~\ref{#1}}

\newcommand{\Sect}[1]{Section~\ref{#1}}

\DeclareSymbolFontAlphabet{\mathcalold}{symbols}

\newcommand{\mc}[1]{{\ensuremath{\mathcal{#1}}}}

\newcommand{\mr}[1]{{\ensuremath{\mathrm{#1}}}}
\newcommand{\mb}[1]{{\ensuremath{\mathbf{#1}}}}
\newcommand{\bs}[1]{\ensuremath{\boldsymbol{#1}}}

\newcommand{\defas}{\triangleq}
\newcommand{\mat}[1]{\ensuremath{\bs{\mr{#1}}}}

\newcommand{\vect}[1]{\ensuremath{\mb{#1}}}
\newcommand{\symvec}[1]{\ensuremath{\bs{#1}}}

\def\argmax{\operatornamewithlimits{arg\,max}}

\newlength{\FramedLength}
{\endequation\endminipage\endSbox
\[\fbox{\TheSbox}\]}

{\endalign\endminipage\endSbox
\[\fbox{\TheSbox}\]}

\newcommand{\pdf}[2][p]{\ensuremath{#1\left(#2\right)}}
\newcommand{\cpdf}[3][p]{\ensuremath{\pdf[#1]{\left.#2\,\right|\,#3}}}

\usepackage[space]{grffile} 
\usepackage{lipsum}
\usepackage{diagbox}

\usepackage{xcolor,colortbl}
\definecolor{coral}{rgb}{0.9570,0.5430,0.2969}
\definecolor{kelly}{rgb}{0.7070,0.7734,0.5352}
\definecolor{alice}{rgb}{0.2617,0.6680,0.7852}
\definecolor{darkgrey}{gray}{0.75}
\definecolor{grey}{gray}{0.9}

\begin{document}

\newcommand{\acro}{\acrodef}\newcommand{\acroindefinite}{\acrodefindefinite}\acro{LOCATA}{LOCalization And TrAcking}
\acro{DICIT}{Distant talking Interfaces for Control of Interactive TV}
\acro{ASLT}{Acoustic Source Localization and Tracking}
\acro{MVDR}{Minimum Variance Distortionless Response}
\acro{MUSIC}{MUltiple SIgnal Classification}
\acro{DNN}{Deep Neural Network}
\acro{BSS}{Blind Source Separation}
\acro{DoA}{Direction-of-Arrival}
\acrodefplural{DoA}{Directions-of-Arrival}
\acro{RTF}{Relative Transfer Function}
\acro{GCC}{Generalized Cross-Correlation}
\acro{PHAT}{PHAse Transform}
\acro{TDoA}{Time Delay of Arrival}
\acro{SRP}{Steered Response Power}
\acro{LMS}{Least-Mean-Square}
\acro{TDE}{Time Delay Estimation}
\acro{MCCC}{Multi-Channel Cross-Correlation}
\acro{EVD}{EigenValue Decomposition}
\acro{GEVD}{Generalized \ac{EVD}}
\acro{AIR}{Acoustic Impulse Response}
\acro{EKF}{Extended Kalman Filter}
\acro{PHD}{Probability Hypothesis Density}
\acro{PSD}{Power Spectral Density}
\acro{ILD}{Interaural Level Difference}
\acro{ITD}{Interaural Time Difference}
\acro{IPD}{Interaural Phase Difference}
\acro{HRTF}{Head-Related Transfer Function}
\acro{SSL}{Sound Source Localization}
\acro{EM}{Expectation-Maximization}
\acro{FAR}{False Alarm Rate}
\acro{TFR}{Track Fragmentation Rate}
\acro{TL}{Track Latency}
\acro{OSPA}{Optimal SubPattern Assignment}
\acro{CNN}{Convolutional Neural Network}
\acro{CSTR}{Centre for Speech Technology Research}
\acro{VCTK}{Voice Cloning ToolKit}
\acro{ESPRIT}{Estimation of Signal Parameters via Rotational Invariance Techniques}
\acro{STFT}{Short-Time Fourier Transform}
\acro{CRB}{Cram\'er-Rao Bound}
\acro{ML}{Maximum Likelihood}
\acro{IR}{Infra-Red}
\acro{MSE}{Mean-Square Error}
\acro{ID}{Identity}
\acro{VAD}{Voice Activity Detector}
\acro{VAP}{Voice-Active Period}
\acro{CDR}{Coherent-to-Diffuse Ratio}
\acro{DRR}{Direct-to-Reverberant Ratio}
\acro{ICA}{Independent Component Analysis}
\acro{TSR}{Track Swap Rate}
\acro{ToA}{Time-of-Arrival}
\acro{CPSD}{Cross-Power Spectral Density}
\acro{pdf}{Probability Density Function}
\acro{SMC}{Sequential Monte Carlo}
\acro{MMSE}{Minimum Mean Square Error}
\acro{MAP}{Maximum \emph{a posteriori}}
\acro{SH}{Spherical Harmonic}
\acro{SLAM}{Simultaneous Localization and Mapping}
\acro{JPDA}{Joint Probabilistic Data Association}
\acro{RFS}{Random Finite Set}
\acro{LS}{Least Squares}
\acro{VAST}{Virtual Acoustic Space Traveler}
\acro{SNR}{Signal-to-Noise Ratio}
\acro{PDAF}{Probabilistic Data Association Filter}
\acro{IMM}{Interacting Multiple Model}
\acro{MHT}{Multiple Hypotheses Tracking}
\acro{CTF}{Convolutive Transfer Function}
\acro{GMM}{Gaussian Mixture Model}
\acro{DPD}{Direct Path Dominance}
\acro{GCF}{Generalized Correlation Function}
\acro{DTFT}{Discrete-Time Fourier Transform}
\acro{ToA}{Time-of-Arrival}
\acrodefplural{ToA}{Times-of-Arrival}
\acro{FD}{Frequency-Domain}
\acro{SVD}{Singular Value Decomposition}

\title{The LOCATA Challenge:\\ Acoustic Source Localization and Tracking}
\author{Christine~Evers,~\IEEEmembership{Senior Member,~IEEE,}
        Heinrich~W.~L\"ollmann,~\IEEEmembership{Senior Member,~IEEE,}\\
        Heinrich~Mellmann,
        Alexander Schmidt,~\IEEEmembership{Member,~IEEE,}
        Hendrik Barfuss,\\
        Patrick~A.~Naylor,~\IEEEmembership{Fellow,~IEEE,}
        and~Walter~Kellermann,~\IEEEmembership{Fellow,~IEEE}\vspace{-.5cm}%
\thanks{C. Evers is with the School of Electronics and Computer Science, University of Southampton, SO17 1BJ, UK (e-mail: c.evers@soton.ac.uk).}
\thanks{H. W. L\"ollmann, A. Schmidt, H. Barfuss, and W. Kellermann are with the Chair of Multimedia Communications and Signal Processing, Friedrich-Alexander University Erlangen-N\"urnberg, Erlangen 91058, Germany (e-mail: heinrich.loellmann@fau.de, walter.kellermann@fau.de).}
\thanks{H. Mellmann is with the Institut f\"ur Informatik, Humboldt-Universit\"at zu Berlin, Berlin 10099,
Germany (e-mail: mellmann@informatik.hu-berlin.de).}
\thanks{P. A. Naylor is with the Dept. Electrical and Electronic Engineering, Imperial College London, Exhibition Road, SW7 2AZ, UK (e-mail: p.naylor@imperial.ac.uk).}%
\thanks{The research leading to these results has received funding from the UK EPSRC Fellowship grant no. EP/P001017/1 while C. Evers was with the Dept. Electrical and Electronic Engineering, Imperial College London, UK.}}

\markboth{IEEE/ACM Transactions on Audio, Speech, and Language Processing}%
{Evers \emph{et al.}: The LOCATA Challenge}

\maketitle

\makeatletter
\def\ps@IEEEtitlepagestyle{
  \def\@oddfoot{\mycopyrightnotice}
  \def\@evenfoot{}
}
\def\mycopyrightnotice{
  {\footnotesize
  \begin{minipage}{\textwidth}
  \centering
  This work is licensed under a Creative Commons Attribution 4.0 License. For more information, see https://creativecommons.org/licenses/by/4.0/
  \end{minipage}
  }
}

\begin{abstract}
The ability to localize and track acoustic events is a fundamental prerequisite for equipping machines with the ability to be aware of and engage with humans in their surrounding environment. However, in realistic scenarios, audio signals are adversely affected by reverberation, noise, interference, and periods of speech inactivity. In dynamic scenarios, where the sources and microphone platforms may be moving, the signals are additionally affected by variations in the source-sensor geometries. In practice, approaches to sound source localization and tracking are often impeded by missing estimates of active sources, estimation errors, as well as false estimates. The aim of the LOCAlization and TrAcking (LOCATA) Challenge is an open-access framework for the objective evaluation and benchmarking of broad classes of algorithms for sound source localization and tracking. This paper provides a review of relevant localization and tracking algorithms and, within the context of the existing literature, a detailed evaluation and dissemination of the LOCATA submissions. The evaluation highlights achievements in the field, open challenges, and identifies potential future directions.
\end{abstract}


\IEEEpeerreviewmaketitle

\section{Introduction}
\label{sec:Introduction}
The ability to localize and track acoustic events is a fundamental prerequisite for equipping machines with awareness of their surrounding environment. Source localization provides estimates of positional information, e.g., \acp{DoA} or source-sensor distance, of acoustic sources in scenarios that are either permanently static, or static over finite time intervals. Source tracking extends source localization to dynamic scenarios by exploiting `memory' from information acquired in the past in order to infer the present and predict the future source locations. It is commonly assumed that the sources can be modelled as point sources.

Situational awareness acquired through source localization and tracking benefits applications such as beamforming \cite{vanVeen1988,vanTrees2004,Benesty2008a}, signal extraction based on \ac{BSS} \cite{Parra2002,Zheng2009,Reindl2014,Markovich2017}, automatic speech recognition \cite{Harding2006}, acoustic \ac{SLAM} \cite{Evers2018,Evers2018a}, and motion planning \cite{Harada2014}, with wide impact 
on applications in acoustic scene analysis, including robotics and autonomous systems, smart environments, and hearing aids.

In realistic acoustic environments, reverberation, background noise, interference and source inactivity lead to decreased localization accuracy, as well as 
missed and false detections of acoustic sources. Furthermore, acoustic scenes are often dynamic, involving moving sources, e.g., human talkers, and moving sensors, such as microphone arrays integrated into mobile platforms, such as drones or humanoid robots. Time-varying source-sensor geometries lead to continuous changes in the direct-path contributions of sources, requiring fast updates of localization estimates.

The performance of localization and tracking algorithms is typically evaluated using simulated data generated by means of the image method \cite{Allen1979,Habets2008} or its variants \cite{Jarrett2011}. Evaluation by real-world data is a crucial requirement to assess the relevant performance of localization and tracking algorithms. However, open-access datasets recorded in realistic scenarios and suitable for objective benchmarking are available only for scenarios involving static sources, such as loudspeakers, and static microphone array platforms. 
To provide such data also for a wide range of dynamic scenarios, and thus foster reproducible and comparable research in this area, the \ac{LOCATA} challenge provides a novel framework for evaluation and benchmarking of sound source localization and tracking algorithms, entailing:
\begin{enumerate}
  \item An open-access dataset \cite{LOCATA2019_Corpus_Zenodo} of recordings from four microphone arrays in static and dynamic scenarios, completely annotated with the ground-truth positions and orientations for all sources and sensors, hand-labelled voice activity information, and close-talking microphone signals as reference.
  \item An open-source software framework \cite{LOCATA2019_EvalFramework_Github} of comprehensive evaluation measures for performance evaluation.
  \item Results for all algorithms submitted to the \ac{LOCATA} challenge for benchmarking of future contributions.
\end{enumerate}
The \ac{LOCATA} challenge corpus aims at providing a wide range of scenarios encountered in acoustic signal processing, with an emphasis on speech sources in dynamic scenarios.
The scenarios represent applications in which machines should be equipped with the awareness of the surrounding acoustic environment and the ability to engage with humans, such that the recordings are focused on human speech sources in the acoustic far-field. All recordings contained in the corpus were made in a realistic, reverberant acoustic environment in the presence of ambient noise from a road in front of the building.
The recording equipment was chosen to provide a variety of sensor configurations. The \ac{LOCATA} corpus therefore provides recordings from arrays with diverse apertures. All arrays integrate omnidirectional microphones in a rigid baffle. The majority of arrays use consumer-type low-cost microphones.


The \ac{LOCATA} corpus was previously described in \cite{LOCATA2018a,LOCATA2018b}, and the evaluation measures were detailed in \cite{LOCATA2018c}. This paper provides the following additional and substantial contributions:
\begin{itemize}
  \item A concise, yet comprehensive literature review, providing the background and framing the context of the approaches submitted to the \ac{LOCATA} challenge.
  \item A detailed discussion of the benchmark results submitted to the \ac{LOCATA} challenge, highlighting achievements, open challenges, and potential future directions.
\end{itemize}

This paper is organized as follows: \Sect{sec:Scope} summarizes the scope of the \ac{LOCATA} challenge. \Sect{sec:Tasks} and \sect{sec:Corpus} summarize the \ac{LOCATA} corpus and challenge tasks. \Sect{sec:Literature} reviews the literature on acoustic source localization and tracking in the context of the approaches submitted to the \ac{LOCATA} challenge. \Sect{sec:Measures} details and discusses the evaluation measures. The benchmarked results are presented in \sect{sec:Results}. Conclusions are drawn and future directions discussed in \sect{sec:Conclusions}.

\section{Scope of the \acs{LOCATA} Challenge and Corpus}
\label{sec:Scope}

Evaluation of localization and tracking approaches is often performed in a two-stage process. In the first stage, microphone signals are generated using simulated room impulse responses in order to control parameters, such as the reverberation time, signal-to-noise ratio, or source-sensor geometries. The second stage validates the findings based on measured impulse responses using a typically small number of recordings in real acoustic environments. 

Since the recording and annotation of data is expensive and time-consuming, available open-access recordings are typically targeted at specific scenarios, e.g., for static sources and arrays \cite{Nielsen2014}, or for moving sources \cite{Lathoud2005}. For comparisons of different algorithms across a variety of scenarios, measurement equipment (notably microphone arrays) should be identical, or at least equivalent in all scenarios. In addition, annotation with ground-truth should be based on the same method, especially for assessing tracking performance.

\subsection{Related Challenges \& Corpora}
Previous challenges related to \ac{LOCATA} include, e.g., the CHiME challenges \cite{Barker2015} for speech recognition, the ACE challenge \cite{Eaton2016} for acoustic parameter estimation, and the REVERB challenge \cite{Kinoshita2016} for reverberant speech enhancement. These challenges provide datasets of the clean speech signals and microphone recordings across a variety of scenarios, sound sources, and recording devices. In addition to the audio recordings, accurate ground-truth positional information of the sound sources and microphone arrays are required for source localization and tracking in \ac{LOCATA}.

Available datasets of audio recordings for source localization and tracking are either limited to a single scenario, or are targeted at audio-visual tracking. For example, the SMARD dataset \cite{Nielsen2014} provides audio recordings and the corresponding ground-truth positional information obtained from multiple microphone arrays and loudspeakers in a low-reverberant room \mbox{$(T_{60}\approx 0.15$\,s)}. Only a static single-source scenario is considered, involving microphone arrays and loudspeakers at fixed positions in an acoustically dry enclosure. The DIRHA corpus \cite{Ravanelli2015} provides multichannel recordings for various static source-sensor scenarios in three realistic, acoustic enclosures. 

For dynamic scenarios, corpora targeted at audio-visual tracking, such as the AV16.3 dataset \cite{Lathoud2005}, typically involve multiple moving human talkers. The RAVEL and CAMIL datasets \cite{Alameda2013,Deleforge2011} provide camera and microphone recordings from a rotating robot head. Annotation of the ground-truth source positions is typically performed in a semi-automatic manner, where humans label bounding boxes on small video segments. Therefore, ground-truth source positions are available only as 2D pixel positions, specified relative to the local frame of reference of the camera. For evaluation of acoustic source localization and tracking algorithms, the mapping from the pixel positions to \acp{DoA} or Cartesian positions is required. In practice, this mapping is typically unknown and depends on the specific camera used for the recordings.

For the CLEAR challenge \cite{Stiefelhagen2007}, pixel positions were interpolated between multiple cameras in the environment in order to estimate the Cartesian positions of the sound sources. The CLEAR challenge provided audio-visual recordings from seminars and meetings involving moving talkers. In contrast to LOCATA, which also involves moving microphone arrays, the CLEAR corpus is based on static arrays only.

Infrared tracking systems are used for accurate ground-truth acquisition in \cite{Krindis2005} and by the DREGON dataset \cite{Strauss2018}. However, the dataset in \cite{Krindis2005} provides recordings from only a static, linear microphone array. DREGON is limited to signals emitted by static loudspeakers. Moreover, the microphone array is integrated in a drone, whose self-positions are only known from the motor data and may be affected by drift due to wear of the mechanical parts \cite{Jazar2010}.

\section{\acs{LOCATA} Challenge Tasks}
\label{sec:Tasks}

The scenarios contained in the \ac{LOCATA} challenge corpus are represented by multichannel audio recordings and corresponding positional data. The scenarios were designed to be representative of practical challenges encountered in human-machine interaction, including variation in orientation, position, and speed of the microphone arrays as well as the talkers. Audio signals emitted in enclosed environments are subject to reverberation. Hence, dominant early reflections often cause false detections of source directions, whilst late reverberation, as well as ambient noise, can lead to decreased localization accuracy. Furthermore, temporally sparse or intermittently active sources, e.g., human speakers, result in missing detections during pauses. Meanwhile, interference from competing, concurrent sources requires multi-source localization approaches to ensure that situational awareness can be maintained. In practice, human talkers are directional and highly spatially dynamic, since head and body rotations and translations can lead to significant changes in the talkers’ positions and orientations within short periods of time. The challenge of localization in dynamic scenarios, involving both source and sensor motion, is to
provide accurate estimates for source-sensor geometries that vary significantly over short time frames.

\begin{table}[tb]
  \centering
  \caption{\acs{LOCATA} Challenge Tasks.}
  \label{table:tasks}
	\begin{tabular}{|c||c|c|c|c|}
		\hline
		\multirow{2}{*}{Array} & \multicolumn{2}{c|}{Static Loudspeakers} & \multicolumn{2}{c|}{Moving Human Talkers} \\\cline{2-5}
		& Single & Multiple & Single & Multiple\\\hline\hline
		Fixed & Task 1 & Task 2 & Task 3 & Task 4\\\hline
		Moving & - & - & Task 5 & Task 6\\\hline
	\end{tabular}
\end{table}

\begin{figure*}
  \centering
  \mbox{\subfloat[Robot head]{
    \label{subfig:arrays_robot}
    \includegraphics[width = \textwidth]{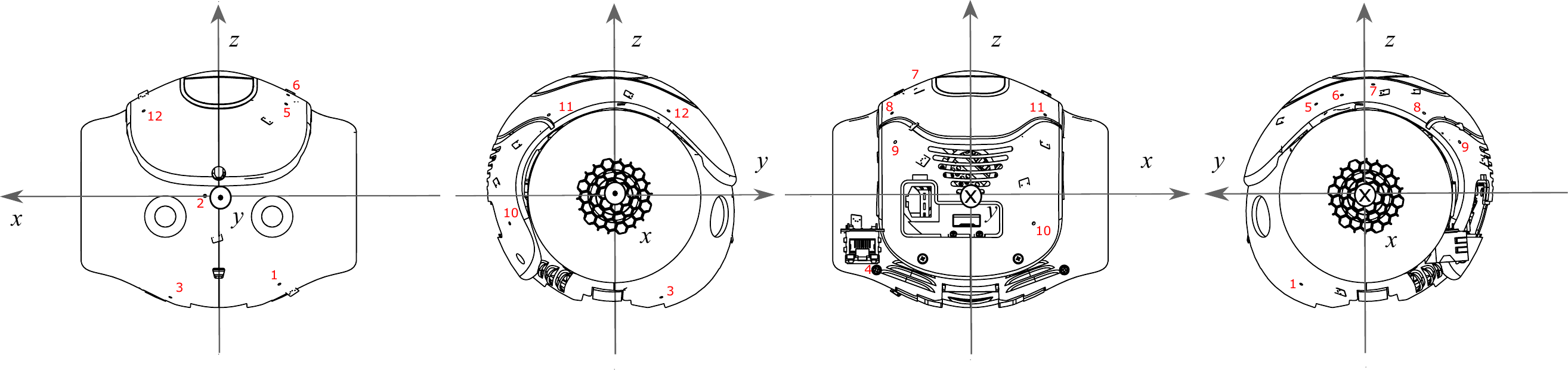}
    }}
    \\
  \mbox{\subfloat[DICIT array]{
    \label{subfig:arrays_dicit}
    \includegraphics[width = .6\textwidth]{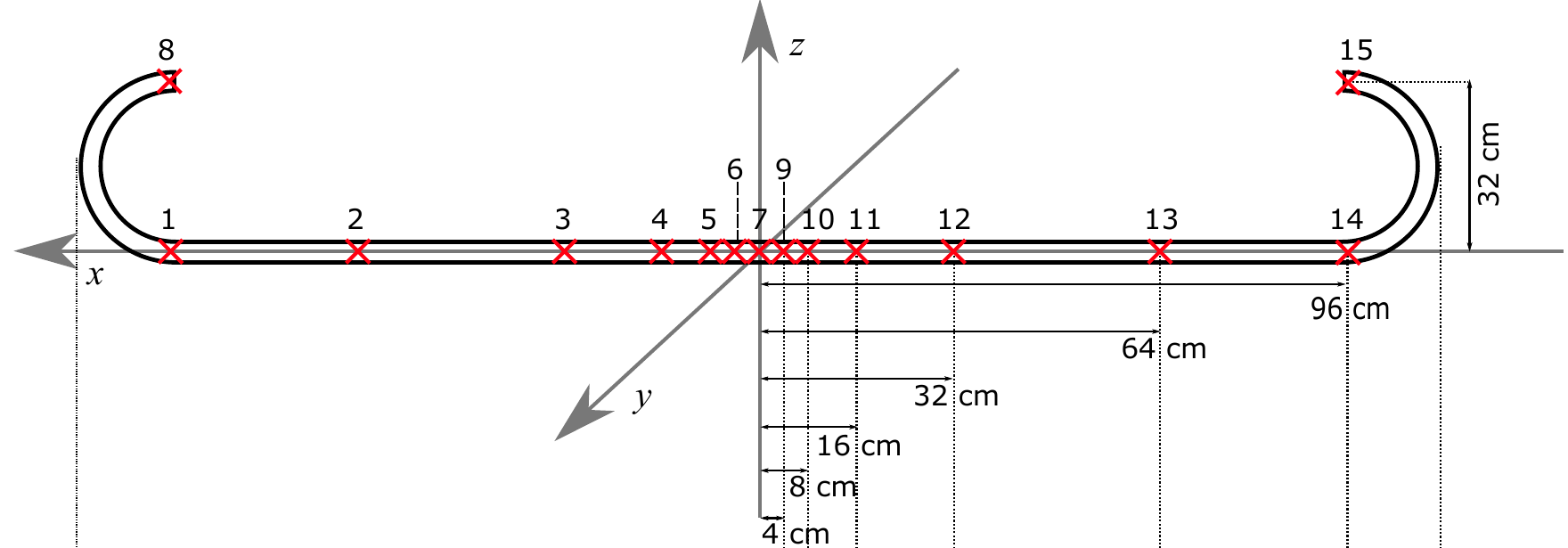}
  }}
  ~
  \mbox{\subfloat[Hearing aids on head-torso simulator]{
    \label{subfig:arrays_hearingaids}
    \includegraphics[width = .28\textwidth]{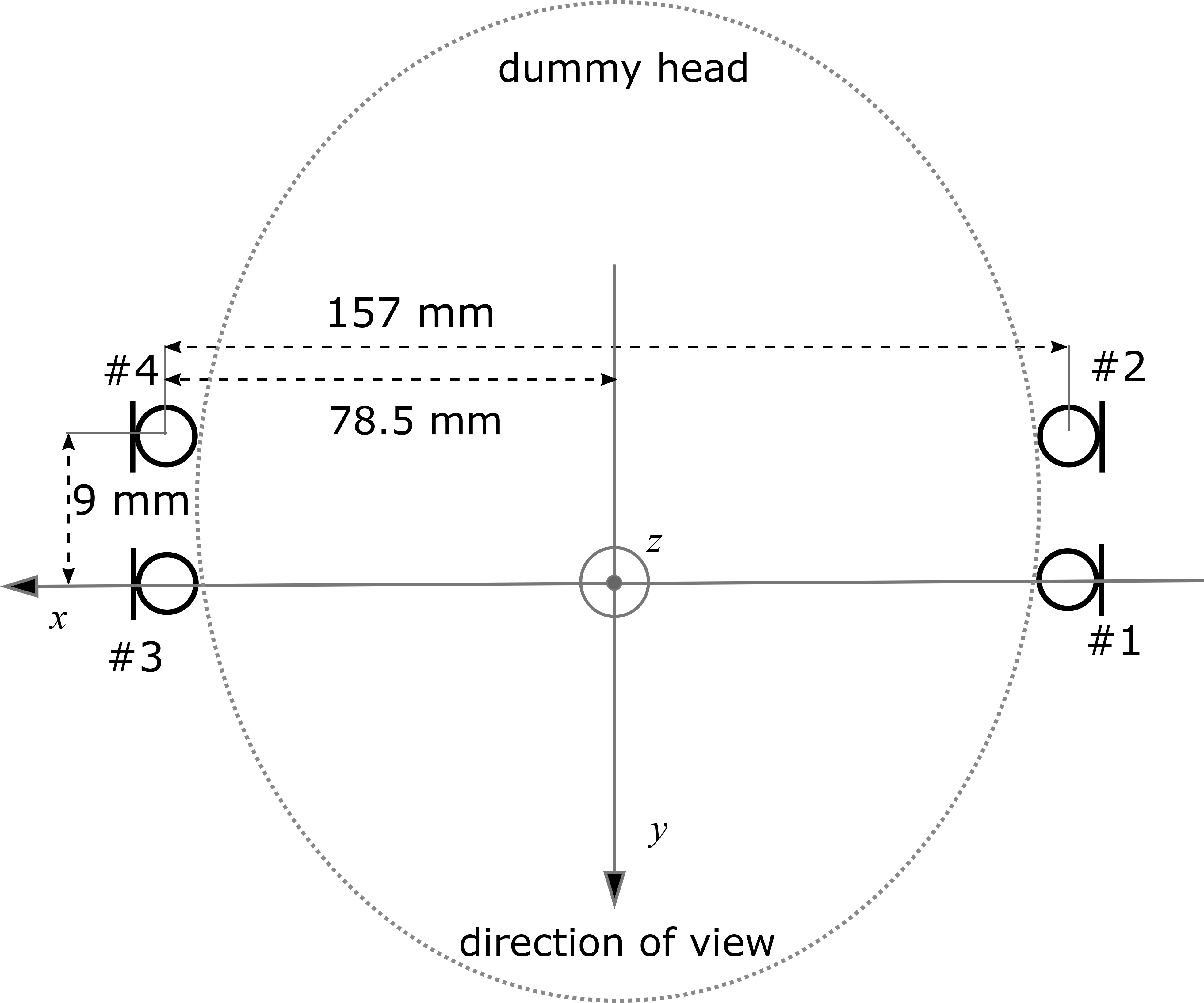}
  }}
  \label{fig:arrays}
  \caption{Schematics of microphone array geometries of (a) the robot head, (b) the DICIT array, (c) the hearing aids  used for the \ac{LOCATA} corpus recordings. Schematics of the Eigenmike can be found in \cite{ManualEigenmike}.}
\end{figure*}

Therefore, machines must be equipped with sound source localization algorithms that prove to be robust against reverberation, noise, interference, and temporal sparsity of sound sources for static as well as time-varying source-sensor geometries. The scenarios covered by the \ac{LOCATA} corpus are therefore aligned with six increasingly challenging tasks, listed in \tabref{table:tasks}.
%

The controlled scenarios of Task~1, involving a single, static sound source, facilitate detailed investigations of the adverse affects of reverberation and noise on source localization. Crucial insights about the robustness against interference and overlapping speech from multiple, simultaneously active sources can be investigated using the static, multi-source scenarios in Task~2. Using the data for Task~3, the impact of source directivity, as well as head and body rotations for human talkers, can be studied. Task~4 provides the recordings necessary to address the ambiguities arising in scenarios involving multiple moving human talkers, such as occlusion and shadowing of crossing talkers, the resolution of individual speakers, and the identification and initialization of new speaker tracks, subject to periods of speech inactivity. The fully dynamic scenarios in Task~5 and Task~6 are designed to bridge the gap between traditional signal processing applications that typically rely on static array platforms, and future directions in signal processing, progressing towards mobile, autonomous systems. Specifically, the data provides the framework required to identify and tackle challenges such as the self-localization of arrays \cite{Evers2018,Evers2018a} and the integration of acoustic data for motion planning \cite{Nguyen2019}.


\section{\acs{LOCATA} Data Corpus}
\label{sec:Corpus}

\subsection{Recording Setup}
The recordings for the \ac{LOCATA} data corpus were conducted in the computing laboratory at the Department of Computer Science at the Humboldt Universit\"at zu Berlin, which is equipped with the optical tracking system OptiTrack \cite{OptTrack2018}. The room size is $7.1 \times 9.8 \times 3$~m$^3$ with a reverberation time of about 0.55\,s.

\subsubsection{Microphone Arrays}
The following four microphone arrays were used for the recordings (see \cite{LOCATA2018b}):
\begin{description}[style=unboxed,leftmargin=0.25cm]
  \item[Robot head:] A pseudo-spherical array with 12 microphones integrated into a prototype head for the humanoid robot NAO (see \fig{subfig:arrays_robot}), developed as part of the EU-funded project `Embodied Audition for Robots (EARS)', \cite{Tourbabin2014a,EARS_TourbabinICR2016}.
  \item[Eigenmike:] The Eigenmike by mh acoustics, which is a spherical microphone array equipped with 32 microphones integrated in a rigid baffle of $84$~mm diameter \cite{ManualEigenmike}.
  \item[\ac{DICIT} array:] A planar array providing a horizontal aperture of width 2.24~m, and sampled by 15 microphones, realizing four nested linear uniform sub-arrays (see \fig{subfig:arrays_dicit}) with inter-microphone distances of 4, 8, 16 and 32\,cm respectively (see also \cite{Brutti2010}).
  \item[Hearing aids:] A pair of non-commercial hearing aids (Siemens Signia, type Pure 7mi) mounted on a head-torso simulator (HMS~II of HeadAcoustics). Each hearing aid (see \fig{subfig:arrays_hearingaids}) is equipped with two microphones (Sonion, type 50GC30-MP2) with an inter-microphone distance of $9$~mm. The Euclidean distance between the hearing aids at the left and right ear of the head-torso simulator is $157$~mm.
\end{description}
The array geometries were selected to sample the diversity of commonly used arrays in a meaningful and representative way. The multichannel audio recordings were performed with a sampling rate of $48$~kHz and synchronized with the ground-truth positional data acquired by the OptiTrack system (see \sect{sec:position_data}).
A detailed description of the array geometries and recording conditions is provided by
\cite{LOCATA2018b}.


\subsection{Speech Material}
For Tasks 1 and 2, involving static sound sources, anechoic utterances from the \ac{CSTR} \ac{VCTK} dataset \cite{VCTK2018} were played back at $48$~kHz sampling rate using Genelec 1029A \& 8020C loudspeakers. For Tasks 3 to 6, involving moving sound sources, 5 non-native human talkers read randomly selected sentences from the \ac{CSTR} \ac{VCTK} dataset. The talkers were equipped with a DPA d:screet SC4060 microphone near their mouth, such that the close-talking speech signals were recorded. The anechoic and close-talking speech signals were provided to participants as part of the development dataset, but were excluded from the evaluation dataset.

\subsection{Ground-Truth Positional Data}
\label{sec:position_data}
For the recordings, a $4 \times 6$~m$^2$ area was chosen within the $7.1 \times 9.8 \times 3$~m$^3$ room. Along the perimeter of the recording area, $10$ synchronized and calibrated \ac{IR} OptiTrack Flex 13 cameras were installed. Groups of reflective markers, detectable by the \ac{IR} sensors, were attached to each source (i.e., loudspeaker or human talker) and microphone array. Each group of markers was arranged with a unique, asymmetric geometry, allowing the OptiTrack system to identify, disambiguate, and determine the orientation of all sources and arrays. 

The OptiTrack system provided estimates of each marker position with approximately $1$~mm accuracy \cite{OptTrack2018} and at a frame rate of $120$~Hz by multilateration using the \ac{IR} cameras. 
Isolated outliers of the marker position estimates, caused by visual occlusions and reflections of the \ac{IR} signals off surfaces, were handled in a post-processing stage that reconstructed missing estimates and interpolated false estimates. Details about the experimental setup are provided in \cite{LOCATA2018b}. 

Audio data was recorded in a block-wise manner and each data block was labeled with a time stamp generated by the global system time of the recording computer. On the the same computer, positional data provided by the OptiTrack system was recorded in parallel. Every position sample was labeled with a time stamp. After each recording was finished, the audio and positional data were synchronized using the time stamps.


For \ac{DoA} estimation, local reference frames were specified relative to each array centre as detailed in \cite{LOCATA2018b}. For convenient transformations of the source coordinates between the global and local reference frames, the corpus provides the translation vectors and rotation matrices for all arrays for each time stamp. Source \acp{DoA} are defined within each array's local reference frame.

\subsection{Voice Activity Labels}
\label{sec:VAD_data}

The \acp{VAP} for the recordings of the \ac{LOCATA} datasets were determined manually using the source signals, i.e., the signals emitted by the loudspeakers (Task~1 and 2) and the close-talking microphone signals (Tasks 3 to 6).
The \ac{VAP} labels for the signals recorded at the distant microphone arrays were obtained from the \ac{VAP} labels for the source signals by accounting for the sound propagation delay between each source and the microphone array as well as the processing delay required to perform the recordings. The propagation delay was determined using the ground-truth positional data. The processing delay was estimated based on the cross-correlation between the source and recorded signals.

The ground-truth \ac{VAP} labels were provided to the participants of the challenge as part of the development dataset but were excluded from the evaluation dataset.



\section{Localization Scenarios, Methods and Submissions}
\label{sec:Literature}
\begin{table*}[tb]
\centering
\caption{Summary of localization and tracking frameworks submitted to the \acs{LOCATA} challenge.}
\label{table:submissions}
\begin{tabular}{|c||c|c||c||c|c||c|c||c|}
\hline
\multirow{2}{*}{ID} & \multirow{2}{*}{Details} & \multirow{2}{*}{Tasks} & \multirow{2}{*}{VAD} & \multicolumn{2}{|c||}{Localization} & \multicolumn{2}{|c||}{Tracking} & \multirow{2}{*}{Arrays} \\
\cline{5-8}
& & & & Algorithm & Section & Algorithm & Section & \\ \hline\hline
 1 & \cite{Agcaer2018} & 1 & - & LDA classification & \ref{sec:literature_localization_multisource_deepLearning} & - & - & Hearing Aids \\\hline
 \multirow{4}{*}{2} & \multirow{4}{*}{\cite{Liu2018}} & \multirow{4}{*}{4} & \multirow{4}{*}{-} & \multirow{4}{*}{\acs{MUSIC}} & \multirow{4}{*}{\ref{sec:literature_localization_multisource_subspace}} & & & Robot Head \\
 & & & & & & Particle PHD filter & \ref{sec:literature_tracking_multisource} & DICIT\\
 & & & & & & + Particle Flow & & Hearing Aids \\
 & & & & & & & & Eigenmike \\ \hline
 3 & \cite{Qian2018} & 1,3,5 & - & GCC-PHAT & \ref{sec:literature_localization_singlesource_TDoA} & Particle filter
 & \ref{sec:literature_tracking_singlesource} & DICIT \\\hline
 \multirow{2}{*}{4} & \multirow{2}{*}{\cite{Li2018}} & \multirow{2}{*}{1-6} & \multirow{2}{*}{Variational EM} & Direct-path RTF + &
  \ref{sec:literature_localization_singlesource_TDoA} & \multirow{2}{*}{Variational EM} & \ref{sec:literature_tracking_multisource} & \multirow{2}{*}{Robot Head}\\
 &&&& GMM &&&& \\ \hline
 \multirow{2}{*}{6} & \multirow{2}{*}{\cite{Lebarbechon2018}} & \multirow{2}{*}{1,3,5} & \multirow{2}{*}{-} & \multirow{2}{*}{SRP-PHAT} &
 \multirow{2}{*}{\ref{sec:literature_localization_singlesource_direct}} & \multirow{2}{*}{-} & - & Eigenmike \\
 & & & & & & & & Robot Head \\\hline
 7 & \cite{Salvati2018a} & 1,3,5 & CPSD trace & SRP Beamformer &
 \ref{sec:literature_localization_singlesource_direct} & Kalman filter &
 \ref{sec:literature_tracking_singlesource} & DICIT \\\hline
 8 & \cite{Mosgaard2018} & 1,3,5 & - & TDE using IPDs & \ref{sec:literature_localization_singlesource_TDoA}, \ref{sec:literature_localization_singlesource_binaural}
  &
 Wrapped Kalman filter & \ref{sec:literature_tracking_singlesource} & Hearing Aids \\\hline
 9 & \cite{Pak2018} & 1 & - & DNN & \ref{sec:literature_localization_multisource_deepLearning} & - & - & DICIT \\ \hline
 \multirow{2}{*}{10} & \multirow{2}{*}{\cite{Kitic2018}} & \multirow{2}{*}{1-4} & \multirow{2}{*}{Noise PSD} & PIVs from &
 \multirow{2}{*}{\ref{sec:literature_localization_singlesource_spherical}} & \multirow{2}{*}{Particle filter}
 & \multirow{2}{*}{\ref{sec:literature_tracking_singlesource}} & \multirow{2}{*}{Eigenmike} \\
 & & & & first-order ambisonics & & & & \\ \hline
 11 & \cite{Madmoni2018} & 1,2 & - & DPD-Test + MUSIC &
 \ref{sec:literature_localization_multisource_subspace} & - & - & Robot Head \\ \hline
 \multirow{2}{*}{12} & \multirow{2}{*}{\cite{Madmoni2018}} & \multirow{2}{*}{1,2} & \multirow{2}{*}{-} & DPD-Test + &
 \ref{sec:literature_localization_multisource_subspace}, & \multirow{2}{*}{-} & \multirow{2}{*}{-} & \multirow{2}{*}{Eigenmike} \\
 & & & & MUSIC in SH-domain & \ref{sec:literature_localization_singlesource_spherical} & & & \\ \hline
 13 & \cite{Nakadai2018} &  1,3 & Zero-crossing rate & MUSIC (SVD) & \ref{sec:literature_localization_multisource_subspace} & Kalman filter &
 \ref{sec:literature_tracking_singlesource} & DICIT \\ \hline
 14 & \cite{Nakadai2018} & 1,3 & Zero-crossing rate & MUSIC (GEVD) & \ref{sec:literature_localization_multisource_subspace} & Kalman filter &
 \ref{sec:literature_tracking_singlesource} & DICIT \\ \hline
 15 & \cite{Moore2018} & 1 & Baseline \cite{Sohn1999} & Subspace PIV & \ref{sec:literature_localization_singlesource_spherical}, \ref{sec:literature_localization_multisource_subspace} & - & - & Eigenmike \\ \hline
 \multirow{2}{*}{16} & \multirow{2}{*}{\cite{Moore2018}} & \multirow{2}{*}{2} & \multirow{2}{*}{Baseline \cite{Sohn1999}} &
Subspace PIV + & \multirow{2}{*}{\ref{sec:literature_localization_singlesource_spherical}, \ref{sec:literature_localization_multisource_subspace}} & \multirow{2}{*}{-} & \multirow{2}{*}{-} & \multirow{2}{*}{Eigenmike} \\
 & & & & Peak Picking &  & & & \\ \hline
\end{tabular}
\end{table*}

Localization systems process the microphone signals either as one batch for offline applications and static source-sensor geometries, or using a sliding window of samples for dynamic scenes. For each window, the instantaneous estimates of the source positions are estimated either directly from the signals, or using spatial cues inferred from the data, such as \acp{TDoA}. To avoid spatial aliasing, nearby microphone pairs or compact arrays are typically used for localization. A few approaches are available to range estimation for acoustic sources, e.g., by exploiting the spatio-temporal diversity of a moving microphone array \cite{Evers2018a,Evers2017c}, or by exploiting characteristics of the room acoustics \cite{Evers2018b,brendel-icassp2018}. Nevertheless, in general, it is typically difficult to obtain reliable range estimates using static arrays. As such, the majority of source localization approaches focus on the estimation of the source \acp{DoA}, rather than the three-dimensional positions. In the following, the term `source localization' will be used synonymously with \ac{DoA} estimation unless otherwise stated.

Due to reverberation, noise, and non-stationarity of the source signals, the position estimates at the output of the localization system are affected by false, missing and spurious estimates, as well as localization errors. Source tracking approaches incorporate spatial information inferred from past observations by applying spatio-temporal models of the source dynamics to obtain smoothed estimates of the source \emph{trajectories} from the instantaneous \ac{DoA} estimates presented by the localization system.\footnote{We note that, within the context of the \ac{LOCATA} challenge, the following discussion focuses on speech, i.e., non-stationary wideband signals corresponding to energy that is concentrated in the lower acoustic frequency bands.}

\begin{figure}[tb]
  \centering
  \includegraphics[width=\columnwidth]{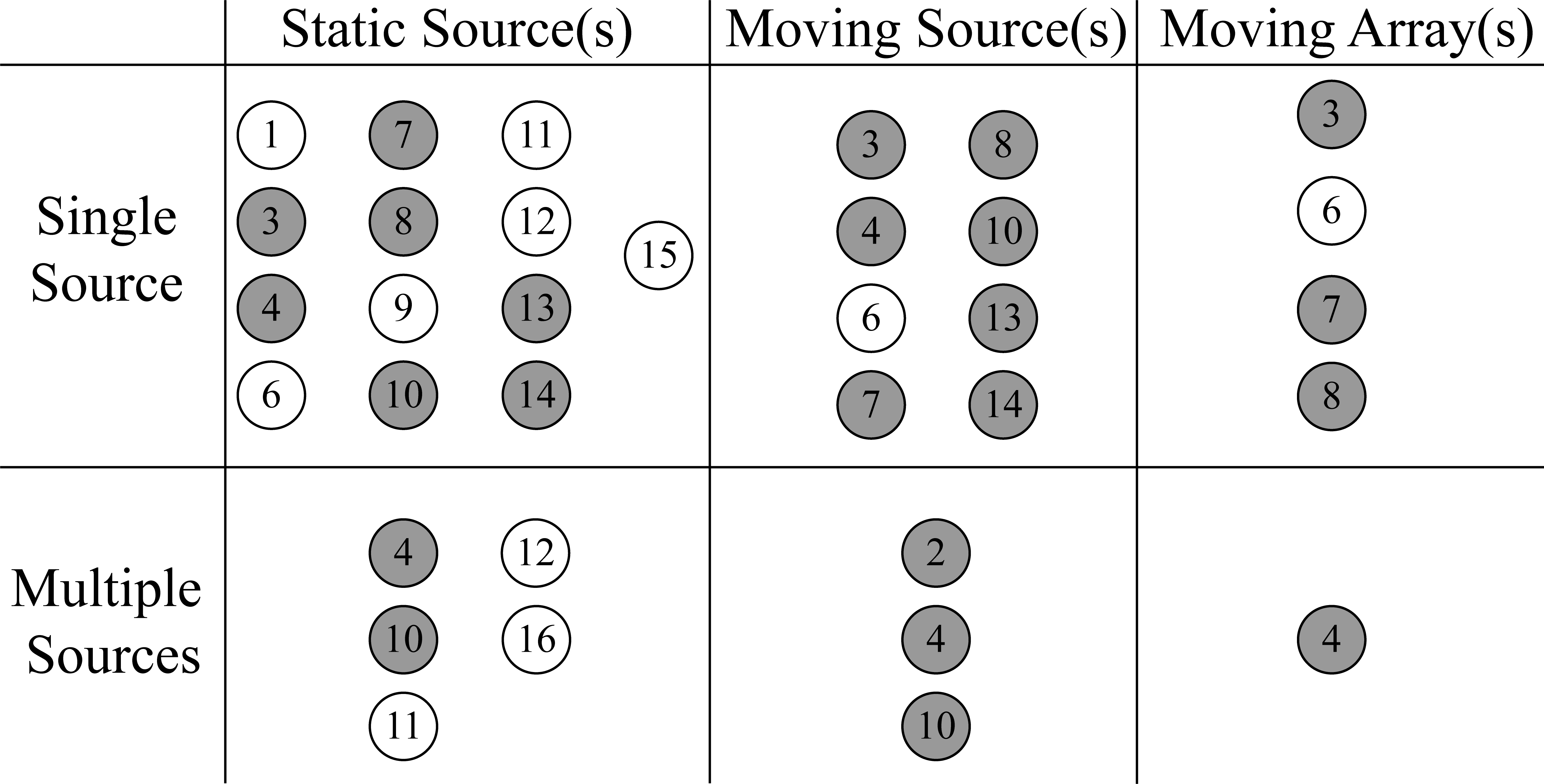}
  \caption{Submissions to the \acs{LOCATA} Challenge, ordered by Challenge Task (see \tabref{table:tasks}). Numbers indicate the submission ID. White shade: approaches incorporating source localization only. Grey shade: Approaches incorporating source localization and tracking.}
  \label{fig:submissions_scenario}
\end{figure}

This section provides the background and context for the approaches submitted to the \ac{LOCATA} challenge so that the submissions can be related to each other and the existing literature in the broad area of acoustic source localization (see \tabref{table:submissions} and \fig{fig:submissions_scenario}). As such, it does not claim the technical depth of surveys like those specifically targeted at sound source localization for robotics, or acoustic sensor networks, e.g., \cite{Argentieri2015,Rascon2017,Cobos2017a}. The structure of the review is aligned with the \ac{LOCATA} challenge tasks as detailed in \sect{sec:Tasks}. Details of each submitted approach are provided in the corresponding \ac{LOCATA} proceedings paper, provided in the references below. Among the 16 submissions to LOCATA, 15 were sufficiently well documented to allow consideration in this paper. 11 were submitted from academic research institutions, 2 from industry, and 2 were collaborations between academia and industry. The global scope of the challenge is reflected by the geographic diversity of the submissions originating from the Asia (3 submissions), Middle East (2 submissions) and Europe (10 submissions).

\subsection{Single-Source Localization}
\label{sec:literature_localization_singlesource}
The following provides a review of approaches for localization of a single, static source, such as a loudspeaker.

\subsubsection{\acl{TDE}}
\label{sec:literature_localization_singlesource_TDoA}
If sufficient characteristics of a source signal are known \emph{a priori}, the time delay between the received signals obtained at spatially diverse microphone positions can be estimated and exploited to triangulate the position of the emitting sound source. \ac{TDE} effectively maximizes the `synchrony' \cite{Souden2010} between time-shifted microphone outputs in order to identify the source position. A brief summary of \ac{TDE} techniques is provided in the following. Details and references can be found in, e.g., \cite[Chap. 9]{Benesty2008a}.

The \ac{TDoA}, $\tau_{m,\ell}(\vect{x}_s)$, of a signal emitted from source position, $\vect{x}_s$, between two microphones, $m$ and $\ell$, at positions $\vect{x}_{m}$ and $\vect{x}_{\ell}$, respectively, is given by:
\begin{align}
 \tau_{m,\ell}(\vect{x}_s) \triangleq \frac{f_s}{c} \left(\| \vect{x}_s - \vect{x}_{m}\| - \|\vect{x}_s - \vect{x}_{\ell}\| \right),
\end{align}
where $f_s$ is the sampling frequency, $c$ is the speed of sound, and $\|\cdot\|$ denotes the Euclidean norm.
If the source signal corresponds to white Gaussian noise and is emitted in an anechoic environment, the \ac{TDoA} between two microphones can be obtained by identifying the peaks in the cross-correlation between microphone pairs.
Since speech signals are often nearly periodic for short intervals, the cross-correlation may exhibit spurious peaks that do not correspond to spatial correlations.
The cross-correlation is therefore typically generalized to include a weighting function in the \ac{DTFT} domain that causes a phase transform to pre-whiten the correlated speech signals, an approach referred to as \ac{GCC}-\ac{PHAT}. The \ac{GCC}, $R_{m,\ell}(\tau)$, is defined as:
\begin{align}
  R_{m,\ell}(\tau) \triangleq \frac{1}{2\pi} \int\limits_{-\pi}^{\pi} \phi_{m,\ell}(e^{\jmath \,\omega})\, S_{m}(e^{\jmath\, \omega})\, S_{\ell}^{\ast}(e^{\jmath\, \omega})\, e^{\jmath\, \omega\, \tau} d\omega,
\end{align}
where $S_{m}(e^{\jmath\, \omega})$ denotes the \ac{DTFT} of the received signal, $s_{m}$, at microphone $m$, and $\ast$ denotes the complex conjugate. The \ac{PHAT} corresponds to a weighting function, $\phi_{m,\ell}(e^{\jmath \,\omega})$, of the \ac{GCC}, where
\begin{align}
  \phi_{m,\ell}(e^{\jmath \,\omega}) \triangleq | S_{m}(e^{\jmath\, \omega})\, S_{\ell}^{\ast}(e^{\jmath\, \omega}) |^{-1}.
\end{align}
The signal models underpinning the \ac{GCC} as well as its alternatives rely on a free-field propagation model of the sound waves. Therefore, in reverberant environments, spectral distortions and temporal correlations due to sound reflections often lead to spurious peaks in the \ac{GCC} function. The presence of multiple, simultaneously active sources can cause severe ambiguities in the distinction of peaks due to the direct path of sources from peaks arising due to reflections.

To explicitly model the reverberant channel, the fact that the \ac{ToA} of the direct-path signal from a source impinging on a microphone corresponds to a dominant peak in the \ac{AIR} can be exploited. The \ac{EVD} \cite{Benesty2000}, realized by, e.g., the gradient-descent constrained \ac{LMS} algorithm, can be applied for estimation of the early part of the relative impulse response.
The work in \cite{Dvorkind2005} extracts the \ac{TDoA} as the main peak in the relative impulse response corresponding to the \ac{RTF} \cite{Gannot2001} for improved robustness against reverberation and stationary noise.
The concept of \acp{RTF} was also used in \cite{Li2016} for a supervised learning approach for \ac{TDoA} estimation.

For localization, it is often desirable to estimate the source directions from \ac{TDoA} estimates, e.g., using multi-dimensional lookup tables \cite{Dmochowski2007}, by triangulation using \ac{LS} optimization if the array geometry is known \emph{a priori} \cite{Berdugo1999,Huang2001}, or by triangulation based on the intersection of interhyperboloidal spatial regions formed by the \ac{TDoA} estimates, e.g., \cite{Cao2017,Sundar2018}.

The following single-source tracking approaches were submitted to the \ac{LOCATA} challenge:
\begin{description}[style=unboxed,leftmargin=0.25cm]
  \item [ID 3 \cite{Qian2018}] combines \ac{TDE} for localization with a particle filter (see \sect{sec:literature_tracking_singlesource}) for tracking using the \ac{DICIT} array for the single-source Tasks 1, 3 and 5.
  \item [ID 4 \cite{Li2018}] combines \ac{DoA} estimation using the direct-path \ac{RTF} approach in \cite{Li2016} with a variational \ac{EM} algorithm \cite{Li2019} (see \sect{sec:literature_tracking_multisource}) for multi-source tracking using the robot head for all Tasks.
  \item [ID 8 \cite{Mosgaard2018}] combines \ac{TDE} (see \sect{sec:literature_localization_singlesource_TDoA}) with binaural features (see \sect{sec:literature_localization_singlesource_binaural}) for localization and applies a wrapped Kalman filter \cite{Traa2013} for source tracking using the hearing aids in the single-source Tasks 1, 3 and 5.
\end{description}

\subsubsection{Binaural Localization}
\label{sec:literature_localization_singlesource_binaural}
The \acp{HRTF} \cite{Blauert1985} at a listener's ears encapsulate spatial cues about the relative source position including \acp{ILD}, \acp{IPD}, and \acp{ITD} \cite{Kuhn1977,Wightman1992,Wang2006}, equivalent to \acp{TDoA}, and are used for source localization in, e.g., \cite{Raspaud2010,Farmani2017,Farmani2018,Benaroya2018,Shashua2005}.

Sources positioned on the `cone of confusion' lead to ambiguous binaural cues that cannot distinguish between sources in the frontal and rear hemisphere of the head \cite{Langendijk2002,Bogaert2011}. Human subjects resolve front-back ambiguities by movements of either their head \cite{Wallach1940,Burger1958,Thurlow1967} or the source controlled by the subject \cite{Wightman1999,Perrett1997}. Changes in \acp{ITD} due to head movements are more significant for accurate localization than changes in \acp{ILD} \cite{Leakey1959}.
In \cite{Hill2000}, the head motion is therefore exploited to resolve front-back ambiguity for localization algorithms. In \cite{Kim2013}, the attenuation effect of an artificial pinna attached to a spherical robot head is exploited in order to identify level differences between signals arriving from the frontal and rear hemisphere of the robot.

The following binaural localization approaches were submitted to the \ac{LOCATA} challenge:
\begin{description}[style=unboxed,leftmargin=0.25cm]
  \item [ID 8 \cite{Mosgaard2018}] combines \ac{TDE} (see \sect{sec:literature_localization_singlesource_TDoA}) with \acp{IPD} for localization and apply a wrapped Kalman filter \cite{Traa2013} (see \sect{sec:literature_tracking_singlesource}) for source tracking using the hearing aids in the single-source Tasks 1, 3 and 5.
\end{description}

\subsubsection{Beamforming and Spotforming}
\label{sec:literature_localization_singlesource_direct}
Beamforming and spotforming techniques can be applied directly to the raw sensor signals in order to `scan' the acoustic environment for positions corresponding to significant sound intensity \cite{Bangs1973,Hahn1973,Wax1983,Taseska2016}.
In \cite{Omologo1998}, a beam is steered in each direction corresponding to a grid, $\mathcal{X}$, of discrete candidate directions. Hence, the \ac{SRP}, $P_{\text{SRP}}(\vect{x}_s)$, is:
\begin{subequations}
  \begin{align}
    P_{\text{SRP}}(\vect{x}) &=\sum\limits_{m=1}^M\sum\limits_{\ell=1}^M R_{m,\ell}(\tau_{m,\ell}(\vect{x}_s)),
  \end{align}
\end{subequations}
where $M$ is the number of microphones. An estimate, $\hat{\vect{x}}_s$, of the source positions is obtained as:
\begin{align}
  \hat{\vect{x}}_s &= \argmax_{\vect{x} \in \mathcal{X}} P_{\text{SRP}}(\vect{x}).
\end{align}
Similar to \ac{GCC}, \ac{SRP} relies on uncorrelated source signals and, hence, may exhibit spurious peaks when evaluated for speech signals. Therefore, \ac{SRP}-\ac{PHAT} \cite{Silvermann2005} applies \ac{PHAT} for pre-whitening of \ac{SRP}.


The following beamforming approaches were submitted to the \ac{LOCATA} challenge:
\begin{description}[style=unboxed,leftmargin=0.25cm]
  \item [ID 6 \cite{Lebarbechon2018}] applies \ac{SRP}-\ac{PHAT} for the single-source Tasks 1, 3, and 5 using the robot head and the Eigenmike.
  \item [ID 7 \cite{Salvati2018a}] combines diagonal unloading beamforming \cite{Salvati2018} for localization with a Kalman filter (see \sect{sec:literature_tracking_singlesource}) for source tracking using a 7-microphone linear subarray of the \ac{DICIT} array for the single-source Tasks 1, 3 and 5.
\end{description}

\subsubsection{Spherical Microphone Arrays}
\label{sec:literature_localization_singlesource_spherical}
Spherical microphone arrays \cite{Rafaely2015} sample the soundfield in three dimensions using microphones that are distributed on the surface of a spherical and typically rigid baffle. The spherical geometry of the array elements facilitates efficient computation based on an orthonormal wavefield decomposition. The response of a spherical microphone array can be described using spherical harmonics \cite{Rafaely2005}. Equivalent to the Fourier series for circular functions, the spherical harmonics form a set of orthonormal basis functions that can be used to represent functions on the surface of a sphere. The sound pressure impinging from the direction, $\mat{\Omega} = \begin{bmatrix} \theta, \phi \end{bmatrix}^T$, on the surface a spherical baffle with radius, $r$, from plane wave with unit amplitude and emitted from the source \ac{DoA}, $\mat{\Phi}_s = \begin{bmatrix} \theta_s, \phi_s\end{bmatrix}^T$, with elevation, $\theta_s$, and azimuth, $\phi_s$, is given by \cite{Rafaely2004}:
\begin{align}
  f_{nm}(k,r,\symvec{\Omega}) &= \sum\limits_{n=0}^{\infty}\sum\limits_{m=-n}^n b_n(kr)\, \left(Y_n^m(\symvec{\Phi})\right)^{\ast}\,Y_n^m(\symvec{\Omega}),
\end{align}
where $k$ is the wavenumber, the weights, $b_n(\cdot)$, are available for many array configurations, and $Y_n^m(\cdot)$ denotes the spherical harmonic of order $n$ and degree $m$.

Therefore, existing approaches to source localization can be extended to the signals in the domain of spherical harmonics. A \ac{MVDR} beamformer \cite{vanTrees2004} is applied for near-field localization in the domain of spherical harmonics in \cite{Kumar2016}.
The work in \cite{Jarrett2011,Jarrett2016} proposes a `pseudo-intensity vector` approach that steers a dipole beamformer along the three principal axes of the coordinate system in order to approximate the sound intensity using the spherical harmonics coefficients obtained from the signals acquired from a spherical microphone array.

The following approaches, targeted at spherical microphone arrays, were submitted to the \ac{LOCATA} challenge:
\begin{description}[style=unboxed,leftmargin=0.25cm]
  \item [ID 10 \cite{Kitic2018}] combines localization using the first-order ambisonics configuration of the Eigenmike with a particle filter (see \sect{sec:literature_tracking_singlesource}) for Tasks 1-4.
  \item [ID 12 \cite{Madmoni2018}] extends \ac{MUSIC} (see \sect{sec:literature_localization_multisource}) to processing in the domain of spherical harmonics of the Eigenmike signals for Tasks 1 and 2.
  \item [ID 15 \cite{Moore2018}] applies the subspace pseudo-intensity approach in \cite{Moore2017} to the Eigenmike signals in the static-source Task 1.
  \item [ID 16 \cite{Moore2018}] extends the approach of ID 15 for the static multi-source Task 2 by incorporating source counting.
\end{description}

\subsection{Multi-Source Localization}
\label{sec:literature_localization_multisource}
This subsection reviews multi-source localization approaches.
Beyond the algorithms submitted to the \ac{LOCATA} challenge, approaches based on, e.g., blind source separation \cite{sawada-ica-ISSPA2003,Lombard2011,Mandel2010,Weiss2011} can be used for multi-source localization.

\subsubsection{Subspace Techniques}
\label{sec:literature_localization_multisource_subspace}
Since spatial cues inferred from the received signals may not be sufficient to resolve between multiple, simultaneously active sources, subspace-based localization techniques rely on diversity between the different sources. Specifically, assuming that the sources are uncorrelated, subspace-based techniques, such as \ac{MUSIC} \cite{Schmidt1986} or \ac{ESPRIT} \cite{Roy1989,Teutsch2005,Teutsch2008} resolve between temporally overlapping signals by mapping the received signal mixture to a space where the source signals lie on orthogonal manifolds.

\ac{MUSIC} \cite{Schmidt1986} exploits the subspace linked to the largest eigenvalues of the correlation matrix to estimate the locations of $N$ sources. The fundamental assumption is that the correlation matrix, $\mat{R}$, of the received signals can be decomposed, e.g., using \ac{SVD} \cite{Golub1965}, into a signal subspace, $\mat{U}_s = \begin{bmatrix} \mat{U}_s^{1}\, \dots, \mat{U}_s^N \end{bmatrix}$, consisting of $N$ uncorrelated plane-wave signals, $\mat{U}_s^{n}$ for $n \in \{1,\dots, N\}$, and an orthogonal noise subspace.
The spatial spectrum from direction, $\mat{\Omega}$, for plane wave, $n \in \{1,\dots,N \}$, is:
\begin{align}
  P_{\text{MUSIC}}(\mat{\Omega}) &= \left( \vect{v}^T(\mat{\Omega}) \left(\mat{I} - \mat{U}_s^{n}\, (\mat{U}_s^n)^H \right) \vect{v}^{\ast}(\mat{\Omega}) \right)^{-1},
\end{align}
where $H$ denotes the Hermitian transpose, $\mat{I}$ denotes the identity matrix, and $\vect{v}$ corresponds to the steering vector.
\ac{MUSIC} extensions to broadband signals, such as speech, can be found in, e.g., \cite{vanTrees2002,Dmochowski2007}. However, the processing of correlated sources remains challenging since highly correlated sources correspond to a rank-deficient correlation matrix, such that the signal and noise space cannot be separated effectively.
This is particularly problematic in realistic acoustic environments, since reverberation corresponds to a convolutive process, in contrast to the additive noise model underpinning \ac{MUSIC}.

For improved robustness in reverberant conditions, \cite{Nadiri2014} introduce a `direct-path dominance' test. The test retains only the time-frequency bins that exhibit contributions of a single source, i.e., whose spatial correlation matrix corresponds to a rank-1 matrix, hence reducing the effects of temporal smearing and spectral correlation induced by reverberation. For improved computational efficiency, \cite{Moore2017} replaces \ac{MUSIC} with the pseudo-intensity approach in \cite{Jarrett2016}.

The following subspace-based localization approaches were submitted to the \ac{LOCATA} challenge:
\begin{description}[style=unboxed,leftmargin=0.25cm]
  \item [ID 2 \cite{Liu2018}] utilizes \ac{DoA} estimates from \ac{MUSIC} as inputs to a \ac{PHD} filter \cite{Ristic2013,Mahler2007} (see \sect{sec:literature_tracking_multisource}) for Task 4, evaluated for all four arrays.
  \item [ID 11 \cite{Madmoni2018}] utilizes the direct-path dominance test \cite{Nadiri2014} and \ac{MUSIC} in the \ac{STFT} domain for the robot head signals for static-source Tasks 1 and 2.
  \item [ID 12 \cite{Madmoni2018}] extends the approach of ID 11 to processing in the domain of spherical harmonics (see \sect{sec:literature_localization_singlesource_spherical}) of the Eigenmike signals for Tasks 1 and 2.
  \item [ID 13 \cite{Nakadai2018}] applies \ac{MUSIC} for localization and a Kalman filter (see \sect{sec:literature_tracking_singlesource}) for tracking to single-source Tasks 1 and 3 using the robot head and the Eigenmike.
  \item [ID 14 \cite{Nakadai2018}] extends the approach of ID 13 to apply the \ac{GEVD} to \ac{MUSIC}.
  \item [ID 15 and 16 \cite{Moore2018}] apply the subspace pseudo-intensity approach in \cite{Moore2017} (see \sect{sec:literature_localization_singlesource_spherical}) to the Eigenmike signals in Tasks 1 and 2, respectively.
\end{description}

\subsubsection{Supervised Learning and Neural Networks}
\label{sec:literature_localization_multisource_deepLearning}
Data-driven approaches can be used to exploit prior information available from large-scale datasets. The work in \cite{Deleforge2012} assumes that frequency-dependent \ac{ILD} and \ac{IPD} values are located on a locally linear manifold. In a supervised learning approach, the mapping between the binaural cues and the source locations is learnt from annotated data using a probabilistic piecewise affine regression model. A semi-supervised approach is proposed in \cite{Laufer2016} that uses \ac{RTF} values input features in order to learn the source locations based on manifold regularization.

To avoid the efforts for hand-crafted signal models, neural network-based ('deep') learning approaches can also be applied to sound source localization. Previous approaches use hand-crafted input vectors including established localization parameters such as \ac{GCC} \cite{Xiao2015,Ferguson2018}, eigenvectors of the spatial coherence matrix
\cite{Takeda2016,Takeda2017} or \acp{ILD} and cross-correlation function in \cite{Ma2017}. \acp{TDoA} were used in, e.g., \cite{Bai2018,Grondin2019}, to reduce the adverse affects of reverberation.
End-to-end learning for given acoustic environments uses either the time-domain signals or the \ac{STFT}-domain signals only as the input for the network.
In \cite{Ma2018}, the \ac{DoA} of a single desired source from a mixture of the desired source and an interferer is estimated by a \ac{DNN} with separate models for the desired source and the interferer. In \cite{Chakrabarty2017b}, \ac{DoA} estimation is considered as a multi-label classification problem, where the range of candidate DoA values is divided into small sectors, each sector representing one class.

The following approaches were submitted to \ac{LOCATA}:
\begin{description}[style=unboxed,leftmargin=0.25cm]
  \item [ID 1 \cite{Agcaer2018}] proposes a classifier based on linear discriminant analysis and trained using features based on the amplitude modulation spectrum of the hearing aid signals for Task 1.
  \item [ID 9 \cite{Pak2018}] uses a \ac{DNN} regression model for localization of the source \ac{DoA} for Task 1 using four microphone signals of the \ac{DICIT} array.
\end{description}

\subsection{Tracking of Moving Sources}
\label{sec:literature_tracking_movingSources}
Source localization approaches provide instantaneous estimates of the source \acp{DoA}, independent of information acquired from past observations. The \ac{DoA} estimates are typically unlabelled and cannot be easily associated with estimates from the past. In order to obtain smoothed source trajectories from the noisy \ac{DoA} estimates, tracking algorithms apply a two-stage process that \begin{inparaenum}[a)] \item predicts potential future source locations based on past information, and \item corrects the localized estimates by trading off the uncertainty in the prediction against the estimation error of the localization system. \end{inparaenum}

\subsubsection{Single-Source Tracking}
\label{sec:literature_tracking_singlesource}
Tracking algorithms based on Bayesian inference aim to estimate the marginal posterior \ac{pdf} of the current state of the source, conditional on the full history of observations. In the context of acoustic tracking, the source state often corresponds to either the Cartesian source position, $\vect{x}(t)$, or the \ac{DoA}, $\mat{\Phi}(t)$, at time stamp, $t$. The state may also contain the source velocity and acceleration. The observations correspond to estimates of either the source position, $\vect{y}(t)$, \acp{TDoA}, $\tau_{m,\ell}(\vect{x}(t))$, or \ac{DoA}, $\symvec{\omega}(t)$ provided by the localization system. Assuming a first-order Markov chain and observations in the form of \acp{DoA}, the posterior \ac{pdf} can be expressed as:
\begin{align}
  \begin{split}
  &\cpdf{\mat{\Phi}(0:t')}{\symvec{\omega}(1:t')} \\
  &= \pdf{\mat{\Phi}(0)}\, \prod\limits_{t=1}^{t'}\cpdf{\mat{\Phi}(t)}{\mat{\Phi}(0:t-1), \symvec{\omega}(1:t)},
  \end{split}
\end{align}
where $\mat{\Phi}(0:t') \defas \begin{bmatrix} \mat{\Phi}^T(0), \dots, \mat{\Phi}^T(t')\end{bmatrix}^T$. Using Bayes's theorem:
\begin{align}
  \label{eqn:bayes}
  \begin{split}
  &\cpdf{\mat{\Phi}(t)}{\mat{\Phi}(0:t-1), \symvec{\omega}(1:t)} \\
  &= \frac{\cpdf{\symvec{\omega}(t)}{\mat{\Phi}(t)}\, \cpdf{\mat{\Phi}(t)}{\mat{\Phi}(t-1)}}{\int\limits_{\mc{P}} \cpdf{\symvec{\omega}(t)}{\mat{\Phi}(t)}\, \cpdf{\mat{\Phi}(t)}{\mat{\Phi}(t-1)} d\mat{\Phi}(t)},
  \end{split}
\end{align}
where $\cpdf{\symvec{\omega}(t)}{\mat{\Phi}(t)}$ is the likelihood function, $\cpdf{\mat{\Phi}(t)}{\mat{\Phi}(t-1)}$ is the prior \ac{pdf}, determined using a dynamical model, and $\mc{P}$ is the support of $\mat{\Phi}(t)$.
For online processing, it is often desirable to estimate sequentially the filtering density, $\cpdf{\mat{\Phi}(t)}{\symvec{\omega}(1:t)}$, instead of \eq{eqn:bayes}.
For linear Gaussian state spaces \cite{Hinrichsen2005}, where the dynamical model and the likelihood function correspond to normal distributions, the filtering density reduces to a Kalman filter \cite{Kalman1960,Ristic2004}.

However, the state space models used for acoustic tracking are typically non-linear and/or non-Gaussian \cite{Evers2018a,Evers2018b}. For example, in \cite{Ward2003,Lehmann2006}, the trajectory of Cartesian source positions is estimated from the \ac{TDoA} estimates. Since the relationship between a source position and the corresponding \acp{TDoA} is non-linear, the integral in \eq{eqn:bayes} is analytically intractable. The particle filter is a widely used sequential Monte Carlo method \cite{Doucet2001} that approximates the intractable posterior \ac{pdf} by importance sampling of a large number of random variates, $\{\symvec{\hat{\phi}}^{(i)}(t)\}_{i=1}^{I}$, - or `particles' -,  from a proposal distribution, $\cpdf[g]{\mat{\Phi}(t)}{\mat{\Phi}(0:t-1), \symvec{\omega}(1:t)}$, i.e.,
\begin{align}
  \cpdf{\mat{\Phi}(t)}{\mat{\Phi}(0:t-1), \symvec{\omega}(1:t)}  \approx \sum\limits_{i=1}^I w^{(i)}(t)\, \delta_{\hat{\mat{\Phi}}^{(i)}(t)}(\mat{\Phi}(t)),
\end{align}
where $\delta$ denotes the Dirac measure, and the importance weights, $w^{(i)}(t)$, are given by:
\begin{align}
  \begin{split}
  &w^{(i)}(t) = w^{(i)}(t-1)\, \frac{\cpdf{\symvec{\omega}(t)}{\mat{\Phi}(t)}\, \cpdf{\mat{\Phi}(t)}{\mat{\Phi}(t-1)}}{\cpdf[g]{\mat{\Phi}(t)}{\mat{\Phi}(0:t-1), \symvec{\omega}(1:t)}}.
  \end{split}
\end{align}
The authors of \cite{Ward2003,Lehmann2006} rely on  prior importance sampling \cite{Doucet2000} from the prior \ac{pdf}. Each resulting particle is assigned a probabilistic weight, evaluated using the likelihood function of the \acp{TDoA} estimates. The work in \cite{Fallon2012} uses the \ac{SRP} function instead of \ac{TDoA} estimates as observations.
Rao-Blackwellized particle filters \cite{Li2004} are applied in \cite{Zhong2014,Levy2011} instead of prior importance sampling.
Resampling algorithms \cite{Li2015,Arulampalam2002,Bolic2003,Halimeh2018,Halimeh2018a} ensure that only stochastically relevant particles are retained and propagated in time.

The tracking accuracy is highly dependent on the specific algorithm used for localization. Moreover, tracking approaches that rely on \ac{TDoA} estimates are crucially dependent on accurate calibration \cite{Plinge2016} and synchronization \cite{Cherkassky2017}. To relax the dependency on calibration and synchronization, \ac{DoA} estimates can be used as observations instead of \ac{TDoA} estimates. To appropriately address the resulting non-Gaussian state-space model, a wrapped Kalman filter is proposed in \cite{Traa2013} that approximates the posterior \ac{pdf} of the source directions by a Gaussian mixture model, where the mixture components account for the various hypotheses that the state at the previous time step, the predicted state at the current time step, or the localized \ac{DoA} estimate may be wrapped around $\pi$. To avoid an exponential explosion of the number of mixture components, mixture reduction techniques \cite{Salmond2009} are required.

Rather than approximating the angular distribution by a Gaussian mixture, a von Mises filter, based on directional statistics \cite{Mardia2009, Mardia2010}, is proposed in \cite{Evers2018b}. The \ac{CDR} \cite{Jeub2011,Braun2018} is used as a measure of reliability of the \ac{DoA} estimates in order to infer the unmeasured source-to-sensor range.

The following single-source tracking approaches were submitted to the \ac{LOCATA} challenge:
\begin{description}[style=unboxed,leftmargin=0.25cm]
  \item [ID 3 \cite{Qian2018}] combines \ac{TDE} (see \sect{sec:literature_localization_singlesource_TDoA}) for localization with a particle filter for tracking using the \ac{DICIT} array for the single-source Tasks 1, 3 and 5.
  \item [ID 7 \cite{Salvati2018a}] combines diagonal unloading beamforming \cite{Salvati2018} (see \sect{sec:literature_localization_singlesource_direct}) for localization with a Kalman filter for source tracking using a 7-microphone linear subarray of the \ac{DICIT} array for Tasks 1, 3 and 5.
  \item [ID 8 \cite{Mosgaard2018}] combines \ac{TDE} (see \sect{sec:literature_localization_singlesource_TDoA}) with \acp{IPD} (see \sect{sec:literature_localization_singlesource_binaural}) for localization and apply a wrapped Kalman filter \cite{Traa2013} for source tracking using the hearing aids for Tasks 1, 3 and 5.
  \item [ID 10 \cite{Kitic2018}] combines localization using the first-order ambisonics configuration (see \sect{sec:literature_localization_singlesource_spherical}) of the Eigenmike with a particle filter for Tasks 1-4.
  \item [ID 13 and ID 14 \cite{Nakadai2018}] apply variants of \ac{MUSIC} (see \sect{sec:literature_localization_multisource_subspace}) for localization and a Kalman filter for tracking the source \acp{DoA} for Tasks 1 and 3 using the robot head and the Eigenmike.
\end{description}

\subsubsection{Multi-Source Tracking}
\label{sec:literature_tracking_multisource}
For multiple sources, not only the source position, but also the number of sources is subject to uncertainty. However, this uncertainty cannot be accounted for within the classical Bayesian framework.

Heuristic data association techniques are often used to associate existing tracks and observations, as well as to initialize new tracks.
Data association partitions the observations into track `gates' \cite{Blackman1990}, or collars, around each predicted track in order to eliminate unlikely observation-to-track pairs. Only observations within the collar are considered when evaluating the track-to-observation correlations. Nearest-neighbour approaches determine a unique assignment between each observation and at most one track by minimizing an overall distance metric. However, in dense, acoustic environments, such as the cocktail party scenario \cite{Cherry1953,Haykin2005}, many pairs between tracks and observations may result in similar distance values, and hence a high probability of association errors.
For improved robustness, probabilistic data association can be used instead of heuristic gating procedures, e.g., the \ac{PDAF} \cite{BarShalom1975,BarShalom2009}, or \ac{JPDA} \cite{Fortmann1983,Gehrig2007}.

To avoid explicit data association, the work in \cite{Li2019} models the observation-to-track associations as discrete latent variables within a variational \ac{EM} approach for multi-source tracking. Estimates of the latent variables provide the track-to-observation associations. The work in \cite{Ban2019} extends the variational \ac{EM} in \cite{Li2019} to a incorporate a von Mises distribution \cite{Evers2018b} for robust estimation of the \ac{DoA} trajectories.

To incorporate track initiation and termination in the presence of false and missing observations, the states of multiple sources can be formulated as realizations of a \ac{RFS} \cite{Mahler2007,Mahler2014}. In contrast to random vectors, \acp{RFS} capture not only the time-varying source states, but also the unknown and time-varying number of sources. Finite set statistics \cite{Mahler2004,Mahler2013} provide the mathematical mechanisms to treat \acp{RFS} within the Bayesian paradigm. Since the \ac{pdf} of \ac{RFS} realizations is combinatorially intractable, its first-order approximation, the \ac{PHD} filter \cite{Mahler2007} provides estimates of the intensity function -- as opposed to the \ac{pdf} -- of the number of sources and their states.

The \ac{PHD} filter was applied in \cite{Ma2006,Pessentheiner2017} for the tracking of the positions of multiple sources from the \ac{TDoA} estimates. Due to the non-linear relationship between the Cartesian source positions and \acp{TDoA} estimates, the prediction and update for each hypothesis within the \ac{PHD} filter is realized using a particle filter as previously detailed in \sect{sec:literature_tracking_singlesource}. A \ac{PHD} filter for bearing-only tracking from the localized \ac{DoA} estimates was proposed in \cite{Markovic2015}, incorporating a von Mises mixture filter for the update of the source directions. The work in \cite{Evers2018a,Evers2018} applies a \ac{PHD} filter in order to track the source positions from \ac{DoA} estimates for \ac{SLAM}.

The following multi-source tracking approaches were submitted to the \ac{LOCATA} challenge:
\begin{description}[style=unboxed,leftmargin=0.25cm]
  \item [ID 2 \cite{Liu2018}] utilizes \ac{DoA} estimates from \ac{MUSIC} (see \sect{sec:literature_localization_multisource_subspace}) as inputs to a \ac{PHD} filter \cite{Ristic2013,Mahler2007} with intensity particle flow \cite{Liu2017} for Task 4, using all four arrays.
  \item [ID 4 \cite{Li2018}] combines \ac{DoA} estimation using the direct-path \ac{RTF} approach in \cite{Li2016} (see \sect{sec:literature_localization_singlesource_TDoA}) with the variational \ac{EM} algorithm in \cite{Li2019} for all Tasks using the robot head.
\end{description}


\section{Evaluation Measures}
\label{sec:Measures}
This section provides a discussion of the performance measures used for evaluation of the \ac{LOCATA} challenge.

\begin{figure}[tb]
  \centering
  \includegraphics[width=.9\columnwidth]{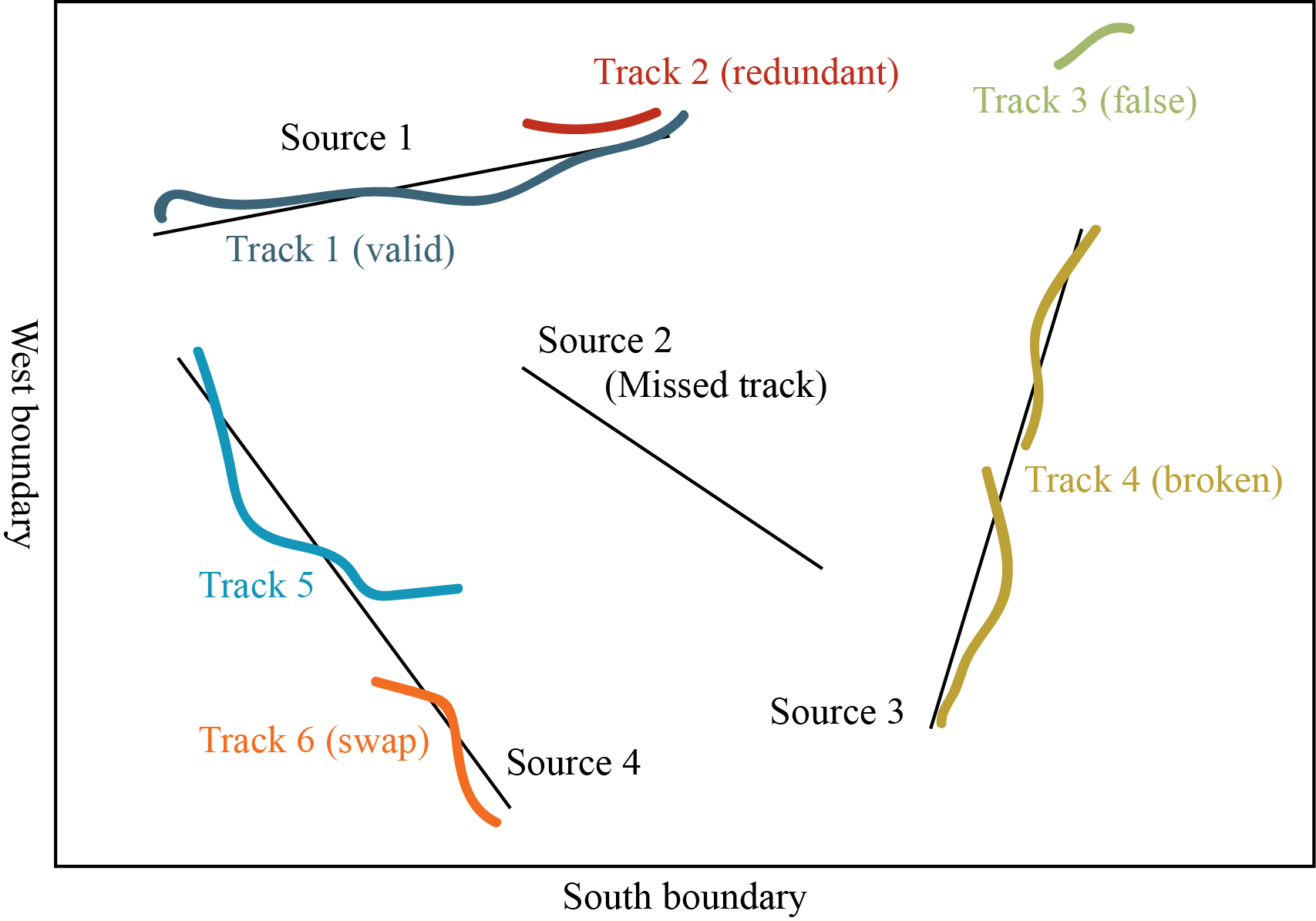}
  \caption{Tracking ambiguities. Colors indicate unique track IDs.}
  \label{fig:illustration_ambiguities}
\end{figure}

\subsection{Source Localization \& Tracking Challenges}
\label{sec:measures_challenges}
In realistic acoustic scenarios, source localization algorithms are affected by a variety of challenges (see \fig{fig:illustration_ambiguities}). Fast localization estimates using a small number of time frames often result in estimation errors for signals that are affected by late reverberation and noise. Sources are often missed, e.g., due to periods of voice inactivity, for distant sources corresponding to low signals levels, or for sources oriented away from the sensors. False estimates arise due to, e.g., strong early reflections mistaken as the direct path of a source signal, or reverberation causing temporal smearing of speech energy beyond the endpoint of a talker's utterance, and due to overlapping speech energy in the same spectral bins for multiple, simultaneously active talkers.

Source tracking algorithms typically use localization estimates as observations. To distinguish inconsistent false estimates from consistent observations, tracking approaches often require multiple, consecutive observations of the same source direction or position before a track is initialized. Furthermore, track termination rules are necessary to distinguish between speech endpoints and missing estimates. To avoid premature track deletions due to short-term missing estimates, track termination rules are often based on the lapsed time since the last track update. Uncertainty due to the onsets and endpoints of speech activity may therefore lead to a latency between the onsets and endpoints of speech and the initialization and termination, respectively, of the corresponding source track.


In practice, uncertainty in the source dynamical model and in the observations may lead to divergence of the track from the ground-truth trajectory of an inactive source. In multi-source scenarios, track divergence may also occur by mistakenly updating a source's track with estimates of a different, nearby source. As a consequence, track swaps may occur due to the divergence of a track to the trajectory of a different source. Furthermore, a track may be broken if the track is not assigned to any source for one or more time steps, i.e., the assignment between a source and its estimates is temporarily `interrupted'.

Measures selected for the objective evaluation are:
\begin{description}
  \item \textbf{Estimation accuracy}: The distance between a source position and the corresponding localized or tracked estimate.
  \item \textbf{Estimation ambiguity}: The rate of false estimates directed away from sound sources.
  \item \textbf{Track completeness}: The robustness against missing detections in a track or a sequence of localization estimates.
  \item \textbf{Track continuity}: The robustness against fragmentations due to track divergence or swaps affecting a track or a sequence of localization estimates.
  \item \textbf{Track timeliness}: The delay between the speech onset and either the first estimate in a sequence of localization estimates, or at track initialization.
\end{description}

The evaluation measures detailed in the following subsections are defined based on the following nomenclature. A single recording of duration $\mc{T}_{\text{rec}}$, including a maximum number of $N_{\text{max}}$ sources, is considered. Each source $n \in\{ 1, \dots, N_{\text{max}}\}$ is associated with $A(n)$ periods of activity of duration $\mc{T}(a,n) = T_{\text{end}}(a,n)-T_{\text{srt}}(a,n)$ for $a \in\{ 1, \dots, A(n)\}$, where $T_{\text{srt}}(a,n)$ and $T_{\text{end}}(a,n)$, respectively, mark the start and end time of the \ac{VAP}. The corresponding time step indices are $t_{\text{srt}}(a,n) \geq 0$ and $t_{\text{end}}(a,n) \geq t_{\text{srt}}(a,n)$. Each \ac{VAP} corresponds to an utterance of speech, which is assumed to include both voiced and unvoiced segments. $\Delta_{\text{valid}}(a,n)$ and $L_{\text{valid}}(a,n)$, respectively, denote the duration and the number of time steps in which source $n$ is assigned to a valid track during \ac{VAP} $a$. Participants were required to submit azimuth estimates of each source for a sequence of pre-specified time stamps, $t$, corresponding to the rate of the optical tracking system used for the recordings. Each azimuth estimate had to be labelled by an integer-valued \ac{ID}, $k = 1, \dots, K_{\text{max}}$, where $K_{max}$ is the maximum number of source IDs in the corresponding recording. Therefore, each source ID establishes an assignment from each azimuth estimate to one of the active sources. 



\subsection{Individual Evaluation Measures}
\label{sec:measures_individual}
To highlight the various scenarios that need to be accounted for during evaluation, consider, for simplicity and without loss of generality, the case of a single-source scenario, i.e., $N_{\text{max}} = 1$, where $N(t) = 1$ during speech activity and $N(t) = 0$ if the source is inactive. A submission either results in $K(t) = 0$, $K(t) = N(t) = 1$ or $K(t) > N(t)$, where $N(t)$ and $K(t)$, respectively, denote the true and estimated number of sources active at $t$. If $K(t) = 0$, the source is either inactive, i.e., $N(t)=0$, or the estimate of an active source is missing, if $N(t) = 1$. For $K(t) = 1$, the following scenarios are possible. \begin{inparaenum}[a)] \item The source is active, i.e., $N(t) = 1$, and the estimate corresponds to a typically imperfect estimate of the ground-truth source direction. \item The source is active, $N(t) = 1$, but its estimate is missing, whereas a false estimate, e.g., pointing towards the direction of an early reflection, is provided. \item The source is inactive, i.e., $N(t) = 0$, and a false estimate is provided\end{inparaenum}. Evaluation measures are therefore required that quantify, per recording, any missing and false estimates as well as the estimation accuracy of estimates in the direction of the source. Prior to performance evaluation, an assignment of each source to a detection must be established by gating and source-to-estimate association, as detailed in \sect{sec:measures_gating} and \sect{sec:measures_assoc}. The resulting assignment is for evaluation of the estimation accuracy, completeness, continuity, and timeliness (see \sect{sec:measures_accuracy} and \sect{sec:measures_ambiguity}).

\subsubsection{Gating between Sources and Estimates}
\label{sec:measures_gating}
Gating \cite{Blackman1999} provides a mechanism to distinguish between estimation errors, missing, and false estimates. Gating removes improbable assignments of a source with estimates corresponding to errors exceeding a preset threshold. Any estimate removed by gating is counted as a false estimate. If no detection lies within the gate of a source, the source is counted as missed. The gating threshold needs to be selected carefully: If set too low, estimation errors may lead to unassociated sources where a distorted estimate along an existing track is classified as a false estimate and the source estimate is considered as missing. In contrast, if the gating threshold is set too high, a source may be incorrectly assigned to a false track.

For evaluation of the \ac{LOCATA} challenge, the gating threshold is selected such that the majority of submissions within the single-source Tasks 1 and 3 is not affected. As will be shown in the evaluation in \sect{sec:Results}, a threshold of $30^{\circ}$ applied to the azimuth error allows to identify systematic false estimates.

\subsubsection{Source-to-Estimate Association}
\label{sec:measures_assoc}
For $K(t) > 1$, source localisation may be affected by false estimates both inside and outside the gate. Data association techniques are used to assign the source to the nearest estimate within the gate. Spurious estimates within the gate are included in the set of false estimates. At every time step, a pair-wise distance matrix corresponding to the angular error between each track and each source is evaluated. The optimum source-to-estimate assignment is established using the Munkres algorithm \cite{Kuhn1995} that identifies the source-to-estimate pairs corresponding to the minimum overall distance. Therefore, each source is assigned to at most one track and \emph{vice versa}.

Source-to-estimate association therefore allows to distinguish estimates corresponding to the highest estimation accuracy from spurious estimates.
Similar to data association discussed in \sect{sec:Literature}, and by extension of the single-source case, gating and association establish a one-to-one mapping of each active source with an estimate within the source gate. Any unassociated estimates are considered false estimates, whereas any unassociated sources correspond to missing estimates.

Based on the assignments between sources and estimates, established by gating and association, the  evaluation measures are defined to quantify the estimation errors and ambiguities as a single value per measure, per recording. For each assignment between a source and an estimate, the measures detailed in the following are applied to quantify, as a single measure per recording, the estimation accuracy, ambiguity, track completeness, continuity, and timeliness (see \sect{sec:measures_challenges}).

For brevity, a `track’ is synonymously used in the following to describe both, the trajectory of estimates obtained from a tracker, as well as a sequence of estimates labelled with the same \ac{ID} by a localization algorithm. The sequence of ground-truth source azimuth values of a source is referred to as the source's ground-truth azimuth trajectory.

\subsubsection{Estimation Accuracy}
\label{sec:measures_accuracy}
The angular errors are evaluated separately in azimuth and elevation for each assigned source-to-track pair for each time stamp during \acp{VAP}. The azimuth and elevation error, $d_{\phi}\left(\phi(t), \hat{\phi}(t)\right)$ and $d_{\theta}\left(\theta(t), \hat{\theta}(t)\right)$, respectively, are defined as:
\begin{subequations}
  \label{eqn:eval_accuracy}
  \begin{align}
    \label{eqn:eval_accuracy_azimuth}
    d_{\phi}\left(\phi(t), \hat{\phi}(t)\right) &= \text{mod}\left( \phi(t) - \hat{\phi}(t) + \pi, 2\pi \right) - \pi,\\
    \label{eqn:eval_accuracy_inclination}
    d_{\theta}\left(\theta(t), \hat{\theta}(t)\right) &= \theta(t) - \hat{\theta}(t),
  \end{align}
\end{subequations}
where $\text{mod}(q,r)$ denotes the modulo operator for the dividend, $q$, and the divisor, $r$; $\phi(t) \in [-\pi,\pi)$ and $\theta(t) \in [0,\pi]$ are the ground-truth azimuth and elevation, respectively; and $\hat{\phi}(t)$ and $\hat{\theta}(t)$ are the azimuth and elevation estimates, respectively.

\subsubsection{Ambiguity, Track Completeness, Continuity, and Timeliness}
\label{sec:measures_ambiguity}
In addition to the angular errors, multiple, complementary performance measures are used to quantify estimation ambiguity, completeness, continuity, and timeliness.

At each time step, the number of valid, false, missing, broken, and swapped tracks are counted. Valid tracks are identified as the tracks assigned to a source, whereas false tracks correspond to the unassociated tracks. The number of missing tracks is established as the number of unassociated sources. Broken tracks are obtained by identifying each source that was assigned to a track at $t-1$, but are unassociated at $t$, where $t$ and $t-1$ must correspond to time steps within the same voice-activity period. Similar to broken tracks, swapped tracks are counted by identifying each source that was associated to track \ac{ID} $j \in \{1,\dots, K_{\text{max}}\}$, and is associated to track \ac{ID}, $\ell \in \{1,\dots,K_{\text{max}}\}$, where $j \neq \ell$.

Subsequently, the following measures of estimation ambiguity, completeness, continuity, and timeliness are evaluated:
\begin{description}[style=unboxed,leftmargin=0.25cm]
  \item \textbf{Probability of detection ($p_d$)} \cite{Blackman1999}: A measure of completeness, evaluating for each source and voice-activity period the percentage of time stamps during which the source is associated with a valid track.
  \item \textbf{\ac{FAR}} \cite{Rothrock2000}: A measure of ambiguity, evaluating the number of false estimates per second. The \ac{FAR} can be evaluated over the duration of each recording \cite{Evers2018b}, in order to provide a gauge of the effectiveness of any \ac{VAD} algorithms that may have been incorporated in a given submitted localization or tracking framework. In addition, the \ac{FAR} is evaluated in this paper over the duration of each \ac{VAP} in order to provide a measure of source counting accuracy of each submission.
  \item \textbf{\ac{TL}} \cite{Rothrock2000}: A measure of timeliness, evaluating the delay between the onset and the first detection of source $n$ in \ac{VAP} $a$.
  \item \textbf{\ac{TFR}} \cite{Gorji2011}: A measure of continuity, indicating the number of track fragmentations per second. The number of fragmentations corresponds to the number of track swaps plus the number of broken tracks.
\end{description}

The evaluation measures defined above therefore quantify errors and ambiguities by single numerical values per measure, per recording. These individual measures can also be used to quantify, across all recordings in each task, the mean of and standard deviation in the estimation accuracy and ambiguity as well as the track completeness, continuity and timeliness.

\begin{table*}[tb]
\centering
\caption{Average azimuth errors during \ac{VAP}. Submissions corresponding to minimum average errors are highlighted in bold font. Column colour indicates type of algorithm, where white indicates frameworks involving only \acs{DoA} estimation (Submission IDs 1, 6, 9, 11, 12, 15, 16 and the baseline (BL)), and grey indicates frameworks that combine \ac{DoA} estimation with source tracking (Submission IDs 2, 3, 4, 7, 8, 10).}
\label{table:azimuth_errors}

\begin{tabular}{|c|c|c||c|>{\columncolor{darkgrey}}c|>{\columncolor{darkgrey}}c|>{\columncolor{darkgrey}}c|c|>{\columncolor{darkgrey}}c|>{\columncolor{darkgrey}}c|c|>{\columncolor{darkgrey}}c|c|c|c|c|c|}
\hline
\multicolumn{2}{|c|}{\multirow{2}{*}{Task}} & \multirow{2}{*}{Array} & \multicolumn{14}{c|}{Submission ID}\\\cline{4-17}
\multicolumn{2}{|c|}{} & & 1 & 2 & 3 & 4 & 6 & 7 & 8 & 9 & 10 & 11 & 12 & 15 & 16 & BL\\\hline\hline
\multirow{12}{*}{\rotatebox[origin=c]{90}{Single Source}} & \multirow{4}{*}{1} & Robot Head & - & - & - & 2.1 & 1.5 & 1.8 & - & - & - & \textbf{0.7} & - & - & - & 4.2 \\
\cline{3-17}
 &  & DICIT & - & - & \textbf{1.0} & - & - & 2.2 & - & 9.1 & - & - & - & - & - & 12.3 \\
\cline{3-17}
 &  & Hearing Aids & \textbf{8.5} & - & - & - & - & - & 8.7 & - & - & - & - & - & - & 15.9 \\
\cline{3-17}
 &  & Eigenmike & - & - & - & - & 6.4 & 7.0 & - & - & 8.9 & - & \textbf{1.1} & 8.1 & - & 10.2 \\
\cline{2-17}
 & \multirow{4}{*}{3} & Robot Head & - & - & - & 4.6 & 3.2 & \textbf{3.1} & - & - & - & - & - & - & - & 9.4 \\
\cline{3-17}
 &  & DICIT & - & - & \textbf{1.8} & - & - & 4.5 & - & - & - & - & - & - & - & 13.9 \\
\cline{3-17}
 &  & Hearing Aids & - & - & - & - & - & - & \textbf{7.2} & - & - & - & - & - & - & 16.0 \\
\cline{3-17}
 &  & Eigenmike & - & - & - & - & \textbf{8.1} & 9.3 & - & - & 11.5 & - & - & - & - & 17.6 \\
\cline{2-17}
 & \multirow{4}{*}{5} & Robot Head & - & - & - & 4.9 & \textbf{2.2} & 3.7 & - & - & - & - & - & - & - & 5.4 \\
\cline{3-17}
 &  & DICIT & - & - & \textbf{2.7} & - & - & 3.4 & - & - & - & - & - & - & - & 13.4 \\
\cline{3-17}
 &  & Hearing Aids & - & - & - & - & - & - & \textbf{11.8} & - & - & - & - & - & - & 14.6 \\
\cline{3-17}
 &  & Eigenmike & - & - & - & - & \textbf{6.3} & 7.5 & - & - & - & - & - & - & - & 12.9 \\
\hline\hline
\multirow{12}{*}{\rotatebox[origin=c]{90}{Multiple Sources}} & \multirow{4}{*}{2} & Robot Head & - & - & - & 3.8 & - & - & - & - & - & \textbf{2.0} & - & - & - & 9.0 \\
\cline{3-17}
 &  & DICIT & - & - & - & - & - & - & - & - & - & - & - & - & - & \textbf{11.0} \\
\cline{3-17}
 &  & Hearing Aids & - & - & - & - & - & - & - & - & - & - & - & - & - & \textbf{15.6} \\
\cline{3-17}
 &  & Eigenmike & - & - & - & - & - & - & - & - & 7.3 & - & \textbf{1.4} & - & 7.1 & 10.2 \\
\cline{2-17}
 & \multirow{4}{*}{4} & Robot Head & - & 9.4 & - & \textbf{6.0} & - & - & - & - & - & - & - & - & - & 9.2 \\
\cline{3-17}
 &  & DICIT & - & 13.5 & - & - & - & - & - & - & - & - & - & - & - & \textbf{12.9} \\
\cline{3-17}
 &  & Hearing Aids & - & 13.8 & - & - & - & - & - & - & - & - & - & - & - & \textbf{13.7} \\
\cline{3-17}
 &  & Eigenmike & - & 12.8 & - & - & - & - & - & - & \textbf{9.0} & - & - & - & - & 11.8 \\
\cline{2-17}
 & \multirow{4}{*}{6} & Robot Head & - & - & - & \textbf{8.1} & - & - & - & - & - & - & - & - & - & 8.5 \\
\cline{3-17}
 &  & DICIT & - & - & - & - & - & - & - & - & - & - & - & - & - & \textbf{13.9} \\
\cline{3-17}
 &  & Hearing Aids & - & - & - & - & - & - & - & - & - & - & - & - & - & \textbf{13.9} \\
\cline{3-17}
 &  & Eigenmike & - & - & - & - & - & - & - & - & - & - & - & - & - & \textbf{12.9} \\
\hline
\end{tabular}
\end{table*}

\begin{table*}[tb]
\centering
\caption{Difference in average azimuth errors with and without gating, evaluated for single-source tasks 1, 3, 5 for all submissions and the baseline (BL). Submissions unaffected by gating, and hence outliers, are highlighted in bold font.}
\label{table:azimuth_errors_gating_offset}

\begin{tabular}{|c|c|c||c|>{\columncolor{darkgrey}}c|>{\columncolor{darkgrey}}c|>{\columncolor{darkgrey}}c|c|>{\columncolor{darkgrey}}c|>{\columncolor{darkgrey}}c|c|>{\columncolor{darkgrey}}c|c|c|c|c|c|}
\hline
\multicolumn{2}{|c|}{\multirow{2}{*}{Task}} & \multirow{2}{*}{Array} & \multicolumn{14}{c|}{Submission ID}\\\cline{4-17}
\multicolumn{2}{|c|}{} & & 1 & 2 & 3 & 4 & 6 & 7 & 8 & 9 & 10 & 11 & 12 & 15 & 16 & BL\\\hline\hline
\multirow{12}{*}{\rotatebox[origin=c]{90}{Single Source}} & \multirow{4}{*}{1} & Robot Head & - & - & - & \textbf{0.0} & \textbf{0.0} & \textbf{0.0} & - & - & - & \textbf{0.0} & - & - & - & 0.2 \\
\cline{3-17}
 &  & DICIT & - & - & \textbf{0.0} & - & - & \textbf{0.0} & - & 0.5 & - & - & - & - & - & 49.6 \\
\cline{3-17}
 &  & Hearing Aids & 42.3 & - & - & - & - & - & \textbf{4.0} & - & - & - & - & - & - & 49.2 \\
\cline{3-17}
 &  & Eigenmike & - & - & - & - & 0.1 & 0.1 & - & - & \textbf{0.0} & - & \textbf{0.0} & \textbf{0.0} & - & 0.4 \\
\cline{2-17}
 & \multirow{4}{*}{3} & Robot Head & - & - & - & \textbf{0.0} & 1.2 & \textbf{0.0} & - & - & - & - & - & - & - & 3.4 \\
\cline{3-17}
 &  & DICIT & - & - & \textbf{0.0} & - & - & \textbf{0.0} & - & - & - & - & - & - & - & 63.2 \\
\cline{3-17}
 &  & Hearing Aids & - & - & - & - & - & - & \textbf{0.4} & - & - & - & - & - & - & 46.8 \\
\cline{3-17}
 &  & Eigenmike & - & - & - & - & 0.6 & \textbf{0.2} & - & - & 1.6 & - & - & - & - & 8.3 \\
\cline{2-17}
 & \multirow{4}{*}{5} & Robot Head & - & - & - & \textbf{0.1} & 0.8 & 1.2 & - & - & - & - & - & - & - & 1.8 \\
\cline{3-17}
 &  & DICIT & - & - & \textbf{0.6} & - & - & 16.7 & - & - & - & - & - & - & - & 53.8 \\
\cline{3-17}
 &  & Hearing Aids & - & - & - & - & - & - & \textbf{12.7} & - & - & - & - & - & - & 43.7 \\
\cline{3-17}
 &  & Eigenmike & - & - & - & - & \textbf{1.1} & 1.9 & - & - & - & - & - & - & - & 14.9 \\
\hline
\end{tabular}
\end{table*}

\subsection{Combined Evaluation Measure}
The \ac{OSPA} metric \cite{Schuhmacher2008} and its variants, e.g., \cite{Rahmathullah2017}, correspond to a comprehensive measure that consolidates the cardinality error in the estimated number of sources and the estimation accuracy across all sources into a single distance metric at each time stamp of a recording. The \ac{OSPA} therefore provides a measure that combines the estimation accuracy, track completeness and timeliness. The \ac{OSPA} selects, at each time stamp, the optimal assignment of the subpatterns between sources and combines the sum of the corresponding cost matrix with the cardinality error in the estimated number of sources. Since the \ac{OSPA} is evaluated independently of the \acp{ID} assigned to the localization and tracking estimates, the measure is agnostic to uncertainties in the identification of track labels.

The \ac{OSPA} \cite{Ristic2013,Ristic2011} is defined as:
\begin{small}
\begin{align}
  \label{eqn:OSPA}
  \begin{split}
    &\text{OSPA}(\hat{\mat{\Phi}}(t),\mat{\Phi}(t)) \defas\\
    &\biggl[ \frac{1}{K(t)} \min_{\pi \in \mat{\Pi}_{K(t)}} \sum\limits_{n=1}^{N(t)} d_c(\phi_n(t), \hat{\phi}_{\pi(n)}(t))^p  + (K(t)-N(t)) c^p \biggr]^{\frac{1}{p}},
  \end{split}
\end{align}%
\end{small}%
for $N(t) \leq K(t)$, where $\hat{\mat{\Phi}}(t) \defas \{ \hat{\phi}_1(t), \dots, \hat{\phi}_{K(t)}(t)\}$ denotes the set of $K(t)$ track estimates;
$\mat{\Phi}(t) \defas \{ \phi_1(t), \dots, \phi_{N(t)}(t)\}$ denotes the set of $N(t)$ ground-truth sources active at $t$; $1 \leq p < \infty$ is the order parameter;
$c$ is the cutoff parameter; $\mat{\Pi}_{K(t)}$ denotes the set of permutations of length $N(t)$ with elements $\{ 1, \dots, K(t) \}$ \cite{Ristic2011}; $d_c(\phi_n(t), \hat{\phi}_{\pi(n)}(t)) \defas \min{\left(c,\text{abs}\left(d_{\phi}(\phi_n(t), \hat{\phi}_{\pi(n)}(t))\right)\right)}$, where $\text{abs}(\cdot)$ denotes the absolute value; $d_\phi(\cdot)$ is the angular error (see \eq{eqn:eval_accuracy}); and $\pi(n)$ denotes the $n^{th}$ element of each subset $\pi \in \mat{\Pi}$. For $N(t) > K(t)$, the \ac{OSPA} distance is evaluated as $\text{OSPA}(\mat{\Phi}(t),\hat{\mat{\Phi}}(t))$ \cite{Ristic2011}. The impact of the choice of $p$ and $c$ is discussed in \cite{Schuhmacher2008}. In this paper, $c = 30^{\circ}$.

To provide further insight into the \ac{OSPA} measure, we note that the term $\frac{1}{K(t)} \min_{\pi \in \mat{\Pi}_{K(t)}} \sum_{n=1}^{N(t)} d_c(\phi_n(t), \hat{\phi}_{\pi(n)}(t))^p$ evaluates the average angular error by comparing each angle estimate against every ground-truth source angle. The \ac{OSPA} is therefore agnostic of the estimate-to-source association. The cardinality error is evaluated as $K(t)-N(t)$. The order parameter, $p$, determines the weighting of the angular error relative to the cardinality error.

Due to the dataset size of the \ac{LOCATA} corpus, a comprehensive analysis of the \ac{OSPA} at each time stamp for each submission, task, array, and recording is impractical. Therefore, the analysis of the \ac{LOCATA} challenge results is predominantly based on the mean and variance of the \ac{OSPA} across all time stamps and recordings for each task. 

\section{Evaluation Results}
\label{sec:Results}

The following section presents the performance evaluation for the \ac{LOCATA} challenge submissions using the measures detailed in \sect{sec:Measures}. The evaluation in \sect{sec:results_135} focuses on the single-source tasks 1, 3 and 5. \sect{sec:results_246} presents the results for the multi-source tasks 2, 4 and 6.

The evaluation framework establishes an assignment between each ground-truth source location and a source estimate for every time stamp during voice-active periods in each recording, submission, task, and array (see \sect{sec:Measures}). The azimuth error in \eq{eqn:eval_accuracy_azimuth} between associated source-to-track pairs is averaged over all time stamps and all recordings. The resulting average azimuth errors for each task, submission, and array are provided in \tabref{table:azimuth_errors}. The baseline (BL) corresponds to the \ac{MUSIC} implementation as detailed in \cite{LOCATA2018c}. One submission (ID 5) is not included in the discussion as details of the method are not available at the time of writing. Two further submissions (ID 13 and ID 14) are also not included due to inconclusive results.

\subsection{Single-Source Tasks 1, 3, 5}
\label{sec:results_135}
\subsubsection{Task 1 - Azimuth Accuracy}
\label{sec:eval_task1_azimuth}
For Task~1, involving a single, static source and a static microphone array, average azimuth accuracies of around $1^{\circ}$ can be achieved (see \tabref{table:azimuth_errors}). Notably, Submission 3 results in $1.0^{\circ}$ using the \ac{DICIT} array by combining \ac{TDE} with a particle filter for tracking; Submission 11 results in an average azimuth accuracy of $0.7^{\circ}$ using the robot head; and Submission 12 achieves an accuracy of $1.1^{\circ}$ using the Eigenmike. Submissions 11 and 12 are \ac{MUSIC} implementations, applied to the microphone signals in the \ac{STFT} domain and domain of spherical harmonics, respectively.

A possible reason for the performance of Submissions 11 and 12 is that \ac{MUSIC} does not suffer from spatial aliasing if applied to arrays that incorporate a large number of microphones. As such, the overall array aperture can be small for low noise levels. Therefore, the performance of the two \ac{MUSIC}-based Submissions 11 (robot head) and 12 (Eigenmike) is comparable. Moreover, for the Eigenmike, Submission 12 ($1.1^{\circ}$) leads to improvements of the \ac{SRP}-based Submissions 6 ($6.4^{\circ}$) and 7 ($7.0^{\circ}$).

For the pseudo-intensity-based approaches that were applied to the Eigenmike, Submission 10 achieves an azimuth accuracy of $8.9^{\circ}$ by extracting pseudo-intensity vectors from the first-order ambisonics and applying a particle filter for tracking. Submission 15, which extracts the pseudo-intensity from the signals in the domain of spherical harmonics and applies subspace-based processing, results in $8.1^{\circ}$. The pseudo-intensity-based Submissions 10 and 15 lead to a performance degradation of approximately $7^{\circ}$, compared to the \ac{MUSIC}-based Submission 12, also applied in the domain of spherical harmonics. The reduced accuracy may be related to the resolution of the spatial spectra provided by the pseudo-intensity-based approaches compared to \ac{MUSIC}. The spatial spectrum is computed using \ac{MUSIC} by scanning each direction in a discrete grid, specified by the steering vector. In contrast, pseudo-intensity-based approaches approximate the spatial spectrum by effectively combining the output of three dipole beamformers, steered along the $x$-, $y$-, and $z$-axis relative to the array. Therefore, compared to \ac{MUSIC}, pseudo-intensity approaches evaluate a coarse approximation of the spatial spectrum, but require reduced computational load.

A performance degradation from the 12-channel robot head to the 32-channel Eigenmike is observed for the submissions that involved both arrays. For ground-truth acquisition using the OptiTrack system, the reflective markers were attached to the shockmount of the Eigenmike, rather than the baffle of the array, to minimize shadowing and scattering effects, see \cite{LOCATA2018a,LOCATA2018b}. Therefore, a small bias in the \ac{DoA} estimation errors is possible due to rotations of the array within the shockmount. Nevertheless, this bias is expected to be significantly smaller than some of the errors observed for the Eigenmike in \tabref{table:azimuth_errors}. Possible reasons are that \begin{inparaenum} \item the irregular array topology of the robot head may lead to improved performance for some of the algorithms, or that \item the performance improvements in localization accuracy may be related to the larger array aperture of the robot head, compared to the Eigenmike\end{inparaenum}. However, with the remaining uncertainty regarding the actual implementation of the algorithms, conclusions remain somewhat speculative at this point.

Submission 6, applying \ac{SRP}-\ac{PHAT} to a selection of microphone pairs, results in average azimuth errors of $1.5^{\circ}$ using the robot head and $6.4^{\circ}$ using the Eigenmike. Similar results of $1.8^{\circ}$ and $7.0^{\circ}$ for the robot head and Eigenmike, respectively, are obtained using Submission 7, which combine an \ac{SRP} beamformer for localization with a Kalman filter for tracking. Therefore, the \ac{SRP}-based approaches in Submissions 6 and 7, applied without and with tracking, respectively, lead to comparably accurate results.

\begin{figure}
  \centering
  \mbox{\subfloat[Azimuth ground-truth and estimates]{
    \label{subfig:az_task3_sub357}
    \includegraphics[width=.99\columnwidth]{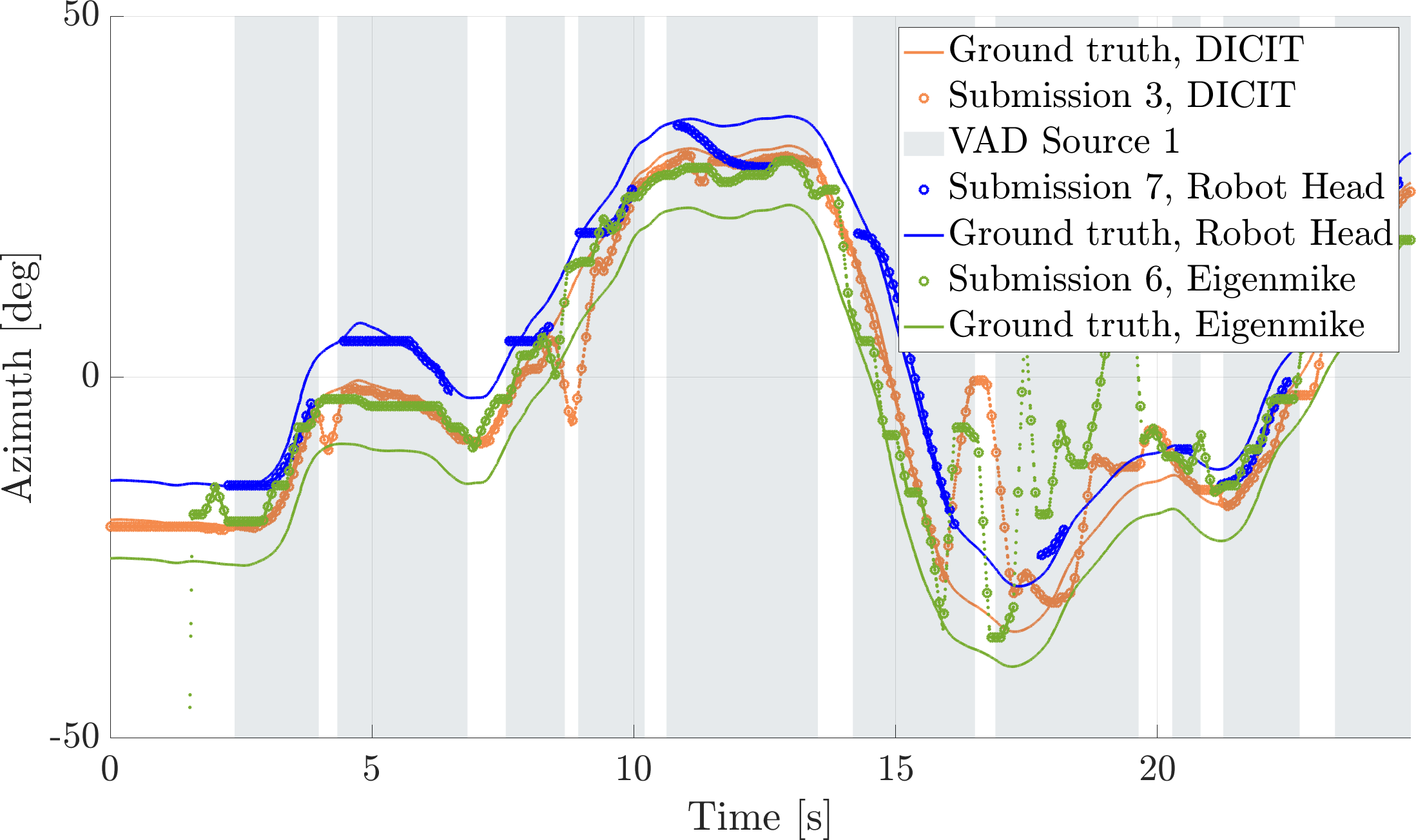}}}
  \\
  \mbox{\subfloat[Ground-truth range between source and robot head]{
    \label{subfig:az_task3_range_robot}
    \includegraphics[width=.99\columnwidth]{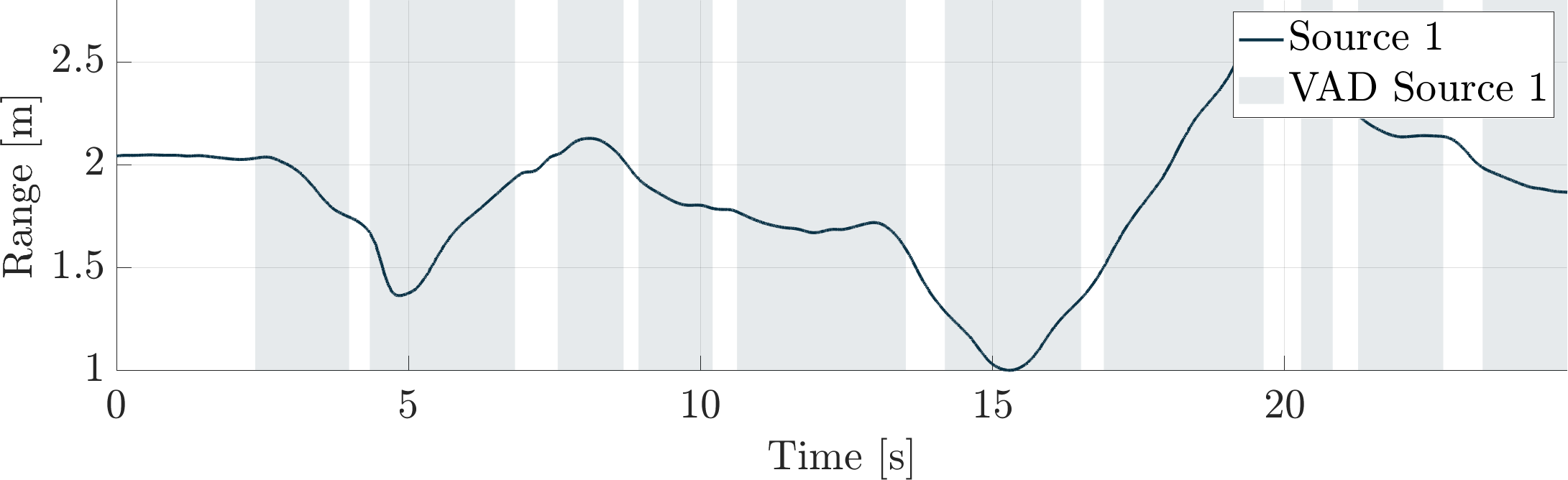}}}
  \caption{Azimuth estimates for Task 3, recording 4 for (a) azimuth estimates for Submissions 3, 6, 7. As a reference, the ground-truth range between the robot head and the source is shown in (b).}
  \label{fig:az_task3_subs735}
\end{figure}

\tabref{table:azimuth_errors} also highlights a significant difference in the performance results between the approaches submitted to Task~1 using the \ac{DICIT} array. Submission 3 achieves an average azimuth accuracy of $1.0^{\circ}$ by combining \ac{GCC}-\ac{PHAT} with a particle filter. Submission 7, combining \ac{SRP} beamforming and a Kalman filter, results in a small degradation to $2.2^{\circ}$ in average azimuth accuracy. Submission 9 leads to a decreased accuracy of $9.1^{\circ}$.
Submission 3 uses the subarray of microphone pairs corresponding to $32$~cm spacings to exploit spatial diversity between the microphones; Submission 7 uses the 7-microphone linear subarray at the array centre; Submission 9 uses three microphones at the centre of the array, with a spacing of $4$~cm, to form two microphone pairs. A reduction of the localization accuracy can therefore be intuitively expected for Submission 9, compared to Submissions 3 and 7, due to \begin{inparaenum}[a)]\item the reduced number of microphones, and \item the reduced inter-microphone spacing, and hence reduced spatial diversity of the sensors\end{inparaenum}.

For the hearing aids in Task~1, both Submissions 1 and 8 result in comparable azimuth errors of $8.5^{\circ}$ and $8.7^{\circ}$ respectively.
The recordings for the hearing aids were performed separately from the remaining arrays, and are therefore not directly comparable to the results for other arrays. Nevertheless, a reduction in azimuth accuracy for the hearing aids is intuitively expected due to the small number of microphones integrated in each of the arrays.

To conclude, we note that the results for the static single-source Task~1 indicate a comparable performance between the submissions that incorporate localization and those submissions that combine localization with source tracking. Since the source is static, long blocks of data can be used for localization. Furthermore, temporal averaging can be applied across data blocks. Therefore, since a dynamical model is not required for the static single-source scenario, localization algorithms can apply smoothing directly to the \ac{DoA} estimate, without the need for explicit source tracking.

\begin{figure}[tb]
  \centering
  \mbox{\subfloat[Task 1]{
    \label{subfig:pd_task_1}
    \includegraphics[width=.99\columnwidth]{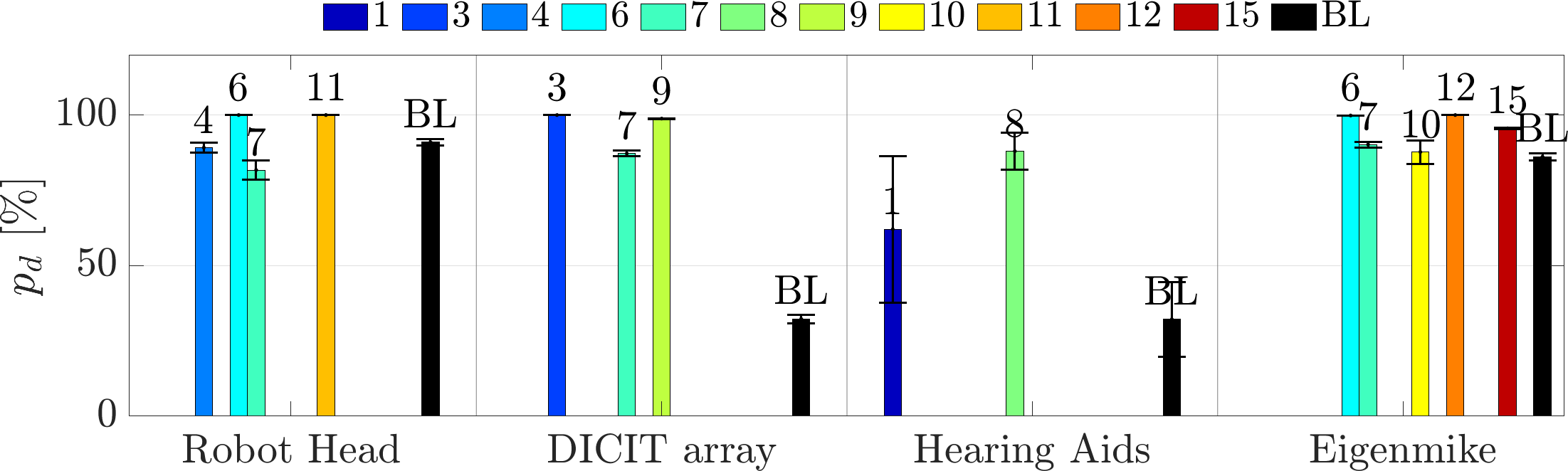}}}
  \\
  \mbox{\subfloat[Task 3]{
    \label{subfig:pd_task_3}
    \includegraphics[width=.99\columnwidth]{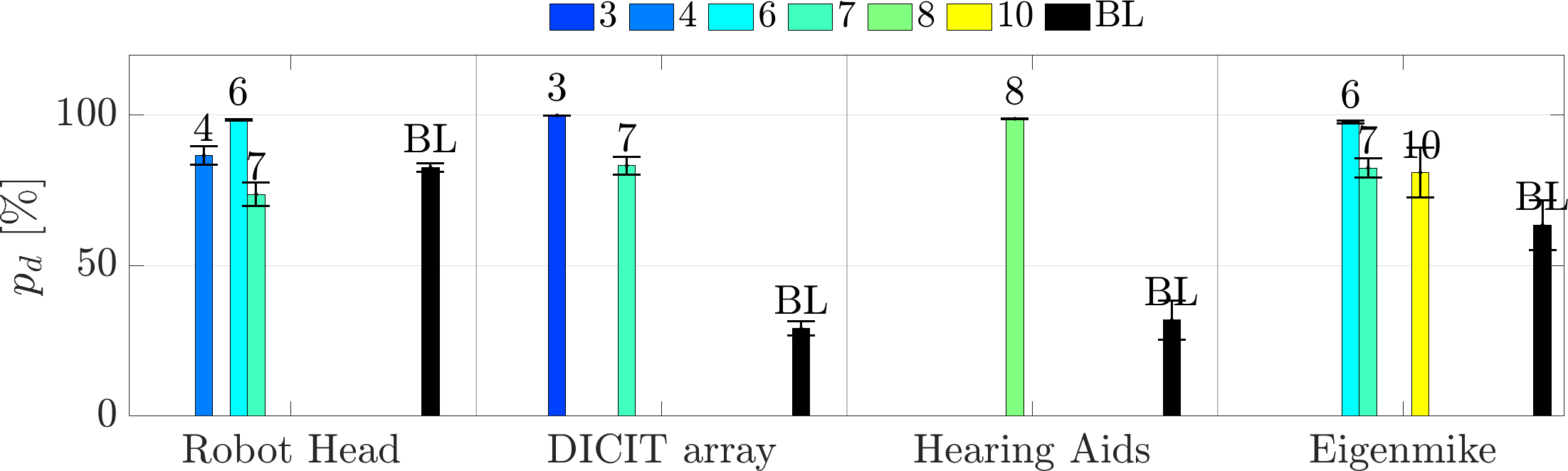}}}
  \\
  \mbox{\subfloat[Task 5]{
    \label{subfig:pd_task_5}
    \includegraphics[width=.99\columnwidth]{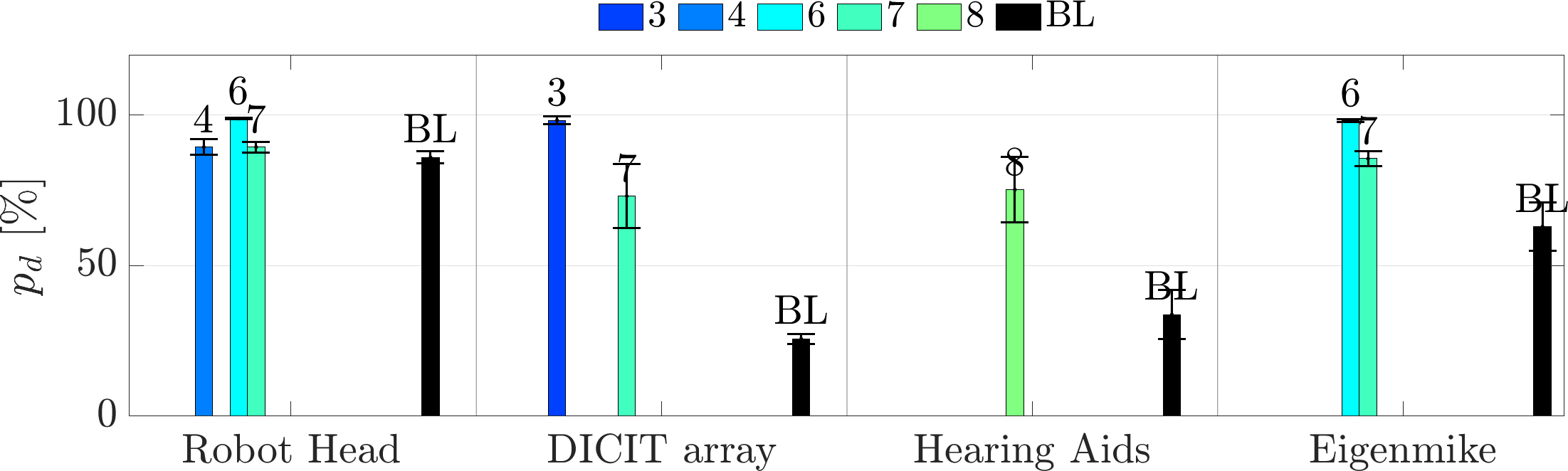}}}
  \caption{Probability of detection (bars) and standard deviation over recordings (whiskers) for Tasks 1, 3, 5, for each submission and array. Legends indicate the submission \acsp{ID} available for each of the tasks.}
  \label{fig:pd_tasks_135}
\end{figure}

\begin{figure}
  \centering
  \mbox{\subfloat[Task 1, for entire recording duration]{
    \label{subfig:FAR_recording_task1}
    \includegraphics[width=.99\columnwidth]{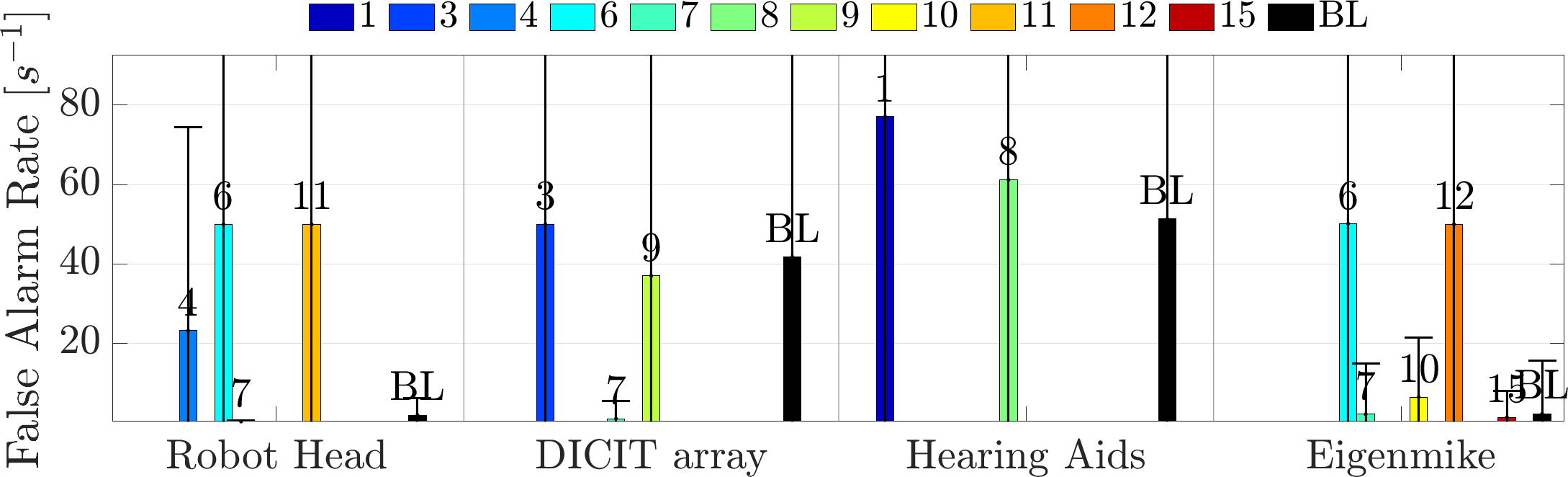}}}
  \\
  \mbox{\subfloat[Task 1, during voice activity only]{
    \label{subfig:FAR_task13_task1}
    \includegraphics[width=.99\columnwidth]{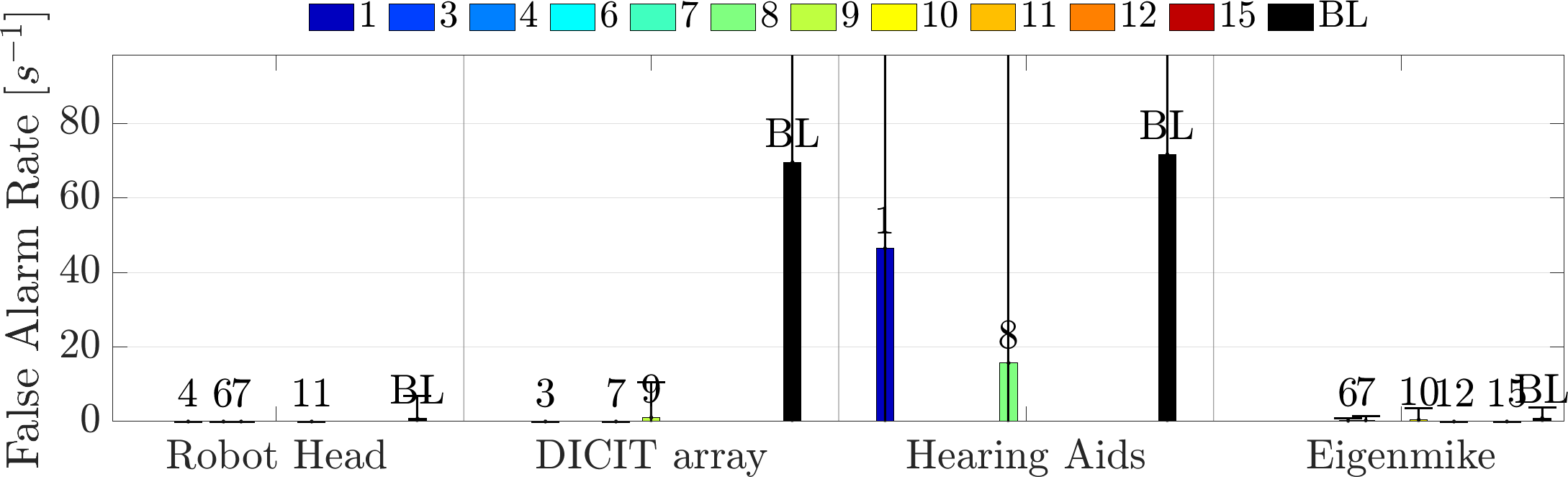}}}
  \caption{\acs{FAR} for Task~1 involving single static loudspeakers (a) for entire recording duration, and (b) during voice-activity periods only.}
  \label{fig:FAR_task13}
\end{figure}

\subsubsection{Task 3 - Azimuth Accuracy}
In the following, $\mc{S}_{135}= \{3, 4, 6, 7, 8\}$ denotes the set of submissions that were evaluated for Tasks~1,~3~and~5. For Task~3, involving a single, moving source, a small degradation is observed in the azimuth error over $\mc{S}_{135}$ from $4.3^{\circ}$ for Task~1 to $5.5^{\circ}$ for Task~3. For example, Submission 7 leads to the lowest average absolute error in azimuth with only $3.1^{\circ}$ for Task~3 using the robot head, corresponding to a degradation of $1.3^{\circ}$ compared to Task~1. The accuracy of Submission 3 reduces from $1.0^{\circ}$ for Task~1 to $1.8^{\circ}$ for Task~3.

The reduction in azimuth accuracy from static single-source Task~1 to moving single-source Task~3 is similar for all submissions. Trends in performance between approaches for each array are identical to those discussed for Task~1. The overall degradation in performance is therefore related to differences in the scenarios between Task~1 and Task~3. Recordings from human talkers are subject to variations in the source orientation and source-sensor distance. The orientation of sources directed away from the microphone array leads to a decreased direct-path contribution to the received signal. Furthermore, with increasing source-sensor distance, the noise field becomes increasingly diffuse. Hence, reductions in the \ac{DRR} \cite{Eaton2016} due to the source orientation, as well as the \ac{CDR} due to the source-sensor distance, result in increased azimuth estimation errors.

To provide further insight into the results for Task~3, \fig{fig:az_task3_subs735} provides a comparison for recording 4 of the approaches leading to the highest accuracy for each array, i.e., Submission 7 using the robot head, Submission 3 using the \ac{DICIT} array, and Submission 6 using the Eigenmike.
For Submission 7, accurate and smooth tracks of the azimuth trajectories are obtained during \acp{VAP}. Therefore, diagonal unloading \ac{SRP} beamforming clearly provides power maps of sufficiently high resolution to provide accurate azimuth estimates whilst avoiding systematic false detections in the directions of early reflections. Moreover, application of the Kalman filter provides smooth azimuth trajectories.

Similar results in terms of the azimuth accuracy are obtained for Submission 3, combining \ac{GCC}-\ac{PHAT} with a particle filter for the \ac{DICIT} array. However, due to the lack of a \ac{VAD}, temporary periods of track divergence can be observed for Submission 3 around periods of voice inactivity, i.e., between [3.9,4.4]~s and [8.5,9.2]~s.

For the voice-active period between [16.9,19.6]~s, the results of Submission 7 are affected by a significant number of missing detections, whilst the results for Submission 3 exhibits diverging track estimates. \fig{subfig:az_task3_range_robot} provides a plot of the range between the source and robot head, highlighting that the human talker is moving away from the arrays between [15.1,20]~s. Therefore, the \ac{CPSD}-based \ac{VAD} algorithm of Submission 7 results in missing detections of voice activity with decreasing \ac{CDR}. For Submission 3 and 6, that do not involve a \ac{VAD}, the negative \ac{DRR} leads to missing and false \ac{DoA} estimates in the direction of early reflections. Therefore, increasing \ac{DoA} estimation errors are observed in voice-active periods during which the source-sensor distance increases beyond 2~m.

\subsubsection{Task 5 - Azimuth Accuracy}
The mean azimuth accuracy over $\mc{S}_{135}$, averaged over the corresponding submissions and arrays, decreases from $5.5^{\circ}$ for Task~3, using static arrays, to $9.7^{\circ}$ for Task~5, using moving arrays.
Despite the reduced number of submissions for Task~5, the overall performance trends are similar to those in Task~1 and Task~3 (see \tabref{table:azimuth_errors}).

The trend of an overall performance degradation is related to the increasingly challenging conditions. Similar to Task~3, the motion of the source and arrays lead to time-varying source-sensor distances and source orientations relative to the array. Furthermore, due to the motion of the array, it is crucial that the microphone signals in Task~5 are processed over analysis windows of sufficiently short duration.

\begin{figure}
  \centering
  \mbox{\subfloat[Azimuth ground-truth for Source 1 and estimates]{
    \label{subfig:az_task3_forFAR_sub67}
    \includegraphics[width=.99\columnwidth]{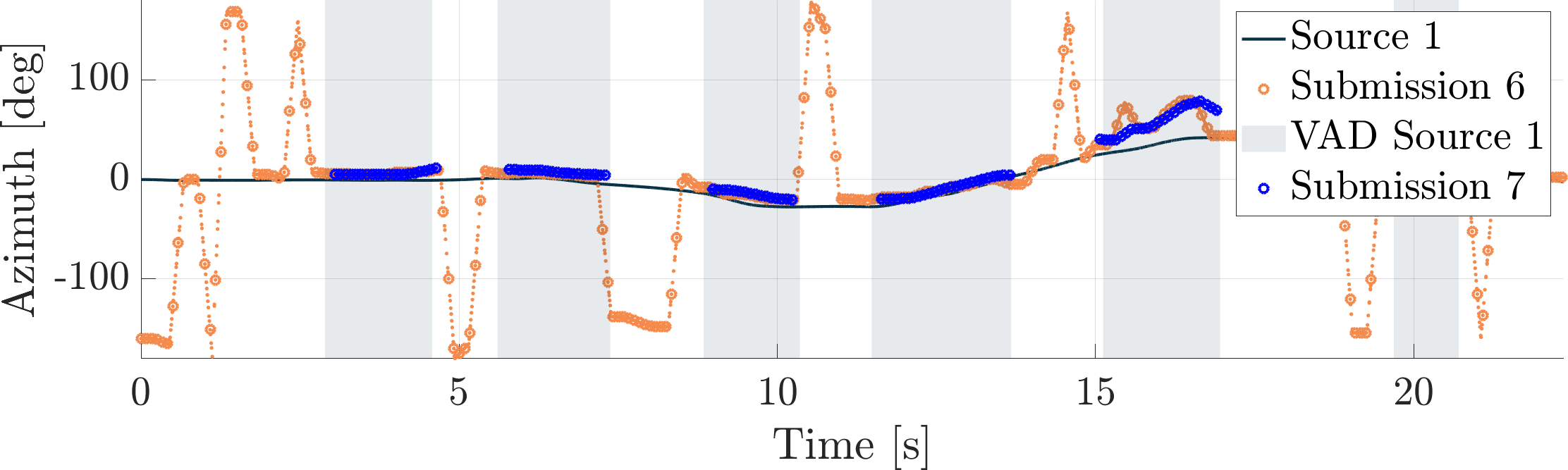}}}
  \\
  \mbox{\subfloat[Ground-truth source-sensor range]{
    \label{subfig:range_task3_forFAR_sub7}
    \includegraphics[width=.99\columnwidth]{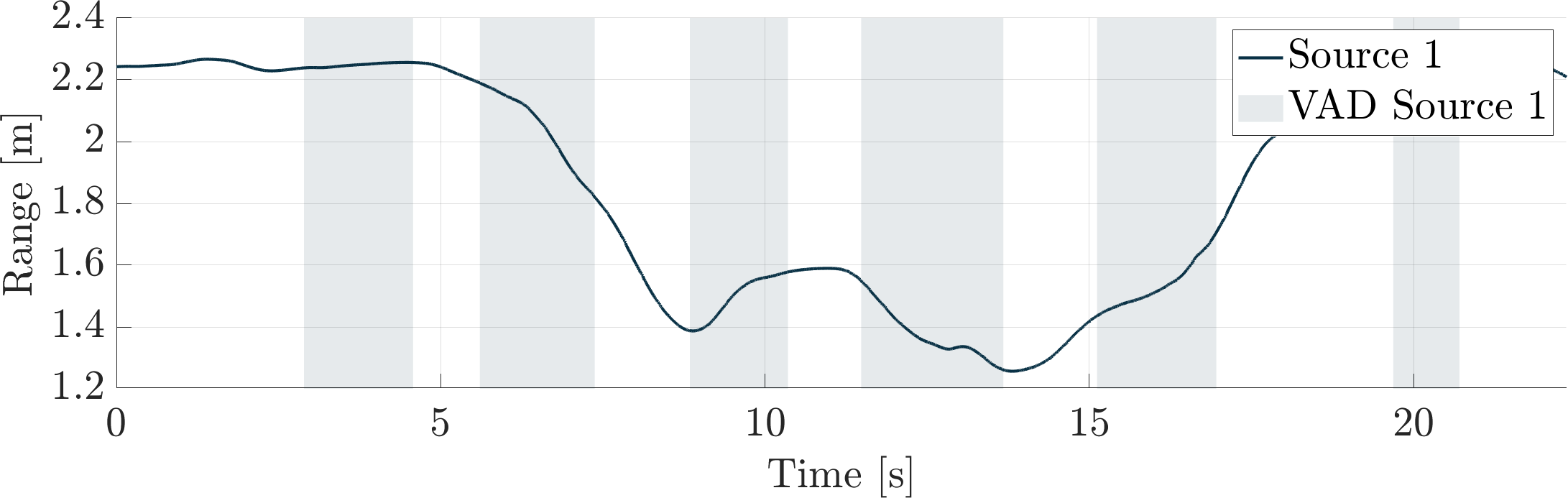}}}
  \caption{Comparison of (a) azimuth estimates for Task~3, recording~2 using the Eigenmike for Submissions 6, 7, and (b) ground-truth range between the source and the Eigenmike array origin. Results indicate outliers during voice inactivity for Submission~6 and temporary track divergence during voice activity between [15.1,17]~s for Submissions 6 and 7.}
  \label{fig:az_task3_forFAR}
\end{figure}

\begin{figure}
  \centering
  \mbox{\subfloat[Task 1]{
    \label{subfig:TL_135_task_1}
    \includegraphics[width=.99\columnwidth]{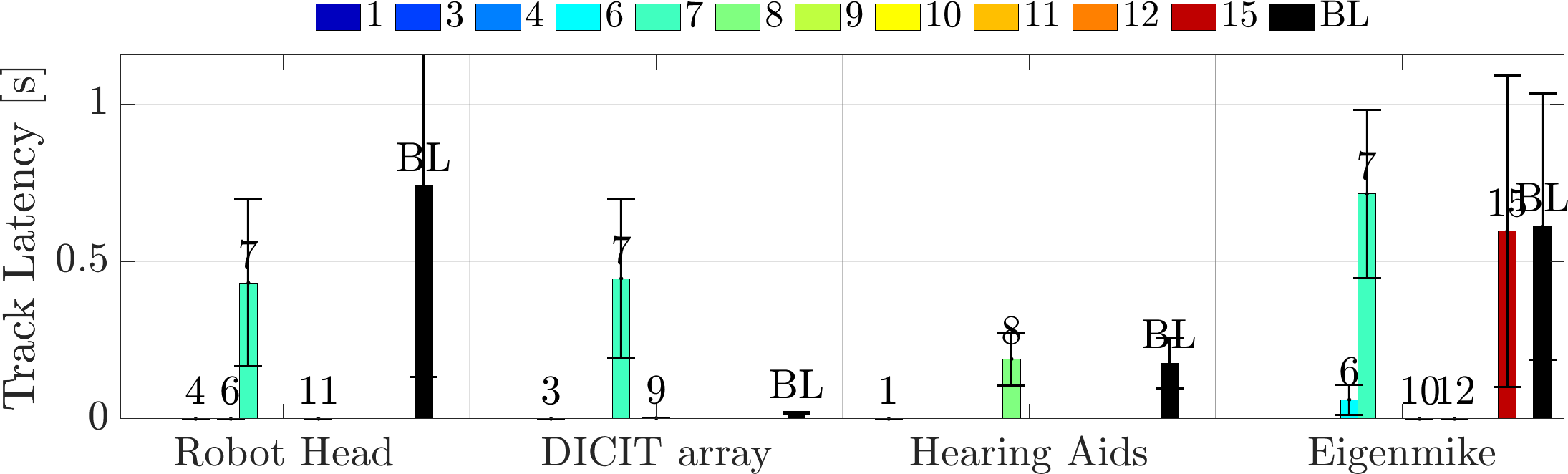}}}
  \\
  \mbox{\subfloat[Task 3]{
    \label{subfig:TL_135_task_3}
    \includegraphics[width=.99\columnwidth]{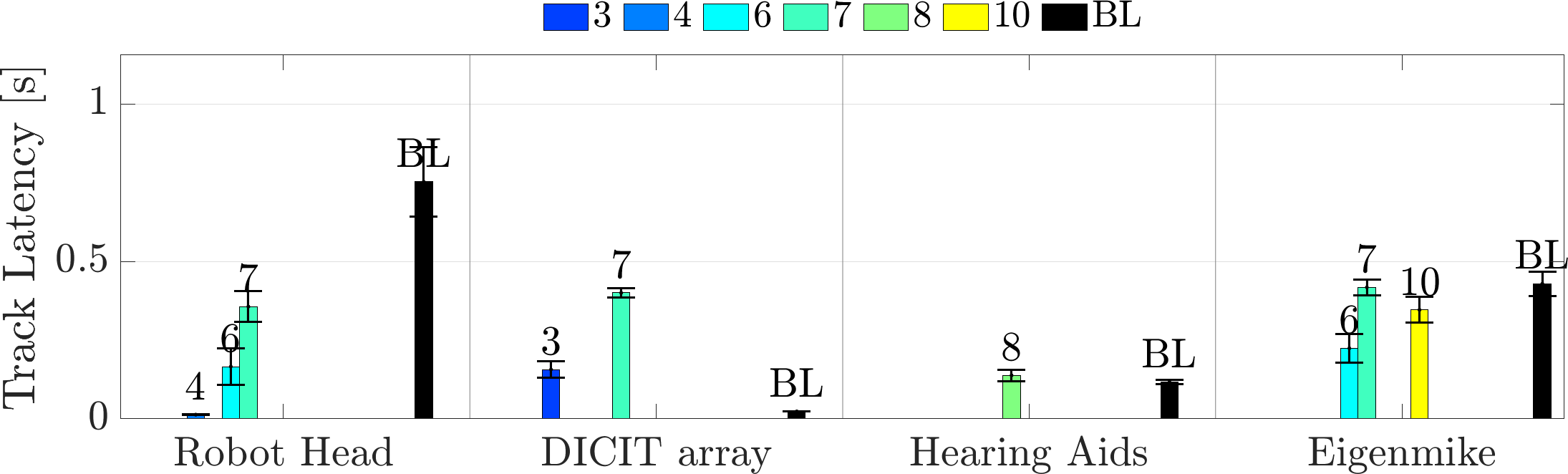}}}
  \\
  \mbox{\subfloat[Task 5]{
    \label{subfig:TL_135_task_5}
    \includegraphics[width=.99\columnwidth]{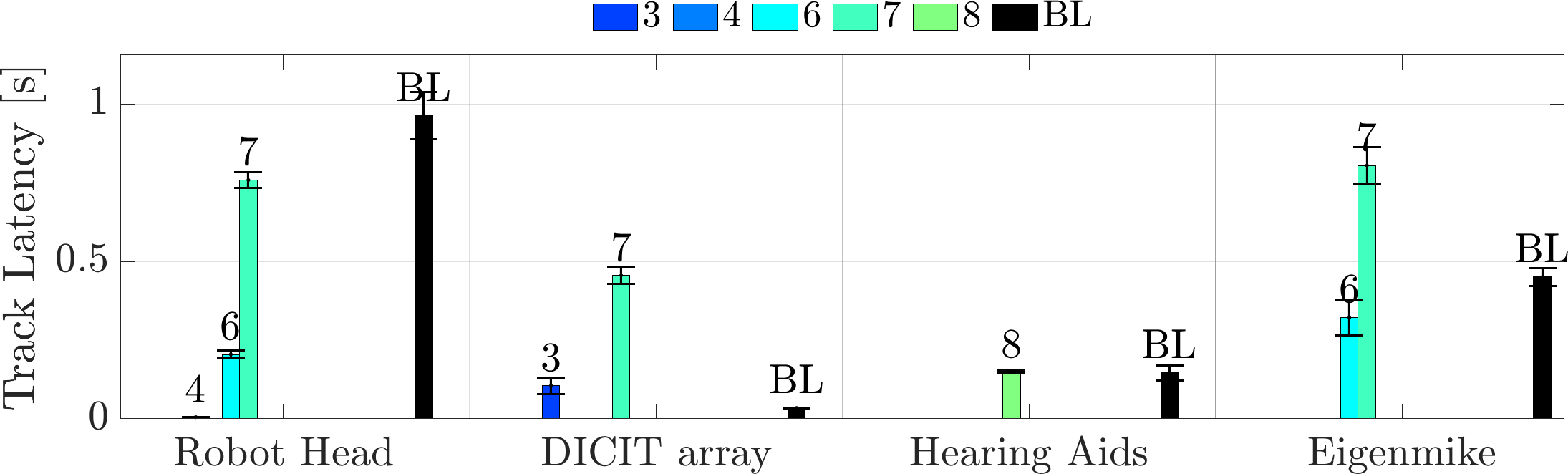}}}
  \caption{Track latency (bars) and standard deviation over recordings (whiskers) for Tasks 1, 3 and 5, for each submission and array. Legends indicate the submission \acsp{ID} available for each of the tasks.}
  \label{fig:TL_135}
\end{figure}

\begin{figure}[tb]
  \centering
  \mbox{\subfloat[Task 2, Track Fragmentation Rate]{
    \label{subfig:TFR_task2}
    \includegraphics[width=.99\columnwidth]{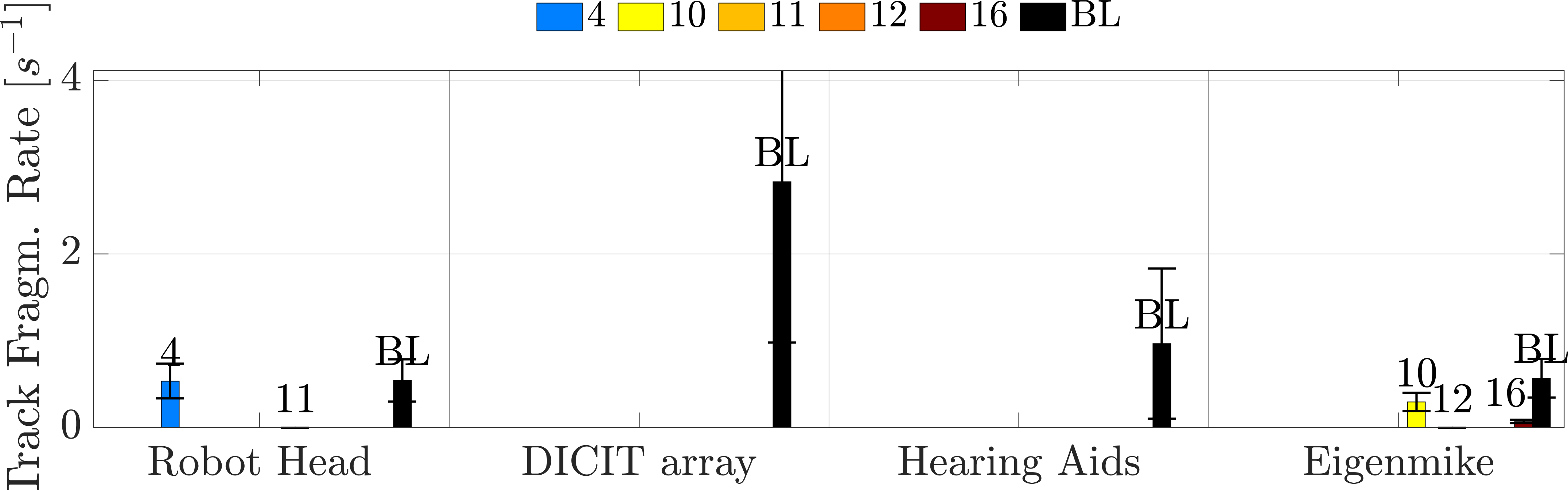}}}
  \\
  \mbox{\subfloat[Task 4, Track Fragmentation Rate]{
    \label{subfig:TFR_task4}
    \includegraphics[width=.99\columnwidth]{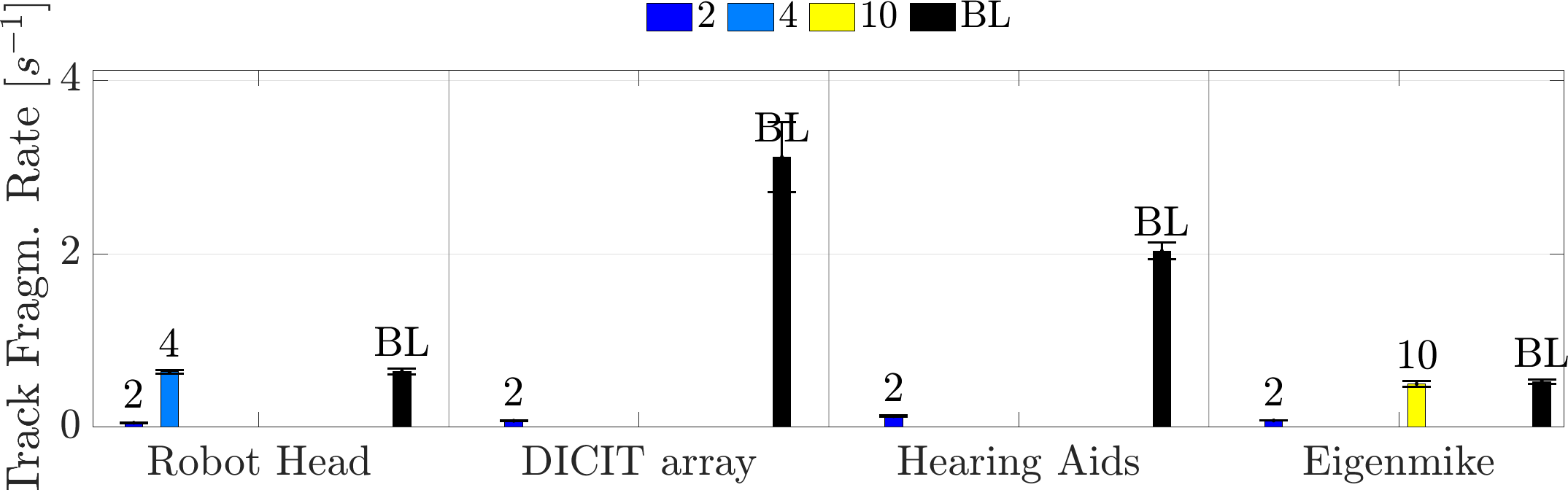}}}
  \\
  \mbox{\subfloat[Task 6, Track Fragmentation Rate]{
    \label{subfig:TFR_task6}
    \includegraphics[width=.99\columnwidth]{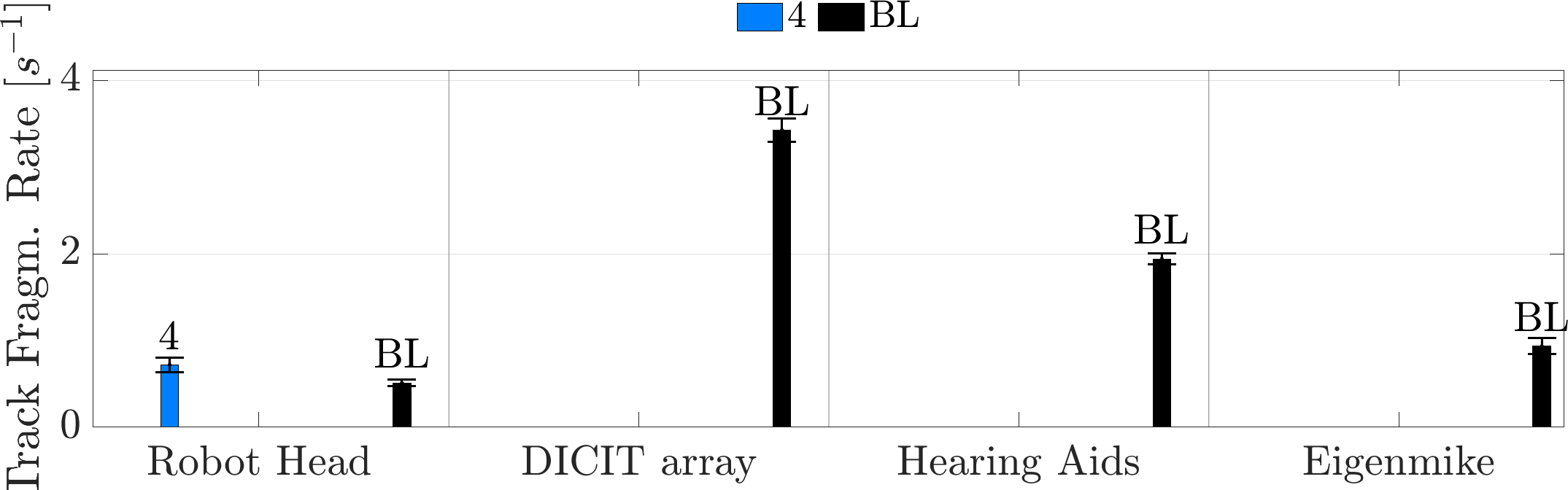}}}
  \caption{Track fragmentation rate (bars) and standard deviation over recordings (whiskers) for Tasks 2, 4, 6, for each submission and array.}
  \label{fig:TSR_TFR_246}
\end{figure}

\subsubsection{Tasks 1, 3, 5: Impact of Gating on Azimuth Accuracy}

To illustrate the effect of gating on the evaluation results, the evaluation was repeated without gating by assigning each source to its closest estimate.\footnote{Even though Tasks 1, 3 and 5 correspond to single-source scenarios, gating and association is required for evaluation, since azimuth estimates corresponding to multiple source \acp{ID} were provided for some submissions.} \tabref{table:azimuth_errors_gating_offset} provides the difference in the average azimuth errors with and without gating. In \tabref{table:azimuth_errors_gating_offset}, entries with value $0.0$ indicate that evaluation with and without gating lead to the same result. Entries with values greater than $0.0$ highlight that the azimuth error increases without gating, i.e., the submitted results are affected by outliers outside of the gating collar.

For the majority of submissions, a gating threshold of $30^{\circ}$ results in improved azimuth accuracies in the range of $0.1^{\circ}$ to $4^{\circ}$ across Tasks 1, 3 5. A significant number of outliers are observed for Submissions 1, 7 and 8. To reflect outliers in the analysis of the results, evaluation measures, such as the \ac{FAR} and probability of detection, are required in addition to the average azimuth error.

\subsubsection{Completeness \& Ambiguity}
\label{sec:eval_task135_completeness}
As detailed in \sect{sec:Measures}, the track cardinality and probability of detection are used as evaluation measures of the track completeness. For single-source scenarios, the track completeness quantifies the robustness of localization and tracking algorithms against changes in the source orientation and source-sensor distance. Furthermore, the \ac{FAR} is used as an evaluation measure of the track ambiguity, quantifying the robustness against early reflections and noise in the case of the single-source scenarios.

The probability of detection and \ac{FAR}, averaged over all recordings in each task, are shown in \fig{fig:pd_tasks_135} and \fig{fig:FAR_task13}, respectively. The results indicate that the probability of detection between Tasks~1, 3 and 5 remains approximately constant, with a trend towards a small reduction in $p_d$, when changing from static to dynamic sources.

The results also highlight that Submissions 11 and 12, corresponding to the highest average azimuth accuracy for Task~1 using the robot head and Eigenmike (see \sect{sec:eval_task1_azimuth}), exhibit $100$\% probability of detection. The same submissions also correspond to a comparatively high \ac{FAR} of 50 false estimates per second, averaged across all recordings for Task~1 and evaluated for the full duration of each recording (see \fig{subfig:FAR_recording_task1}). These results are indicative of the fact that Submissions 11 and 12 do not incorporate \ac{VAD} algorithms. For comparison, \fig{subfig:FAR_task13_task1} depicts the average \acp{FAR} for Task~1 evaluated during voice-activity only. The results in \fig{subfig:FAR_task13_task1} clearly highlight a significant reduction in the \ac{FAR} for Submissions 3, 6, 11, which do not incorporate \ac{VAD}.

\fig{subfig:az_task3_forFAR_sub67}, selected from Submission 6 for Task~3 and recording~2, shows that estimates during periods of voice inactivity are affected by outliers, which are removed from the measure for azimuth accuracy due to the gating process, and are accounted for in the \ac{FAR}. The majority of \ac{DoA} estimates provided during voice-activity correspond to smooth tracks near the ground-truth source azimuth. In the time interval  [15.1,17]~s, the estimates exhibit a temporary period of track divergence. The results for Submission 7 in \fig{subfig:az_task3_forFAR_sub67} highlight that outliers during voice inactivity are avoided since the submission incorporates \ac{VAD}. The results also indicate diverging track estimates in the interval [15.1,17]~s. The track divergence affecting both submissions is likely caused by the time-varying source-sensor geometry due to the motion of the source. \fig{subfig:range_task3_forFAR_sub7} highlights that the source is moving away from the array after 13~s. As the source orientation is directed away from the array, the contribution of the direct-path signal decreases, resulting in reduced estimation accuracy in the source azimuth. The reduction in azimuth accuracy eventually results in false estimates outside of the gating threshold.

\subsubsection{Timeliness}
The track latency is used as an evaluation measure of the timeliness of localization and tracking algorithms. Therefore, the track latency quantifies the sensitivity of algorithms to speech onsets, and the robustness against temporal smearing at speech endpoints.

\fig{fig:TL_135} shows the track latency, averaged across all recordings for Tasks~1, 3 and 5. Submissions 1, 3, 6, 8, 9, 11 and 12 do not incorporate \ac{VAD}. Hence, estimates are provided at every time stamp for all recordings. Submissions 3 and 8 incorporate tracking algorithms, where the source estimates are propagated through voice-inactive periods by track prediction. Submissions 1, 11 and 12, submitted for only the static tasks, estimate the average azimuth throughout the full recording duration and extrapolate the estimates across all time steps.

Therefore, for Task~1, Submissions 1, 3, 11 and 12 correspond to $0$~s track latency throughout. However, these algorithms also correspond to high \acp{FAR}, when the \ac{FAR} is evaluated across voice-active and inactive periods (see \fig{subfig:FAR_recording_task1}). Submissions 3 and 8, which do not involve a \ac{VAD} and were submitted to the tasks involving moving sources, result in track latencies of below $0.2$~s for Tasks 3 and 5, where the extrapolation of tracks outside of \acp{VAP} is non-trivial.

Submission 4 incorporates a \ac{VAD} that estimates voice activity as a side-product of the variational \ac{EM} algorithm for tracking. The results show that Submission 4 effectively detects speech onsets, leading to negligible track latencies across Tasks~1, 3 and 5. Submission 10, incorporating the noise \ac{PSD}-based \ac{VAD} of \cite{Gerkmann2012}, detects speech onsets accurately in the static source scenario in Task~1. However, the track latency for Task~3, involving a moving source, increases to $0.35$~s.
It is important to note that Submissions 7 and 10 incorporate Kalman or particle filters with heuristic approaches to track initialization. Therefore, it is likely that track initialization rules - rather than the \ac{VAD} algorithms - lead to delays in the confirmation of newly active sources.


\begin{figure}[tb]
  \centering
  \mbox{\subfloat[Azimuth estimates: Task 2, Recording 5]{
    \label{subfig:TSR_task2_sub4}
    \includegraphics[width=.94\columnwidth]{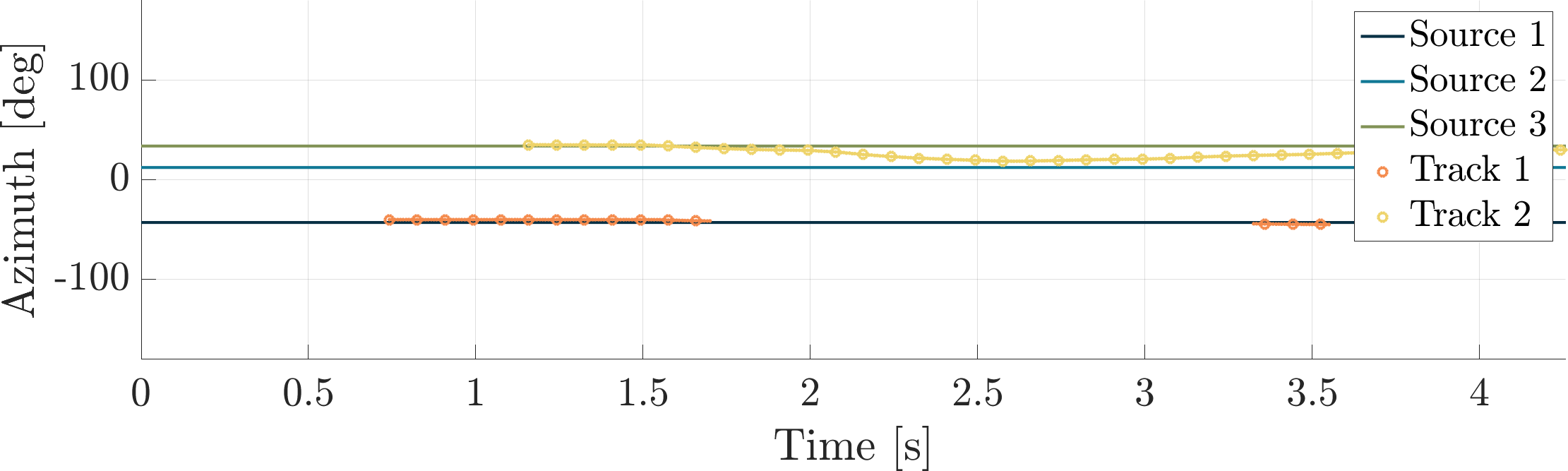}}}
  \\
  \mbox{\subfloat[VAD: Task 2, Recording 5]{
    \label{subfig:TSR_task2_sub4_VAD}
    \includegraphics[width=.94\columnwidth]{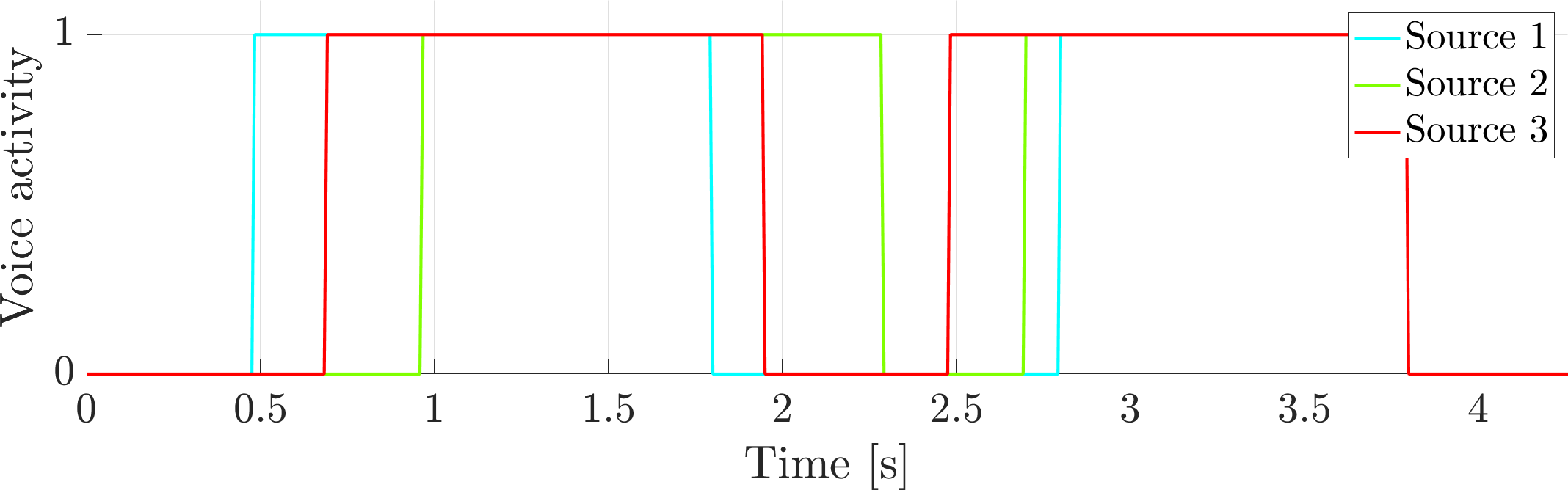}}}
  \\
  \mbox{\subfloat[Azimuth estimates: Task 4, Recording 4]{
    \label{subfig:TSR_task4_sub4}
    \includegraphics[width=.94\columnwidth]{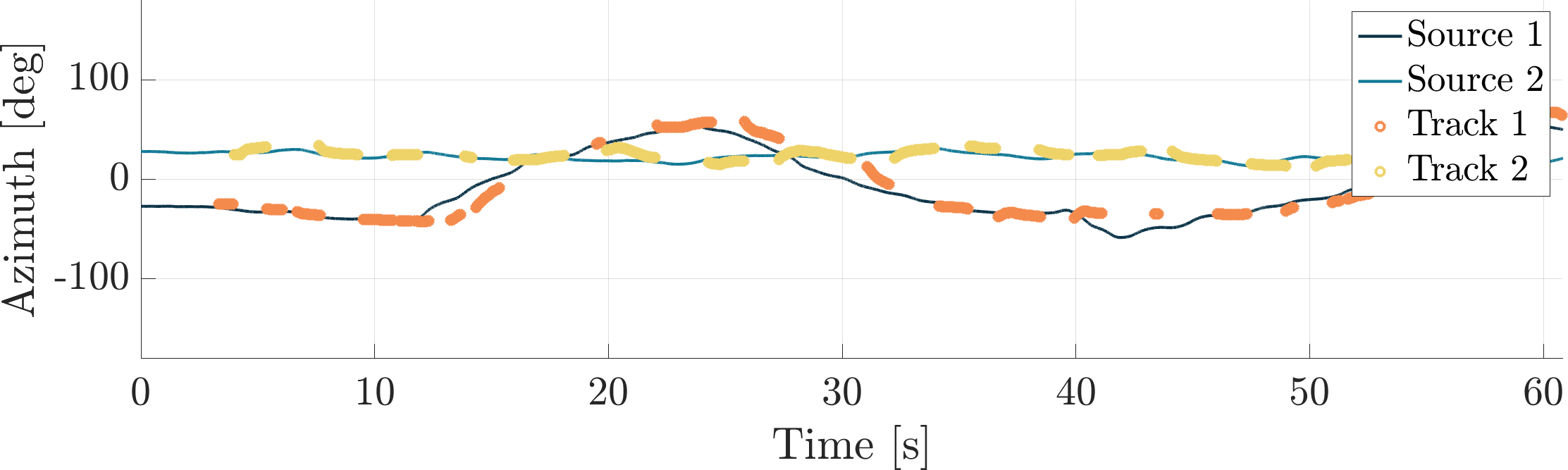}}}
  \\
  \mbox{\subfloat[VAD: Task 4, Recording 4]{
    \label{subfig:TSR_task4_sub4_VAD}
    \includegraphics[width=.94\columnwidth]{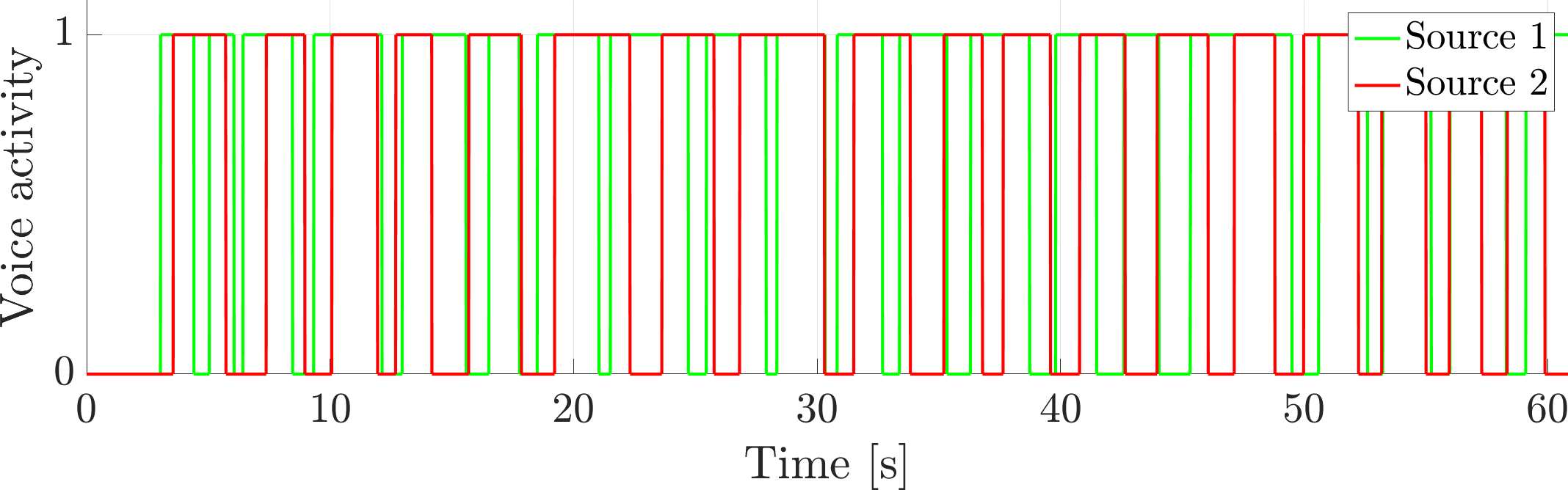}}}
  \\
  \mbox{\subfloat[Azimuth estimates: Task 6, Recording 2]{
    \label{subfig:TSR_task6_sub4}
    \includegraphics[width=.94\columnwidth]{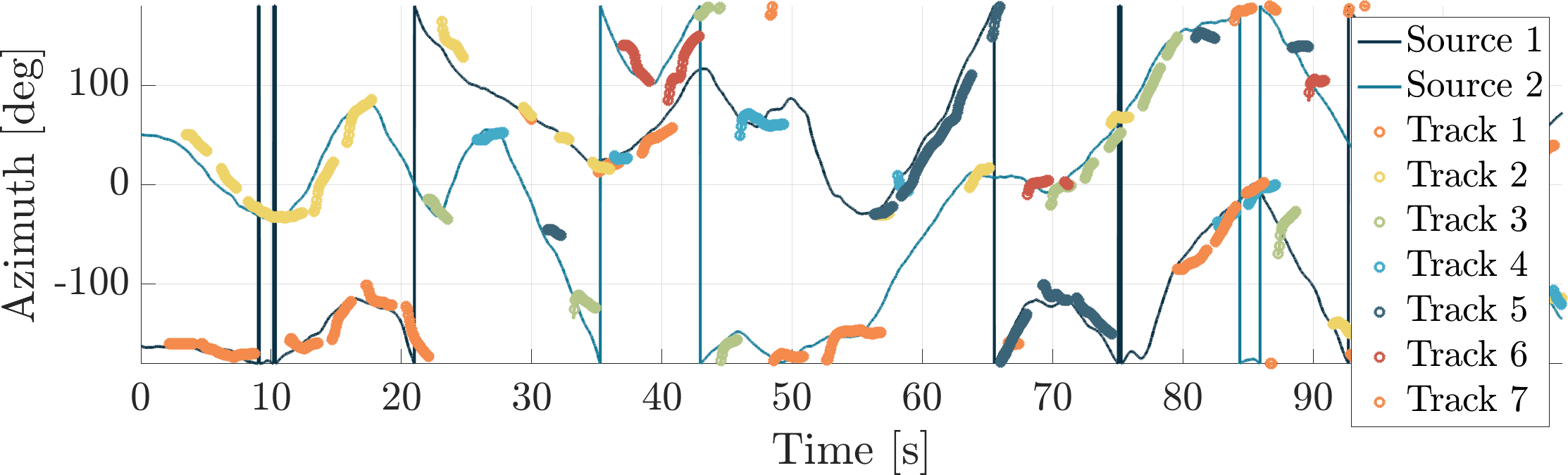}}}
  \\
  \mbox{\subfloat[VAD: Task 6, Recording 2]{
    \label{subfig:TSR_task6_sub4_VAD}
    \includegraphics[width=.94\columnwidth]{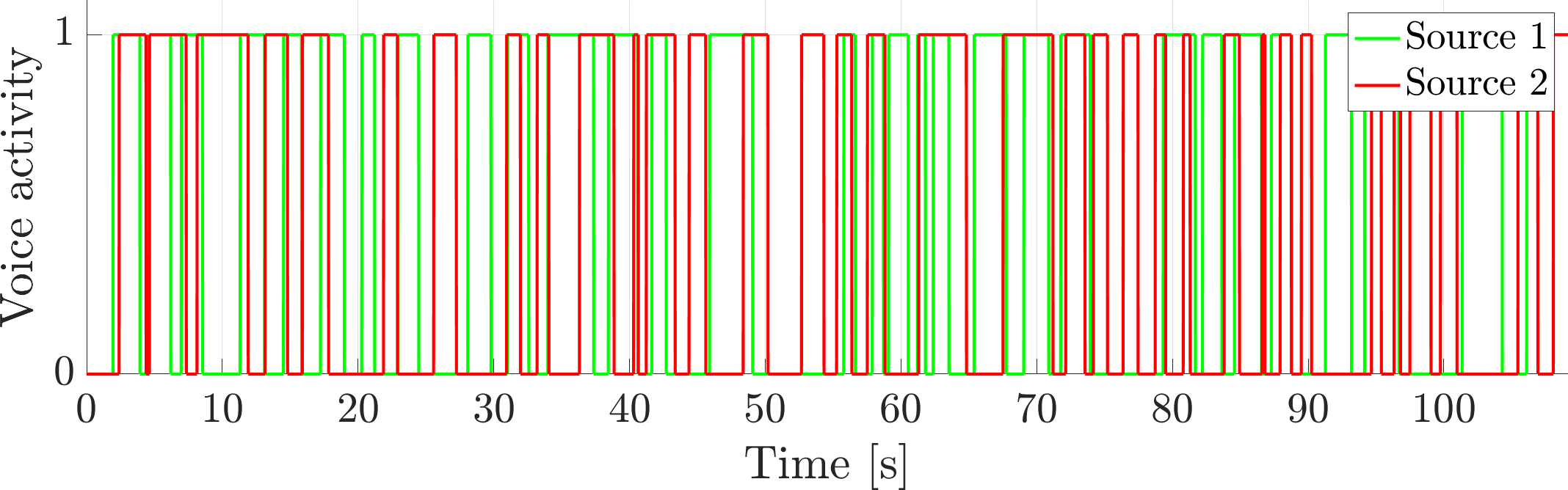}}}
  \caption{Azimuth estimates and \acs{VAD} for Submission 4 using the robot head for (a)-(b) Task~2, (c)-(d) Task~4, and (e)-(f) Task~6.}
  \label{fig:TSR_task246_sub4}
\end{figure}

\begin{figure}
  \centering
  \mbox{\subfloat[Submission 2, Eigenmike]{
    \label{subfig:OSPA_task4_rec1_sub2_emaz}
    \includegraphics[width=.99\columnwidth]{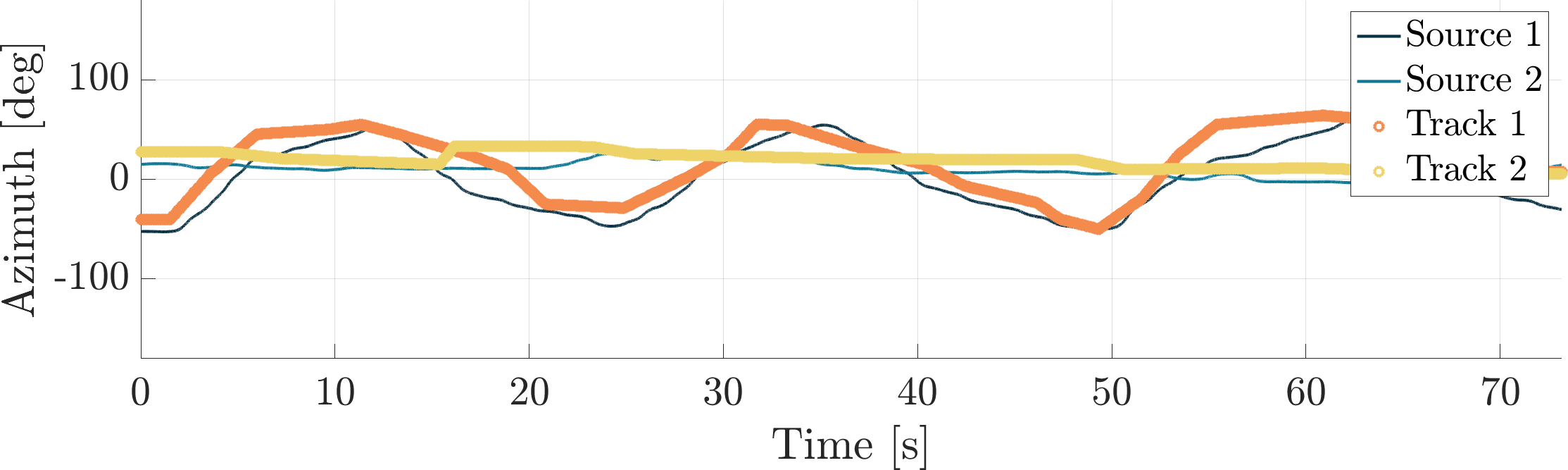}}}
  \\
  \mbox{\subfloat[Submission 2, Eigenmike]{
    \label{subfig:OSPA_task4_rec1_sub2_emOSPA}
    \includegraphics[width=.99\columnwidth]{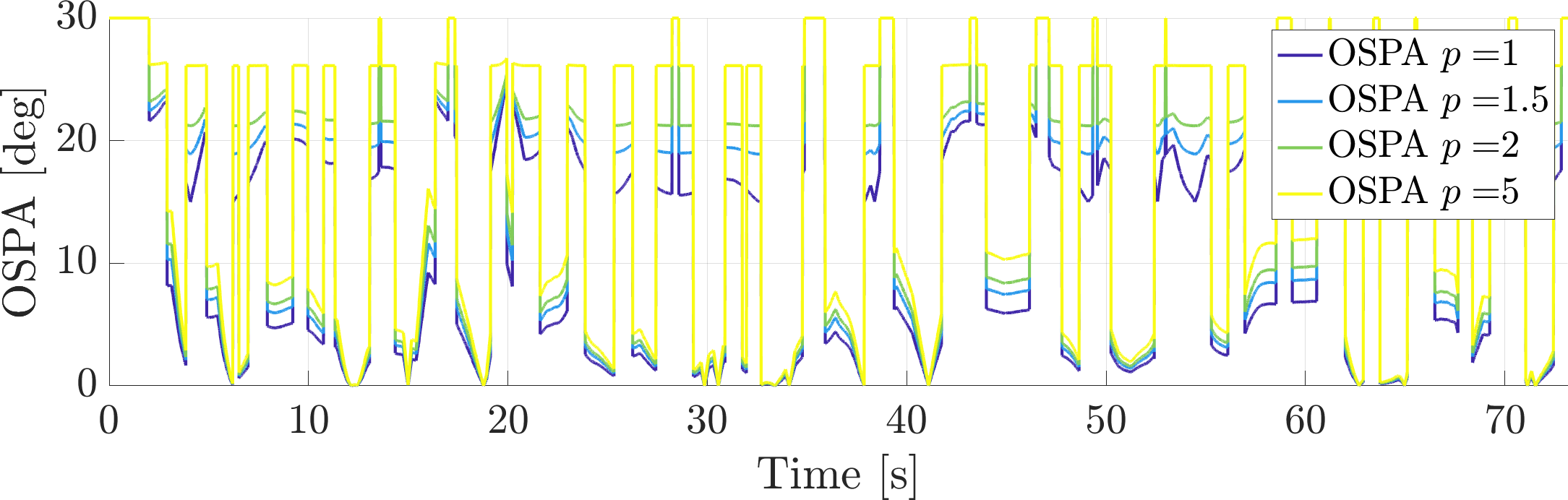}}}
  \\
  \mbox{\subfloat[Submission 10, Eigenmike]{
    \label{subfig:OSPA_task4_rec1_sub10_az}
    \includegraphics[width=.99\columnwidth]{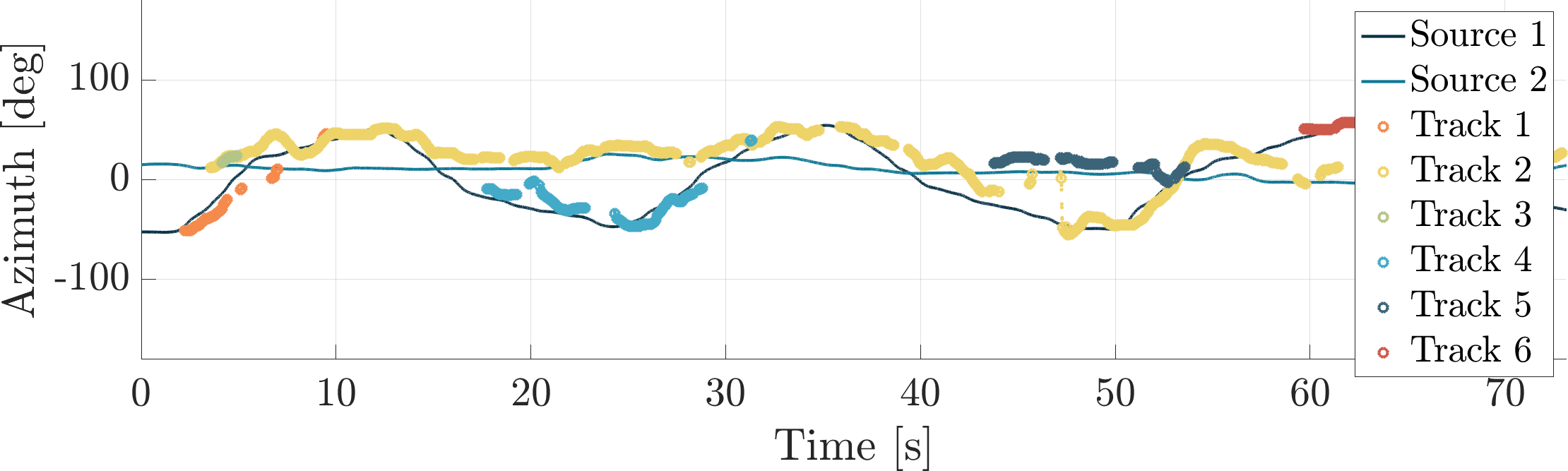}}}
  \\
  \mbox{\subfloat[Submission 10, Eigenmike]{
    \label{subfig:OSPA_task4_rec1_sub10_OSPA}
    \includegraphics[width=.99\columnwidth]{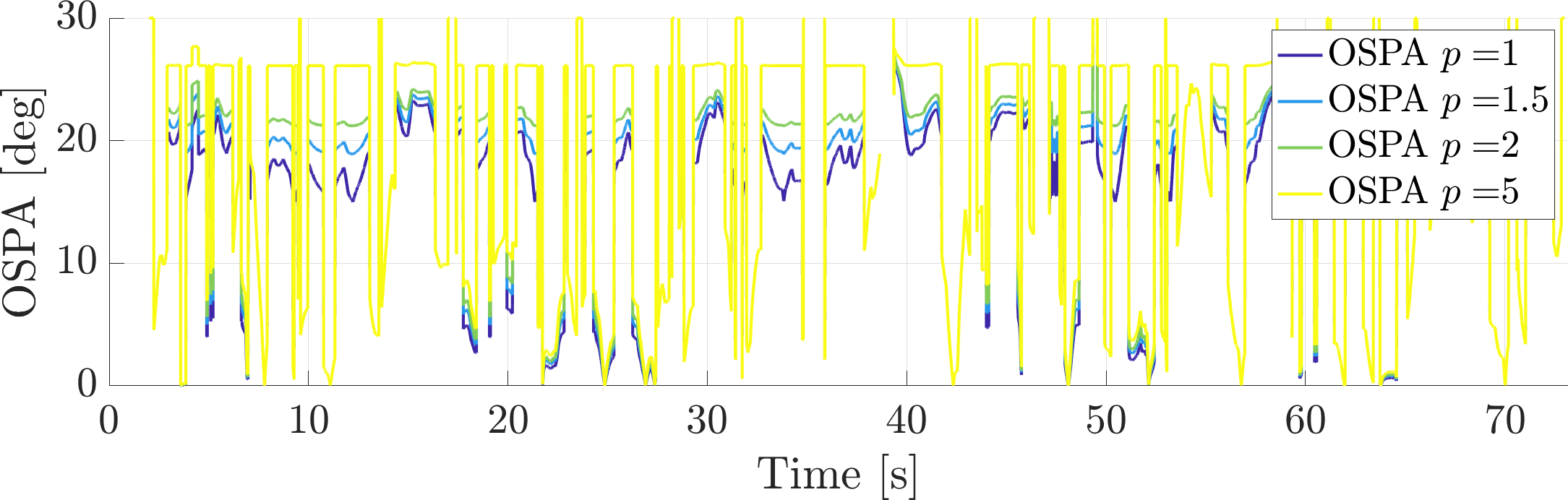}}}
  \\
  \mbox{\subfloat[VAD, Eigenmike]{
    \label{subfig:OSPA_task4_rec1_sub2_OSPA_VAD_em}
    \includegraphics[width=.99\columnwidth]{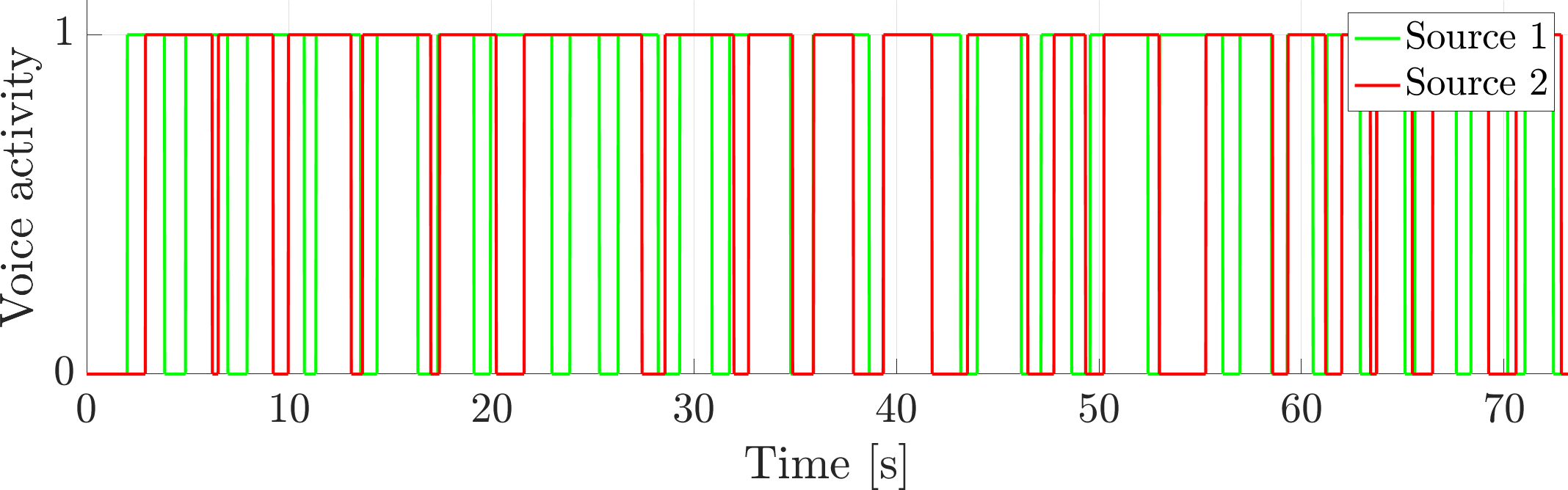}}}
  \caption{Azimuth trajectories and corresponding \acs{OSPA} metric for recording 1 of Task~4 for (a)-(b) Submission 2 using the Eigenmike, (c)-(d) Submission 10 using the Eigenmike. The VAD periods are shown in (e).}
  \label{fig:OSPA_task4}
\end{figure}

\subsection{Multi-Source Tasks 2, 4, 6}
\label{sec:results_246}
\subsubsection{Accuracy}
For the multi-source Tasks 2, 4 and 6, the results in \tabref{table:azimuth_errors} indicate similar trends as discussed for the single-source Tasks 1, 3 and 5. However, the overall performance of all submissions for Tasks 2, 4 and 6 is decreased compared to Tasks 1, 3 and 5.

The reduction in azimuth accuracy is due to the adverse effects of interference from multiple simultaneously active sound sources. Due to the broadband nature of speech, the speech signals of multiple talkers often correspond to energy in the overlapping time-frequency bins, especially for talkers with similar voice pitch. Therefore, localization approaches that rely on the $W$-disjoint orthogonality of speech may result in biased estimates of the \ac{DoA} (see, e.g., Submission 4).

Robustness against interference can be achieved by incorporating time-frequency bins containing the contribution of a single source only, e.g., at the onset of speech. For example, Submission 11 and 12 incorporate the \ac{DPD}-test in \cite{Nadiri2014}, and result in azimuth accuracies of $2.0^{\circ}$ and $1.4^{\circ}$, respectively, for the robot head and Eigenmike in Task~2, compared to $0.7^{\circ}$ and $1.1^{\circ}$ in Task~1.

An increasing number of sources also results in an increasingly diffuse sound field in reverberant environments. For data-dependent beamforming techniques \cite{vanVeen1988}, the directivity pattern of the array is typically evaluated based on the signal and noise levels. For increasing diffuse noise, it is therefore expected that the performance of beamforming techniques decreases in multi-source scenarios.

In addition to a reduction in the angular accuracy, ambiguities arising in scenarios involving multiple, simultaneously active sound sources result in missing and false \ac{DoA} estimates, affecting the completeness, continuity, and ambiguity of localization and tracking approaches.

\subsubsection{Continuity}
The \ac{TFR} is used as an evaluation measure for track continuity (see \sect{sec:Results}). \fig{fig:TSR_TFR_246} provides the \acp{TFR} for Tasks 2, 4 and 6 for each array and submission and averaged over the recordings.

The results indicate that the subspace-based Submissions 11, 12 and 16 are robust to track fragmentation. Although the submissions rely on the assumption of $W$-disjoint orthogonal sources, localization is performed only on a subset of frequency bins that correspond to the contribution of a single source. In contrast, \ac{BSS}-based approaches assume that the $W$-disjoint orthogonality applies to all frequency bands required for the reconstruction of the source signals.

The advantage of subspace-based processing for robustness against track fragmentation is reinforced when comparing the results for Submission 10, based on pseudo-intensity vectors for ambisonics, against Submission 16, using subspace pseudo-intensity vectors in the domain of spherical harmonics. The azimuth accuracies of both submissions are comparable, where Submission 10 results in an average azimuth error of $7.3^{\circ}$ and Submission 16 leads to $7.1^{\circ}$ in Task~2. In contrast, Submission 10 leads to $0.3$ fragmentations per second, whereas Submission 16 exhibits only $0.07$ fragmentations per second.

Comparing the results for static Task~2 against the moving-source Task~4 and the fully dynamic Task~6, the results in \fig{fig:TSR_TFR_246} highlight increasing \acp{TFR} across submissions. For example, Submission 4, the only approach that was submitted for all three multi-source tasks, corresponds to 0.53 fragmentations per second for Task~2, involving multiple static loudspeakers, to 0.64 fragmentations per second for Task~4, involving multiple moving human talkers, and to 0.71 fragmentations per second for Task~6 involving multiple moving human talkers and moving arrays. The increasing \ac{TFR} is due to the increasing spatio-temporal variation of the source azimuth between the three tasks. Task~2 corresponds to constant azimuth trajectories of the multiple static loudspeakers, observed from static arrays (see  \fig{subfig:TSR_task2_sub4}, showing the azimuth estimates for Task~2, recording 5).
The motion of the human talkers that are observed from static arrays in Task~4 correspond to time-varying azimuth trajectories within limited intervals of azimuth values. For example, for Task~4, recording 4 shown in \fig{subfig:TSR_task4_sub4}, source 1 is limited to azimuth values in the interval between $[6,24]^{\circ}$, whilst source 2 is limited between $[-66,50]^{\circ}$. The motion of the moving sources and moving arrays in Task~6 result in azimuth trajectories that vary significantly between $[-180,180]^{\circ}$ (see \fig{subfig:TSR_task6_sub4} for the azimuth estimates provided for Task~6, recording~2). Furthermore, the durations of recordings for Task~4 and Task~6 are substantially longer than those for Task~2. As to be expected, periods of speech inactivity and the increasing time-variation of the source azimuth relative to the arrays result in increasing \acp{TFR} when comparing Task~2, Task~4, and Task~6.

\begin{table*}[tb]
\centering
\caption{Average \acs{OSPA} results. Column colour indicates type of algorithm, where white indicates frameworks involving only \acs{DoA} estimation (Submission IDs 11, 12, 16 and the baseline (BL)), and grey indicates frameworks that combine \ac{DoA} estimation with source tracking (Submission IDs 2, 4, 10).}
\label{table:OSPA}
\begin{tabular}{|c|c||>{\columncolor{darkgrey}}c|>{\columncolor{darkgrey}}c||>{\columncolor{darkgrey}}c|>{\columncolor{darkgrey}}c||>{\columncolor{darkgrey}}c|>{\columncolor{darkgrey}}c||c|c||c|c||c|c||c|c|}
\hline
\multirow{4}{*}{\rotatebox[origin=c]{90}{Task}} & \multirow{4}{*}{Array} & \multicolumn{14}{c|}{Submission ID}\\\cline{3-16}
 & & \multicolumn{2}{|c|}{2} & \multicolumn{2}{|c|}{4} & \multicolumn{2}{|c|}{10} & \multicolumn{2}{|c|}{11} & \multicolumn{2}{|c|}{12} & \multicolumn{2}{|c|}{16} & \multicolumn{2}{|c|}{BL}\\\cline{3-16}
& & \multicolumn{2}{c|}{$p$} & \multicolumn{2}{c|}{$p$} & \multicolumn{2}{c|}{$p$} & \multicolumn{2}{c|}{$p$} & \multicolumn{2}{c|}{$p$} & \multicolumn{2}{c|}{$p$} & \multicolumn{2}{c|}{$p$} \\\cline{3-16}
& & $1$ & $5$ & $1$ & $5$ & $1$ & $5$ & $1$ & $5$ & $1$ & $5$ & $1$ & $5$ & $1$ & $5$ \\\hline\hline

\multirow{4}{*}{2} & Robot Head & - & - & 17.5 & 22.4 & - & - & 12.4 & 17.6 & - & - & - & - & 19.5 & 23.8 \\
\cline{2-16}
 & DICIT & - & - & - & - & - & - & - & - & - & - & - & - & 26.6 & 28.0 \\
\cline{2-16}
 & Hearing Aids & - & - & - & - & - & - & - & - & - & - & - & - & 26.1 & 27.7 \\
\cline{2-16}
 & Eigenmike & - & - & - & - & 17.5 & 22.3 & - & - & 12.2 & 17.3 & 12.4 & 18.2 & 21.5 & 25.0 \\
\cline{1-16}
\multirow{4}{*}{4} & Robot Head & 13.8 & 18.9 & 13.5 & 16.4 & - & - & - & - & - & - & - & - & 16.3 & 18.9 \\
\cline{2-16}
 & DICIT & 15.6 & 20.0 & - & - & - & - & - & - & - & - & - & - & 25.8 & 26.6 \\
\cline{2-16}
 & Hearing Aids & 15.2 & 19.6 & - & - & - & - & - & - & - & - & - & - & 27.7 & 28.1 \\
\cline{2-16}
 & Eigenmike & 14.6 & 19.3 & - & - & 13.1 & 16.4 & - & - & - & - & - & - & 18.4 & 20.8 \\
\cline{1-16}
\multirow{4}{*}{6} & Robot Head & - & - & 13.8 & 15.0 & - & - & - & - & - & - & - & - & 14.8 & 15.8 \\
\cline{2-16}
 & DICIT & - & - & - & - & - & - & - & - & - & - & - & - & 24.8 & 25.2 \\
\cline{2-16}
 & Hearing Aids & - & - & - & - & - & - & - & - & - & - & - & - & 25.2 & 25.8 \\
\cline{2-16}
 & Eigenmike & - & - & - & - & - & - & - & - & - & - & - & - & 21.1 & 21.7 \\
\hline
\end{tabular}
\end{table*}

\subsubsection{\acs{OSPA} - Accuracy vs. Ambiguity, Completeness and Continuity}
The results for the \ac{OSPA} measure, averaged over all recordings for the multi-source Tasks 2, 4 and 6, is summarized for order parameters $p = \{1, 5\}$ (see \eq{eqn:OSPA}) in \tabref{table:OSPA}. In contrast to the averaged azimuth errors in \tabref{table:azimuth_errors}, the \ac{OSPA} results trade off the azimuth accuracy against cardinality errors, and hence false and missing track estimates. For example, the results for Task~2 in \tabref{table:azimuth_errors} indicate a significant difference in the results for Submission 12 ($1.4^{\circ}$) and Submission 16 ($7.1^{\circ}$). In contrast, due to false track estimates during periods of voice inactivity, \tabref{table:OSPA} highlights only a small difference between the \ac{OSPA} for Submissions 12 and 16.

To provide intuitive insight into the \ac{OSPA} results and the effect of the order parameter, $p$, \fig{fig:OSPA_task4} compares the azimuth estimates obtained using Submissions 2 and 10 for the Eigenmike, Task 4, Recording 1.

The results highlight distinct jumps of the \ac{OSPA} between periods during which a single source is active and the onsets of periods of two simultaneously active sources.
During periods of voice inactivity, detection errors in the onsets of speech lead to errors corresponding to the cutoff threshold of $c = 30^{\circ}$. Therefore, the cardinality error dominates the \ac{OSPA} when $N(t) = 0$ and $K(t) > 0$.
During \acp{VAP} where $N(t) = K(t)$, the \ac{OSPA} is dominated by the angular error between each estimate and the ground-truth direction of each source, resulting in values in the range of $[0,20]^{\circ}$. For $N(t) = K(t)$, the order parameter, $p$, does not affect the results since the cardinality error is $K(t) - N(t) = 0$.
During periods where $K(t) < N(t)$, the cardinality error causes the \ac{OSPA} to increase to between $[15,30]^{\circ}$. The \ac{OSPA} increases with the order parameter $p$.

The results highlight that both approaches are affected by cardinality errors, indicated by jumps in the \ac{OSPA}. For Submission~10, which incorporates \ac{VAD}, the cardinality errors arise predominantly due to missing detections and broken tracks (see \fig{subfig:OSPA_task4_rec1_sub10_OSPA}). In contrast, Submission~2 is mainly affected by false estimates during voice inactivity. Since Submission~2 does not involve a \ac{VAD}, tracks are propagated through periods of voice inactivity using the prediction step of the tracking filter. Temporary periods of track divergence therefore lead to estimates that are classified as false estimates by gating and data association.

\section{Discussion and Conclusions}
\label{sec:Conclusions}

The open-access \ac{LOCATA} challenge data corpus of real-world, multichannel audio recordings and open-source evaluation software provides a framework to objectively benchmark state-of-the-art localization and tracking approaches. The challenge consists of six tasks, ranging from the localization and tracking of a single static loudspeaker using static microphone arrays to fully dynamic scenes involving multiple moving sources and microphone arrays on moving platforms. Sixteen state-of-the-art approaches were submitted for participation in the \ac{LOCATA} challenge, one of which needed to be discarded for evaluation due to the lack of documentation. Seven submissions corresponded to sound source localization algorithms, obtaining instantaneous estimates at each time stamp of a recording. The remaining  submissions combined localization algorithms with source tracking, where spatio-temporal models of the source motion are applied in order to exploit constructively knowledge of the history of the source trajectories. The submissions incorporated localization algorithms based on time-delay estimation, subspace processing, beamforming, classification, and deep learning. Source tracking submissions incorporated the Kalman filter and its variants, particle filters, variational Bayesian approaches and \ac{PHD} filters.

The controlled scenarios of static single-source in Task~1 are used to evaluate the robustness of the submissions against reverberation and noise. The results highlighted azimuth estimation accuracies of up to approximately $1.0^{\circ}$ using the pseudo-spherical robot head, spherical Eigenmike and planar DICIT array. For the hearing aids, recorded separately but in the same environment, the average azimuth error was $8.5^{\circ}$. Interference from multiple static loudspeakers in Task~2 leads to only small performance degradations of up to $3^{\circ}$ compared to Task~1. Variations in the source-sensor geometries due to the motion of the human talkers (Tasks 3 and 4), or the motion of the arrays and talkers (Tasks 5 and 6) affect predominantly the track continuity, completeness and timeliness.

The evaluation also provides evidence for the intrinsic suitability of a given approach for particular arrays or scenarios. For static scenarios (i.e., Tasks 1 and 2), subspace approaches demonstrated particularly accurate localization using the Eigenmike and the robot head incorporating a large number of microphones. Time delay estimation combined with a particle filter resulted in the highest azimuth estimation accuracy for the planar \ac{DICIT} array. Tracking filters were shown to reduce \acp{FAR} and missing detections by exploiting models of the source dynamics. Specifically, the localization for moving human talkers in Tasks 3-6 benefits from the incorporation of tracking in dynamic scenarios, resulting in azimuth accuracies of up to $1.8^{\circ}$ using the \ac{DICIT} array, $3.1^{\circ}$ using the robot head, and $7.2^{\circ}$ using the hearing aids.

Results for the Eigenmike highlighted that localization using spherical arrays benefits from signal processing in the domain of spherical harmonics. The results also indicated that the number of microphones in an array, to some extent, can be traded off against the array aperture. This conclusion is underpinned by the localization results for the 12-microphone robot head that consistently outperformed the 32-microphone Eigenmike for approaches evaluated for both arrays. Nevertheless, increasing microphone spacings also lead to increasingly severe effects of spatial aliasing. As a consequence, all submissions for the 2.24~m-wide \ac{DICIT} array used subarrays of at most 32~cm inter-microphone spacings.

Several issues remain open challenges for localization and tracking approaches. Intuitively, localization approaches benefit from accurate knowledge of the onsets and endpoints of speech to avoid false estimates during periods of speech inactivity. Several approaches therefore incorporated voice activity detection based on power spectral density estimates, zero-crossing rates, or by implicit estimation of the onsets and endpoints of speech from the latent variables estimated within a variational Bayesian tracking approach. For the single-source scenarios, particularly low track latency was achieved by the submission based on implicit estimation of the voice activity periods. However, for the multi-source scenarios, approaches incorporating voice activity detection led to increased track fragmentation rates.

Morover, whereas sufficiently long frames are required to address the non-stationarity of speech, dynamic scenes involving moving sources and/or sensors require sufficiently short frames to accurately capture the spatio-temporal variation of the source positions. Therefore, in dynamic scenes, estimation errors due to the non-stationarity of speech must be traded off against biased \ac{DoA} estimates due to spatio-temporal variation in the source-sensor geometries when selecting the duration of the microphone signals used for localization. In combination with the adverse effects of reverberation and noise, non-stationary signals in dynamic scenes therefore often lead to erroneous, false, missing, spurious \ac{DoA} estimates in practice.


To conclude, current research is predominantly focused on static scenarios. Only a small subset of the approaches submitted to the \ac{LOCATA} challenge address the difficult real-world tasks involving multiple moving sources. The challenge evaluation highlighted that there is significant room for improvement, and hence substantial potential for future research. Except for localizing a single static source in not too hostile scenarios none of the problems is robustly solved to the extent desirable for, e.g., informed spatial filtering with high spatial resolution.
Therefore, research on appropriate localization and tracking techniques remains an open challenge and the authors hope that the \ac{LOCATA} dataset and evaluation tools will be found useful to also evaluate future progress.

Inevitably, there are substantial practical limitations in setting up data challenges. In the case of \ac{LOCATA}, it has resulted in the use of only one acoustic environment because of the need for spatial localization of the ground-truth. Future challenges may beneficially explore variation in performance across different environments.

\section*{Acknowledgement}
The authors would like to thank all participants of the \ac{LOCATA} challenge for their submissions and feedback; Claas-Norman Ritter for his contributions to the recordings of the \ac{LOCATA} corpus; Prof. Verena Hafner for providing access to the facilities at Humboldt-Universit\"at zu Berlin; the anonymous reviewers for their positive and helpfully detailed comments that led to significant improvements of this manuscript; and the IEEE SPS Technical Committee on Audio and Acoustic Signal Processing and the IEEE SPS Challenges and Data Collections subcommittee for the support of the \ac{LOCATA} challenge.

\ifCLASSOPTIONcaptionsoff
  \newpage
\fi

\bibliographystyle{Templates/IEEEtranBST/IEEEtran}
\bibliography{references_LOCATA.bib}

\begin{IEEEbiography}[{\includegraphics[width=1in,height=1.25in,clip,keepaspectratio]{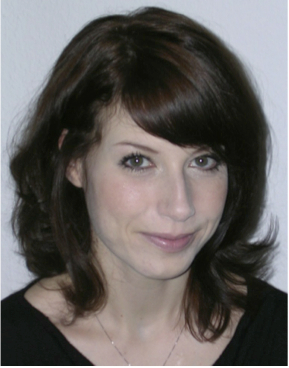}}]{Christine Evers} (M'14-SM'16) is a lecturer in Computer Science at the University of Southampton. She was the recipient of an EPSRC Fellowship, hosted at Imperial College London between 2017-2019. She worked as a research associate at Imperial College London between 2014-2017; as a senior systems engineer at Selex Electronic Systems between 2010-2014; and as a research fellow at the University of Edinburgh between 2009-2010. She received her PhD from the University of Edinburgh in 2010; her MSc degree in Signal Processing and Communications from the University of Edinburgh in 2006; and her BSc degree in Electrical Engineering and Computer Science from Jacobs University, Germany, in 2005. Her research focuses on Bayesian learning for machine listening. She is currently member of the IEEE SPS Technical Committee on Audio and Acoustic Signal Processing and serves as an associate editor of the EURASIP Journal on Audio, Speech, and Music Processing.
\end{IEEEbiography}

\begin{IEEEbiography}[{\includegraphics[width=1.05in,height=1.3in,clip,keepaspectratio]{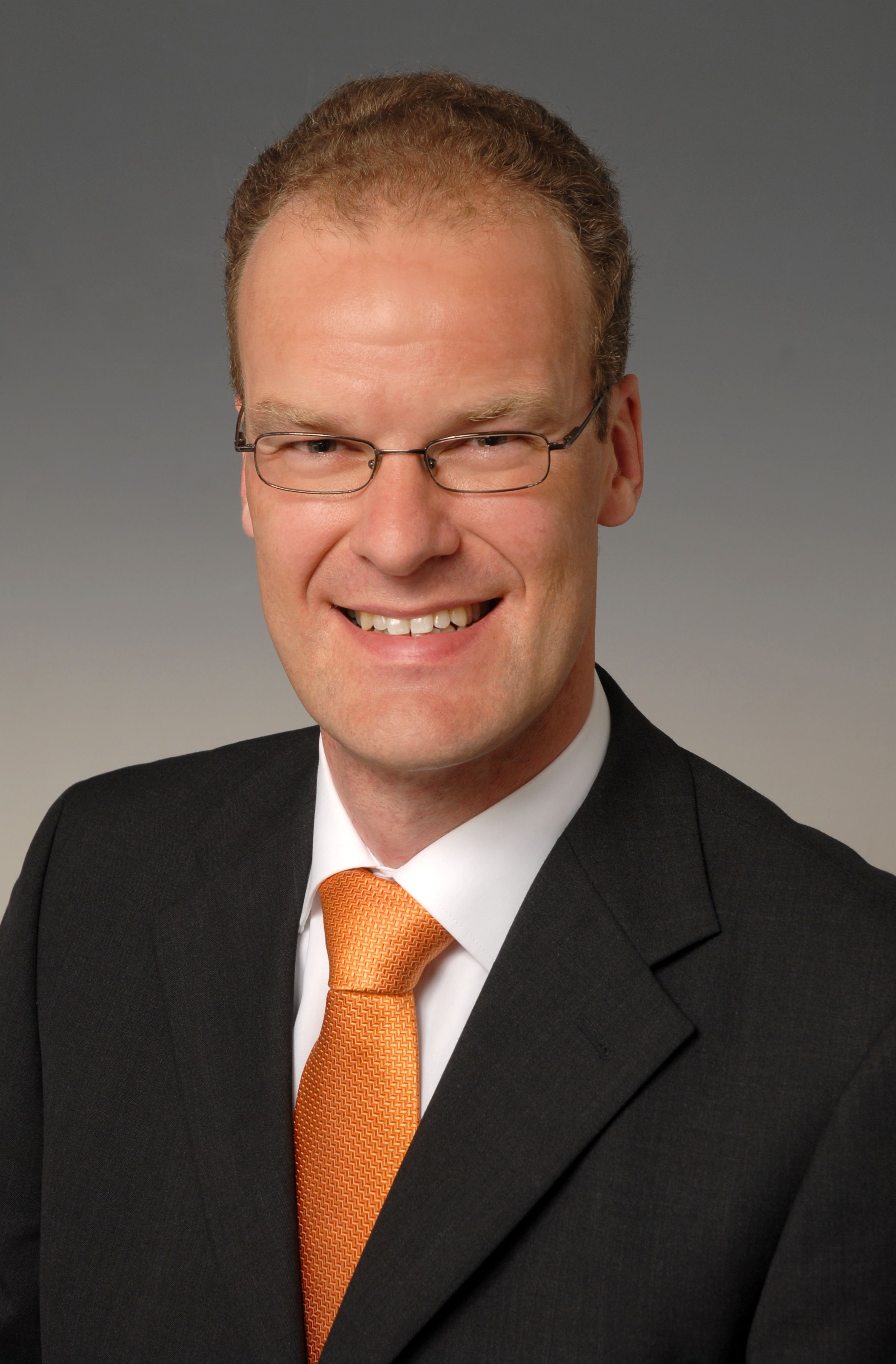}}]{Heinrich W. L\"ollmann} (M'08-SM'18) received the Dipl.-Ing. (univ) degree in electrical engineering in 2001 and the Dr.-Ing. degree in 2011 at RWTH Aachen University. He worked as scientific co-worker at the Institute of Communication Systems and Data Processing of RWTH Aachen University from 2001 to 2012. Since 2012, he is a senior researcher at the Chair of Multimedia Communications and Signal Processing at the Friedrich-Alexander University Erlangen-N{\"u}rnberg (FAU). Heinrich L{\"o}llmann has authored one book chapter and more than 40 refereed papers. His research focuses on speech and audio signal processing, including filter-bank design, speech dereverberation and noise reduction, estimation of room acoustical parameters, and algorithms for robot audition. He is currently member of the IEEE SPS Technical Committee on Audio and Acoustic Signal Processing.
\end{IEEEbiography}

\begin{IEEEbiography}[{\includegraphics[width=1in,height=1.25in,clip,keepaspectratio]{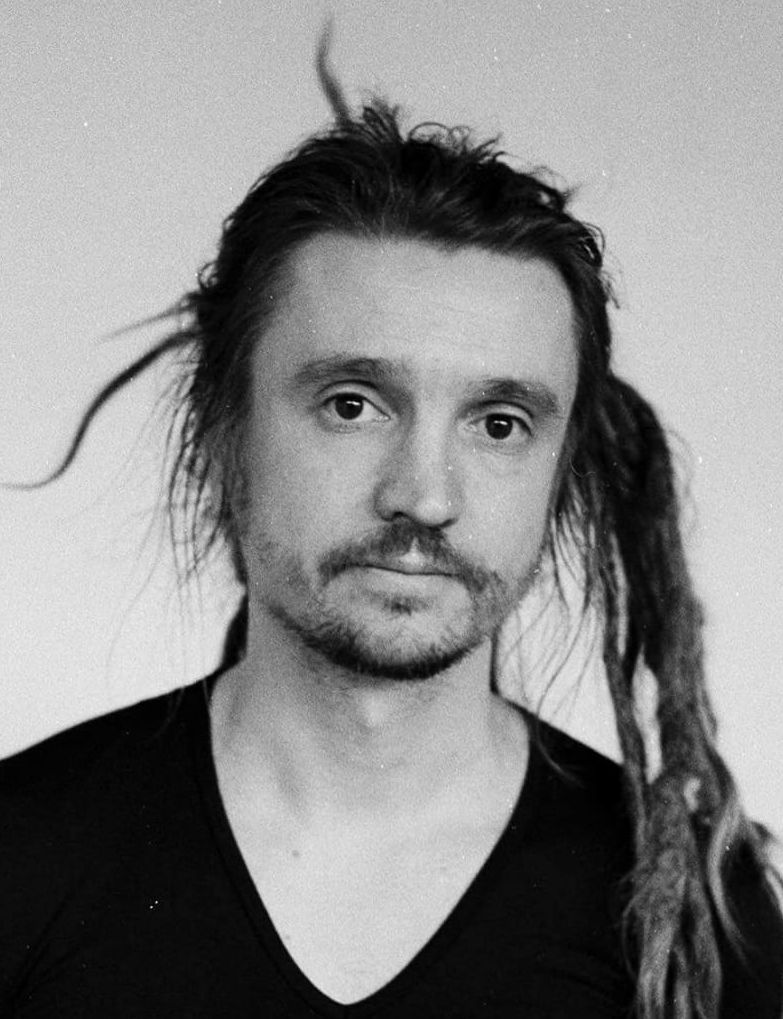}}]{Heinrich Mellmann} studied Computer Science and Mathematics at the Humboldt-Universit\"at zu Berlin (HUB), and received his degrees Dipl.-Inf. in 2011 and Dipl.-Math. in 2017 respectively. He is currently a research assistant at the chair of Adaptive Systems, HUB, pursuing a PhD in the area of cognitive robotics. Over the past years, he was involved in a number of research projects in cognitive robotics. Heinrich Mellmann has been actively involved in the RoboCup initiative since 2005, and has been leading the RoboCup team `Berlin United' since 2009. His research interests within cognitive robotics focus on spatial perception and decision making in autonomous humanoid robots.
\end{IEEEbiography}

\begin{IEEEbiography}[{\includegraphics[width=1in,height=1.25in,clip,keepaspectratio]{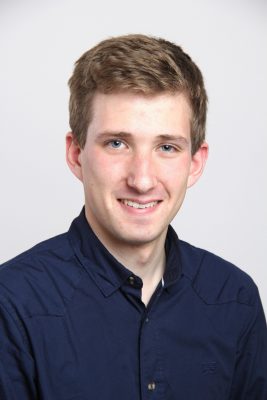}}]{Alexander Schmidt} (M'19) studied Electrical Engineering at Friedrich-Alexander-University Erlangen-N\"urnberg (FAU), Germany, and received his MSc degree in 2015. He is currently a research assistant at the Chair of Multimedia Communications and Signal Processing, FAU, pursuing a PhD in the area of multichannel signal enhancement for robot audition. He was with the EU FP7 Project ‘Embodied Audition for Robots (EARS)’. His special interests lie in the area of (sparse) dictionary learning for signal representation combined with physical-mechanical models.
\end{IEEEbiography}

\begin{IEEEbiography}[{\includegraphics[width=1in,height=1.25in,clip,keepaspectratio]{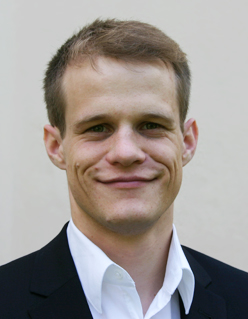}}]{Hendrik Barfuss}
received a Master degree from Friedrich-Alexander-University Erlangen-N\"urnberg (FAU), Germany, in 2013. He is currently pursuing a Dr.-Ing. degree in Electrical, Electronic and Communication Engineering at the Chair of Multimedia Communications and Signal Processing, FAU. His research interests are in microphone array signal processing, spatial filtering, speech enhancement, and localization.
\end{IEEEbiography}

\begin{IEEEbiography}[{\includegraphics[width=1in,height=1.25in,clip,keepaspectratio]{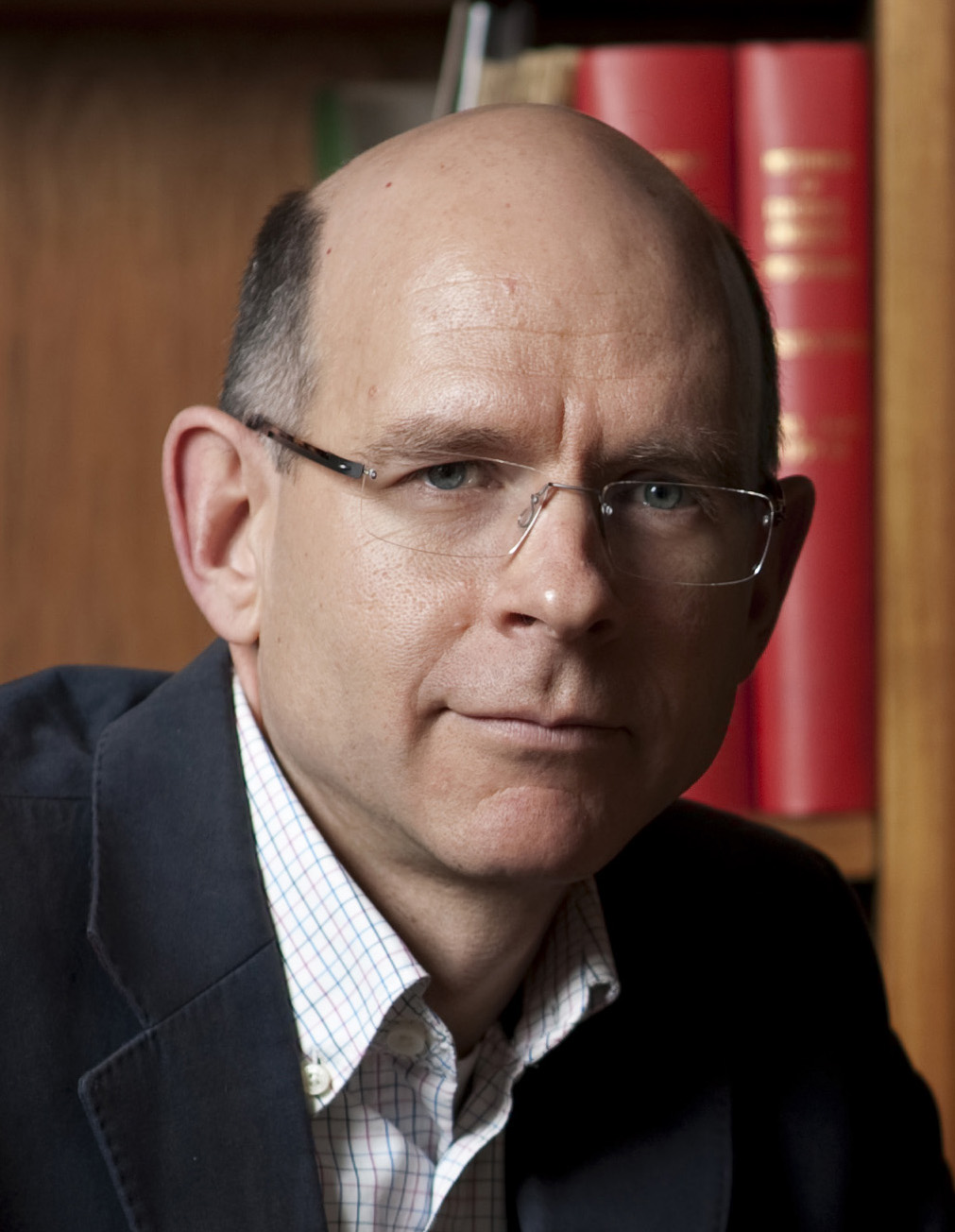}}]{Patrick A. Naylor} (M’89, SM'07, F’20) is Professor of Speech and Acoustic Signal Processing at Imperial College London. He received the BEng degree in Electronic and Electrical Engineering from the University of Sheffield, UK, and the PhD degree from Imperial College London, UK. His research interests are in speech, audio and acoustic signal processing. His current research addresses microphone array signal processing, speaker diarization, and multichannel speech enhancement. He has also worked on speech dereverberation including blind multichannel system identification and equalization, acoustic echo control, non-intrusive speech quality estimation, and speech production modelling with a focus on the analysis of the voice source signal. In addition to his academic research, he enjoys several collaborative links with industry. He is currently a member of the Board of Governors of the IEEE Signal Processing Society and President of the European Association for Signal Processing (EURASIP). He was formerly Chair of the IEEE Signal Processing Society Technical Committee on Audio and Acoustic Signal Processing. He has served as an associate editor of IEEE Signal Processing Letters and is currently a Senior Area Editor of IEEE Transactions on Audio Speech and Language Processing.
\end{IEEEbiography}

\begin{IEEEbiography}[{\includegraphics[width=1in,height=1.25in,clip,keepaspectratio]{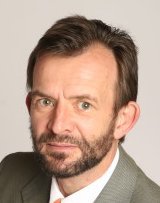}}]{Walter Kellermann}
is a professor for communications at the Friedrich-Alexander-University University Erlangen-N{\"u}rnberg (FAU), Germany, since 1999. He received the Dipl.-Ing. (univ.) degree in Electrical Engineering from FAU, in 1983, and the Dr.-Ing. degree from the Technical University Darmstadt, Germany, in 1988. From 1989 to 1990, he was a postdoctoral Member of Technical Staff at AT\&T Bell Laboratories, Murray Hill, NJ. In 1990, he joined Philips Kommunikations Industrie, Nuremberg, Germany, to work on hands-free communication in cars. From 1993 to 1999, he was a Professor at the Fachhochschule Regensburg, where he also became Director of the Institute of Applied Research in 1997. In 1999, he cofounded DSP Solutions, a consulting firm in digital signal processing, and he joined FAU as a Professor and Head of the Audio Research Laboratory. He authored or coauthored 21 book chapters, 300+ refereed papers in journals and conference proceedings, as well as 70+ patents, and is a co-recipient of ten best paper awards. His current research interests include speech signal processing, array signal processing, adaptive and learning algorithms, and its applications to acoustic human–machine interfaces. Dr. Kellermann served as an Associate Editor and  Guest Editor to various journals, including the IEEE Transactions on Speech and Audio Processing from 2000 to 2004, the  IEEE Signal Processing Magazine in 2015, and presently serves as Associate Editor to the EURASIP Journal on Applied Signal Processing. He was the General Chair of seven mostly IEEE-sponsored workshops and conferences. He served as a Distinguished Lecturer of the IEEE Signal Processing Society (SPS) from 2007 to 2008. He was the Chair of the IEEE SPS Technical Committee for Audio and Acoustic Signal Processing from 2008 to 2010, a Member of the IEEE James L. Flanagan Award Committee from 2011 to 2014, a Member of the SPS Board of Governors (2013-2015), Vice President Technical Directions of the IEEE Signal Processing Society (2016-2018) and is currently a member of the SPS Nominations Appointments Committee (2019-2020). He was awarded the Julius von Haast Fellowship by the Royal Society of New  Zealand in 2012 and the Group Technical Achievement Award of the European Association for Signal Processing (EURASIP) in 2015. In 2016, he was a Visiting Fellow at Australian National University, Canberra, Australia. He is an IEEE Fellow.
\end{IEEEbiography}

\end{document}